\documentclass[nofootinbib,showpacs,preprintnumbers,amsmath,amssymb,floatfix]
{revtex4}
\usepackage{epsf}

\begin{document}


\title{Overview of the spin
structure function $g_1$ at arbitrary $x$ and $Q^2$}

\vspace*{0.3 cm}

\author{B.I.~Ermolaev}
\affiliation{Ioffe Physico-Technical Institute, 194021
 St.Petersburg, Russia}
\author{M.~Greco}
\affiliation{Department of Physics and INFN, University Rome III,
Rome, Italy}
\author{S.I.~Troyan}
\affiliation{St.Petersburg Institute of Nuclear Physics, 188300
Gatchina, Russia}

\begin{abstract}
In the present paper we summarize our results on the structure function $g_1$
and present explicit expressions for the non-singlet and singlet
components of
$g_1$ which can be used at arbitrary $x$ and $Q^2$. These
expressions combine the well-known DGLAP-results for the anomalous
dimensions and coefficient functions with the total resummation of
the leading logarithmic contributions and the shift of $Q^2 \to
Q^2 +
 \mu^2$, with $\mu/\Lambda_{QCD} \approx 10$ $(\approx 55)$ for the non-singlet (singlet) components of
 $g_1$ respectively.
 In contrast to DGLAP, these expressions do not
require the introduction of   singular parameterizations  for the initial parton densities.
We also apply our results to describe the experimental data in the kinematic regions beyond the reach
of DGLAP.

\end{abstract}


\maketitle

\section{Introduction}

As it is well-known, the spin structure function $g_1$ is introduced through
the following conventional parametrization of the
spin-dependent part $W^{spin}_{\mu\nu}$ of the hadronic tensor of the Deep Inelastic
lepton- hadron Scattering:
\begin{equation}\label{wmunu}
W^{spin}_{\mu\nu} = \imath M_h \varepsilon_{\mu\nu\lambda\rho} \frac{q_{\lambda}}{Pq}
\Big[ S_{\rho}g_1(x,Q^2) + \Big(S_{\rho} -
P_{\rho}\frac{Sq}{q^2} \Big)g_2(x,Q^2)\Big]
 \end{equation}
where we have used the standard notations: $P$ is the hadron momentum, $M_h$ is the
hadron mass, $S$ is the hadron spin, $q$ is the virtual photon momentum. Traditionally,
$- q^2\equiv Q^2  > 0$ and $x = Q^2/2Pq$. The scalar
functions $g_1(x,Q^2)$ and $g_2(x,Q^2)$ are called the spin structure functions. Both of
them contribute to the asymmetry between the DIS cross sections when the lepton and hadron spins
are antiparallel and parallel. In particular, $g_1$ describes such asymmetry when  both spins
are longitudinal, i.e. they lie in the plane formed by $P$ and $q$. Obviously, in order to
calculate the structure functions
$g_1(x,Q^2)$ and $g_2(x,Q^2)$ in the framework of QCD, one should know the QCD behaviour at large and small
momenta of virtual particles, i.e. one should be able to account for the perturbative and non-perturbative
effects. At present this is impossible, so the standard description involves the factorization hypothesis:
$W^{spin}_{}\mu\nu$ is represented as a convolution:
\begin{equation}\label{wconv}
W^{spin}_{\mu\nu} = \widetilde{W}^{q}_{\mu\nu} \otimes \Psi_q
+ \widetilde{W}^{g}_{\mu\nu} \otimes \Psi_g
\end{equation}
where $\Psi_q$ and $\Psi_g$ are the probabilities  to find a polarized quark
or gluon in the polarized hadron, while $\widetilde{W}^{q,g}_{\mu\nu}$ describe the DIS of the
quarks and gluons.  There is no rigorous proof of this factorization, especially at small $x$, in the literature.
However, discussing this point is beyond the scope of our paper.
Also, there is no model-independent theoretical description of the probabilities $\Psi_q$ and $\Psi_g$
in the literature because QCD at small momenta is not known. On the contrary,
$\widetilde{W}^{q}_{\mu\nu}$ and $\widetilde{W}^{g}_{\mu\nu}$ can be calculated with the methods of
Pertubative QCD, by summing the contributions of the involved Feynman graphs.
So, the standard procedure is to replace  $\Psi_q$ and $\Psi_g$ by the initial parton densities
$\delta q$ and $\delta g$. Both of them are found by fitting the  experimental data at large $x$ and not
very large momenta $Q^2$ ($Q^2 = \mu^2 \sim 1~$ GeV$^2$ ).
Therefore,
\begin{equation}\label{gconv}
g_1 = g_1^{q} \otimes \delta q + g_1^{g} \otimes \delta g .
\end{equation}
It is well-known that in the Born approximation
\begin{equation}\label{gborn}
g_1^{Born} = (e^2_q/2) \delta(1-x)\otimes \delta q
\end{equation}
where $e_q$ is the electric charge of the quark interacting with the virtual photon. Accounting for the QCD
radiative corrections to $g_1^{Born}$ and other DIS structure functions, especially by trying to perform a complete
resummation of the
corrections, has been  the subject of great interest in recent years. Surely, such resummation cannot be performed precisely, so
 it would be important to  resum, in the first place, the most essential corrections. They
are different for different values of $x$ and $Q^2$. For example for describing $g_1$ in
the region of $x \sim 1$ and large $Q^2$, the contributions $\sim \ln^k x$ are negligibly small
compared to $\ln^k (Q^2/\mu^2)$. In contrast, $\ln^k x$ becomes quite important at $x \ll 1$ and
should be accounted for.

The goal of obtaining an universal description of the structure function
$g_1$, which could be used for arbitrary $x$ and arbitrary $Q^2$
would be appealing both for theorists and experimentalists. Most generally, one encounters
various kinematic regions where the DIS structure functions have been thoroughly studied.
The first kinematic region  is the so-called hard  region \textbf{A} of large $x$ and large
$Q^2$:
\begin{equation}\label{rega}
\textbf{A:}~~~~~ w \gtrsim Q^2 \gg \mu^2,~~~~~x \lesssim 1
\end{equation}
where $w = 2pq$ and $\mu^2$ is the starting point of the $Q^2$ -evolution. Usually, the value of
$\mu$ is chosen $\approx 1$~GeV or so. Through the paper we use the standard notations: $q$ is
the virtual photon momentum, $p$ is the initial parton momentum, and  $x= Q^2/w$.
The region\textbf{ A} was described first by  the LO DGLAP
evolution equations obtained in Ref.~\cite{dglap}:
\begin{eqnarray}\label{dglap}
&&\frac{d \Delta q}{d t} = P_{qq} \otimes \Delta q + P_{qg}
\otimes \Delta g~, \\ \nonumber &&\frac{d \Delta g}{d t} = P_{gq}
\otimes \Delta q + P_{gg} \otimes \Delta g
\end{eqnarray}
where we have used the standard notation $t = \ln(Q^2/\mu^2)$ and $P_{ik}$ (with
$i,k = q,g$) are the splitting
functions. $\Delta q$ and $\Delta g$ are the evolved (with respect to $Q^2$) parton
distributions. The splitting functions $P_{ik}$ in the DGLAP evolution equations (\ref{dglap}) include
the QCD coupling $\alpha_s$. In order to account for the running $\alpha_s$ -effects,
one should define the argument of $\alpha_s$. The DGLAP -prescription is
\begin{equation}\label{alphadglap}
\alpha_s = \alpha_s(Q^2)~.
\end{equation}
 Through the paper we will address the parametrization of $\alpha_s$ in 
Eq.~(\ref{alphadglap}) as the DGLAP -parametrization.
The DGLAP Eqs.~(\ref{dglap}) describe the $Q^2$-evolution of the parton distributions
from $Q^2 = \mu^2$, with $\mu \sim 1$ GeV, to larger $Q^2$.
When general solutions to  Eqs.~(\ref{dglap}) are obtained, one needs to specify appropriate
 initial conditions. Conventionally, the initial conditions to Eqs.~(\ref{dglap}) are
\begin{equation}\label{ipd}
\Delta q|_{t = 0} = \delta q~,\qquad \Delta g|_{t = 0} = \delta
g~,
\end{equation}
with $\delta q, \delta g$ being called the initial parton
densities. They are found by fitting the experimental data
.
After $\Delta q $ and $\Delta q $ have been fixed, the
DIS structure functions, including $g_1$, are found by convoluting
them with the coefficient functions $C_q,~C_g$:
\begin{equation}\label{convg1}
g_1(x,Q^2) = C_q (x/y)\otimes \Delta q (y,Q^2)+ C_g (x/y) \otimes
\Delta g (y,Q^2)~.
\end{equation}
The Mellin transformations of $P_{ik}$ are called the anomalous dimensions.
The splitting functions $P_{ik}$ and coefficient functions $C_k$ for the unpolarized DIS
were calculated with LO accuracy in Ref.~\cite{dglap}.
The LO expressions for $P_{ik}$ and $C_k$ for the polarized DIS were obtained in Ref.~ \cite{ar}.
Later, the LO expressions of Refs~\cite{dglap,ar} for $P_{ik}$ and $C_k$ were complemented by the NLO results\cite{nlodglap}.
A detailed review on that subject can be found in Ref.~\cite{ael}.
From pure theoretical grounds, this approach should  not be used outside the region \textbf{A}.
However, introducing special fits\cite{fitsa, fits} for the initial parton densities DGLAP has been
extended  to the region \textbf{B} of large $Q^2$ and small $x$:
\begin{equation}\label{regb}
\textbf{B:} \qquad w \gg Q^2 \gg \mu^2~,\qquad x \ll 1~.
\end{equation}
Indeed the parameterizations   for $\delta q,~\delta g$ of Ref.~\cite{fitsa,fits} contain singular factors $x^{-a}$, and
 used in Eq.~(\ref{ipd}), they provide $g_1$ with a fast growth at small $x$.
As a result, combining the LO  evolution equations of Ref.~\cite{dglap} and NLO DGLAP results
of Ref.~\cite{nlodglap} with the standard fits of Ref.~\cite{fitsa,fits} it has been possible
to describe the available experimental data on $g_1$
in regions \textbf{A} and \textbf{B}, i.e. for large $Q^2$ and arbitrary $x$.
In the present paper we refer to this  as the Standard Approach (SA).  In
addition to regions \textbf{A} and \textbf{B}, there are two more interesting
kinematic regions:
 \begin{equation}\label{regc}
\textbf{C:} \qquad 0 \leq Q^2 \lesssim \mu^2~,\qquad x \ll 1~,
\end{equation}
\begin{equation}\label{regd}
\textbf{D:} \qquad 0 \leq Q^2 \lesssim \mu^2~,\qquad x \lesssim
1~.
\end{equation}
Besides a purely theoretical interest, the  knowledge of
$g_1$ in the regions \textbf{C} and \textbf{D} is needed because they correspond
to the kinematic region investigated experimentally
by the COMPASS collaboration. Obviously, the regions \textbf{C} and \textbf{D}
are beyond the reach of SA. Strictly speaking, the same could be said  about
the region \textbf{B}: In fact the expressions for $P_{ik}$
are obtained  (see Refs.~\cite{dglap,nlodglap} for detail) under the assumption of the ordering
\begin{equation}\label{dglapord}
\mu^2 < k^2_{1~\perp}< k^2_{2~\perp} <... < Q^2
\end{equation}
where $k_{i~\perp}$ are the transverse momenta of virtual ladder
partons and they are numbered from the bottom of the ladders to
the top. Once this ordering is kept one is led inevitably to  neglect the
double-logarithmic (DL) contributions $\sim \alpha_s \ln^2(1/x)$
and other contributions independent of $Q^2$. Such contributions
are small in the region \textbf{A} where they are  correctly
neglected in the SA. However, they become essential in the region
\textbf{B}.
  In order to
account for them, the DGLAP-ordering of Eq.~(\ref{dglapord})
should be replaced by the other ordering:
\begin{equation}\label{dlord}
\mu^2 < \frac{k^2_{1~\perp}}{\beta_1}< \frac{k^2_{2~\perp}}{\beta_2} <... < w
\end{equation}
where $\beta_j$ are the longitudinal Sudakov variables
\footnote{Sudakov variables were introduced in Ref.~\cite{sud}} for the virtual
parton momenta $k_j$  as follows:
\begin{equation}\label{sud}
k_j = -\alpha_j (q + x p) + \beta_j p + k_{j\perp}~.
\end{equation}
In order to account for such logarithmic contributions,
Eq.~(\ref{dlord}) should be implemented by the ordering for $\beta_i$:
\begin{equation}\label{dlordbeta}
1 > \beta_1 > \beta_2 >..> \mu^2/w~.
\end{equation}
This ordering does not exist in  DGLAP because in this approach
 $\beta_i \geq x \sim 1$.
The ordering (\ref{dlord},\ref{dlordbeta}) was first introduced  in Ref.~\cite{ggfl} in the
 context of QED but it applies  in QCD as well.  Replacing the ordering (\ref{dglapord})
by Eqs.~(\ref{dlord}, \ref{dlordbeta}) makes possible to sum up all DL contributions, regardless of their
argument, to all orders in $\alpha_s$, i.e. to perform calculations in the double-logarithmic approximation
(DLA). Explicit expressions for $g_1$ in DLA were obtained in Ref.~\cite{ber}.
The drawback of those expressions is that $\alpha_s$ is kept fixed at
an unknown scale. The effect of running $\alpha_s$  were taken into account in Ref.~\cite{egtsns}.
The parametrization of $\alpha_s$ in Refs.~\cite{egtsns} differs from the DGLAP-
parametrization. The theoretical grounds for this new parametrization were given in
Ref.~\cite{egtalpha} and a numerical comparison with the standard parameterizations
can be found in Ref.~\cite{kotl2}. On the other hand,  the reason why, in spite of the
lack of  the resummation
of quite important contributions, the SA turned out to be working well in the region \textbf{B}
remained unclear until in Refs.~\cite{egtinp,egtsing} we proved that  the factors $x^{-a}$
in the DGLAP-fits for the initial parton densities mimic the total resummation of the
leading logarithms of $x$.  Besides, in Ref.~\cite{egtinp} we suggested to combine the DGLAP-
results for $g_1$ with the results of Ref.~\cite{egtsns} in order to obtain an unified
description of $g_1$ in the regions\textbf{ A} and \textbf{B},  without singular  initial parton densities.
A prescription for extending  $g_1$ into regions \textbf{C} and \textbf{D} was given in Ref.~\cite{egtsmallq}.

In the present paper we present a unified description of $g_1$ valid in all of the regions
\textbf{A-D}
.    The paper is organized as follows:  in Sect.~II we briefly remind the
DGLAP-description of $g_1$. For the sake of simplicity we consider in more detail,
throughout the paper,  the non-singlet
component of $g_1$ , and summarize the singlet results only. As the expressions
for $g_1$ involve convolutions, they look simpler when an integral transform has been
applied. The conventionally used transform is the Mellin one. However, in the small-$x,$
 region, it is more convenient the use of the Sommerfeld-Watson transform, whose asymptotics
 partly coincides with the Mellin transform. This formalism is the content of Sect.~III.
Before dealing explicitly with $g_1$, we consider in Sect.~IV the appropriate treatment of the QCD coupling
 and compare it with the DGLAP-parametrization. The total resummation of the
leading logarithms of $x$ is quite essential in the small-$x$ region \textbf{B}.
We discuss it in Sect.~V by composing and solving appropriate Infrared Evolution Equations (IREE).
Such equations involve new anomalous dimensions and coefficient functions and contain
the total resummation of the leading logarithms of $x$. The singlet and non-singlet
anomalous dimensions are calculated in Sect.~VI. The non-singlet coefficient
function is obtained in Sect.~VII and is used to write down the explicit expression for
the non-singlet $g_1$ in the region\textbf{ B}.  The singlet $g_1$ in the region \textbf{B} is
obtained in Sect.~VIII. In Sect.~IX  the small-$x$ asymptotics of the non-singlet $g_1$ is discussed,
whereas the singlet asymptotics is considered in Sect.~X. Both asymptotic results are of the
Regge type and their intercepts are found not so small. This may lead to the wrong conclusion
that the IREE method cannot be applied safely to $g_1$. In order to make this point clear,
 we discuss the applicability of our method in Sect.~XI. In Sect.~XII
we compare our results for $g_1$ in the region\textbf{ B} to the DGLAP expressions.
We show that DGLAP works well in the region\textbf{ B} only because of the singular factors
present in the parameterizations for the initial parton densities. On the other hand, when such fits
are used, $g_1^{DGLAP}$  also behaves asymptotically as a  sum of Reggeon contributions.
The Regge behaviour in the two approaches is discussed in Sect.~XIII. Combining the total resummation of the logarithms with
the DGLAP results, we give in Sec.~XIV  the interpolation expressions
describing $g_1$ in the unified region $\textbf{A}\oplus \textbf{B}$. Furthermore we show in Sects.~XV
and XVI
how it is possible to describe $g_1$ in the small-$Q^2$ regions \textbf{C} and
\textbf{D} and arrive thereby to the interpolation expressions for $g_1$ which can
be used in the whole region
$\textbf{A} \oplus \textbf{B}\oplus \textbf{C}\oplus \textbf{D}$. In particular
 the small-$Q^2$ regions \textbf{C} and \textbf{D} are described by a
 shift of $Q^2$. Such shift is a source of new power $Q^2$- corrections and we
discuss them in Sect.~XVII. Due to the experimental investigation of the singlet $g_1$
by the COMPASS collaboration, in Sect.~XVIII we give an interpretation to the
recent COMPASS data. Finally Sect.~XIX contains our concluding remarks.

\section{DGLAP -expressions for $g_1$}

The Standard Approach to $g_1$ is based on the DGLAP evolution
equations and also involves some standard parameterizations for the initial
parton densities $\delta q$ and $\delta g$. As the notations for
the anomalous dimensions, the coefficient functions and the fits for the parton
densities
 vary widely  in the literature,  we explain
below the notation we use trough the present paper.

We will  denote $g_1^{NS~DGLAP}$ and $g_1^{S~DGLAP}$
 the non-singlet and singlet parts of $g_1$ when the SA is invoked.
As the expressions for $g_1$ involve convolutions, it is convenient to
write them down in the Mellin integral form. In particular, the
non-singlet $g_1$ is:
\begin{eqnarray}\label{g1nsdglap}
g_1^{NS~DGLAP}(x, Q^2) = (e^2_q/2) \int_{-\imath \infty}^{\imath
\infty} \frac{d \omega}{2\imath \pi}(1/x)^{\omega}
C^{NS~DGLAP}(\omega,\alpha_s(Q^2)) \delta q(\omega) \\ \nonumber
\exp
\Big[\int_{\mu^2}^{Q^2} \frac{d k^2_{\perp}}{k^2_{\perp}}
\gamma^{NS~DGLAP}(\omega, \alpha_s(k^2_{\perp}))\Big]
\end{eqnarray}
where $C^{NS~DGLAP}(\omega,\alpha_s(Q^2))$ is the non-singlet
coefficient function, $\gamma^{NS~DGLAP}(\omega,\alpha_s(Q^2))$ is
the non-singlet anomalous dimension  and $\delta q(\omega)$ is the
initial quark density in the Mellin (momentum) space. With the
one-loop accuracy (NLO) (see e.g. Ref.~\cite{nlodglap}), the
expression for $C^{NS~DGLAP}$ is

\begin{equation}\label{cnsdglap}
C^{NS~DGLAP}= C^{NS~DGLAP}_{LO}+~ \frac{\alpha_s(Q^2)}{2\pi}~
C^{NS~DGLAP}_{NLO},
\end{equation}
with
\begin{equation}\label{cnsnlo}
C^{NS~DGLAP}_{LO} = 1,~~ C^{NS~DGLAP}_{NLO}= C_F
\Big[\frac{1}{n^2} + \frac{1}{2n}+\frac{1}{2n+1}-\frac{9}{2}  +
\Big(\frac{3}{2} - \frac{1}{n(1+n)}\Big)S_1(n) + S^2_1(n)-
S_2(n)\Big].
\end{equation}
Similarly, with two-loop accuracy,
\begin{equation}\label{gnsdglap}
\gamma^{NS~DGLAP}   = \frac{\alpha_s(Q^2)}{2\pi}\gamma^{(0)}(n)  +
\Big(\frac{\alpha_s(Q^2)}{2\pi}\Big)^2\gamma^{(1)}(n)
\end{equation}
where
\begin{equation}\label{gnslo}
\gamma^{(0)}(n) = C_F\Big[\frac{1}{n(1+n)}  +
\frac{3}{2}-S_2(n)\Big].
\end{equation}
We have used the standard notations $S_{1,2}$ in
Eqs.~(\ref{cnsnlo},\ref{gnslo}):
\begin{equation}\label{s12}
S_1(n) =  \sum_{j=1}^{j=n}  \frac{1}{j}~,~~S_2(n) =
\sum_{j=1}^{j=n} \frac{1}{j^2}~.
\end{equation}
They  are defined for integer $n$. Their generalization for
arbitrary $n$ is well-known:
\begin{equation}\label{s12psi}
S_1(n) = \textbf{C}+  \psi(n-1),~~S_2(n-1)=  \frac{\pi^2}{6}+
\psi'(n)~,
\end{equation}
with $\textbf{C}$ being  the Euler constant.
The standard fits for the initial parton densities include the normalization constants
$N_{q,g}$, the power factors $x^{-a}$, with $a > 0$ and more complicated structures; for
example,
\begin{equation}\label{fita}
\delta q (x) = N_q x^{-\alpha}(1-x)^{\beta} (1 + \gamma x^{\delta}) \equiv
N_q x^{-\alpha} \varphi(x).
\end{equation}
All parameters $N_q,~\alpha,~\beta,~\gamma,~\delta$ in Eq.~(\ref{fita}) are fixed by fitting the experimental
data at large $x$ and $Q^2 \approx 1~$ GeV$^2$.

The expressions for $g_1^{S~DGLAP}$ are similar but more involved and we do not discuss them in
detail in the present paper. For the sake of simplicity through the paper we use $g_1$ non-singlet
for illustration, when it is possible, but we present the final expressions for both the non-singlet
and singlet explicitly.

\section{Sommerfeld-Watson transform}

 As it is well known, the DGLAP expressions for $g_1$ involve  convolutions and in our approach we use them too.
 The standard way  is to use an appropriate integral transform. Traditionally, the SA uses
    the Mellin transform. We will proceed slight differently. Our goal is to obtain expressions for $g_1$ at small $x$ and  will start by
    considering the spin-dependent forward Compton amplitude $T_{\mu\nu}$
 related to $W^{spin}_{\mu\nu}$ as follows:
 \begin{equation}\label{wt}
 W^{spin}_{\mu\nu} = \frac{1}{2 \pi} \Im T_{\mu\nu}
 \end{equation}
 where the symbol $\Im$ means the discontinuity (imaginary part) of $T_{\mu\nu}$
 with respect to the invariant total energy $s = (p+q)^2$) of the Compton scattering.
 At large $s$, when hadron masses can be neglected,
 \begin{equation}\label{ws}
 s \approx 2pq (1-x) \equiv w (1-x)~,
 \end{equation}
 so $s \approx w$ at small $x$. The amplitude $T_{\mu\nu}$ can be parameterized similarly to
  $W^{spin}_{\mu\nu}$:
  \begin{equation}\label{tmunu}
T_{\mu\nu} = \imath M_h \varepsilon_{\mu\nu\lambda\rho} \frac{q_{\lambda}}{pq}
\Big[ S_{\rho}T_1(x,Q^2) + \Big(S_{\rho} -
p_{\rho}\frac{Sq}{q^2} \Big)T_2(x,Q^2)\Big]
 \end{equation}
 so that
 \begin{equation}\label{tg}
 g_1 = \frac{1}{2\pi} \Im T_1~,\qquad g_2 = \frac{1}{2\pi} \Im T_2~.
 \end{equation}
 We call $T_{1,2}$ the invariant amplitudes. Exploiting the factorization, $T_{1,2}$ can be represented as the convolution
 of the perturbative and non-perturbative contributions (cf. Eq.~(\ref{gconv})). In particular,
 \begin{equation}\label{tconv}
 T_1 = T_q \otimes \widetilde{\delta} q +T_g \otimes \widetilde{\delta} g
 \end{equation}
 where $\widetilde{\delta} q$ and $\widetilde{\delta} g$ are related to $\delta q$ and $\delta g$ through Eq.~(\ref{tg}).
 In the Born approximation (cf. Eq.~(\ref{gborn})),
 \begin{equation}\label{tborn}
 T_q^{Born} = e^2_q\, \frac{s}{w - Q^2 + \imath \epsilon}~,\qquad T^{Born}_g = 0~.
 \end{equation}
From the mathematical point of view, Eq.~(\ref{tconv}) as well as the DGLAP equations (\ref{dglap})
and the expressions
(\ref{convg1}) for $g_1$ are convolutions, so an appropriate integral transform can be used.
 On the other hand, the
phenomenological Regge theory (see e.g. \cite{col}) states that in order to study accurately
the scattering amplitudes at high energies, one should use
the Sommerfeld-Watson (SW) transform\cite{sw}.
The asymptotic form of the SW transform partly coincides with the Mellin transform and  often
this form is especially convenient to  account for  the logarithmic
radiative corrections. The SW transform is actually related to
 the signature invariant amplitudes $T^{(\pm)}$ defined, in the context of DIS, as follows:
\begin{equation}\label{tsign}
T^{(+)} =  \frac12 [T(s, Q^2) + T(-s, Q^2)]~,\qquad T^{(-)} =
\frac12 [T(s, Q^2) - T(-s, Q^2)]
\end{equation}
so that
\begin{equation}\label{tsigninv}
T(s, Q^2) = T^{(+)} + T^{(-)}~,\qquad T(-s, Q^2) = T^{(+)} -
T^{(-)}.
\end{equation}
Let us demonstrate that the signature of the Compton invariant amplitude $T_1$ in Eq.~(\ref{tmunu}) is negative.
Using Eq.~(\ref{tsigninv}), we can represent $T_1$ in Eq.~(\ref{tmunu}) as the sum of the signature
amplitudes $T_1^{(\pm)}$.
In order to satisfy the Bose statistics, $T_{\mu\nu}$  should be invariant to the
permutation of the incoming and outgoing photons in Eq.~(\ref{tmunu}), i.e. to the replacement combining
 $\mu \rightleftharpoons \nu$ and $q \leftrightarrows -q$.
 On the other hand, in the limit of large $s$, where the SW transform makes sense, the proton spin remains
 unchanged under such replacement because $S_{\rho} \approx P_{\rho}/M$ whereas $s \approx 2pq \to -s$.
 It immediately allows one to conclude that the amplitude $T_1^{(+)}$ should not be present in Eq.~(\ref{tmunu}).
 Therefore,

\begin{equation}\label{tsigng}
g_1(x, Q^2) = \frac{1}{\pi}\; \Im T^{(-)}_1(s, Q^2)~.
\end{equation}

The Compton amplitudes with the positive signature contribute to the structure
functions $F_1$ and  $F_2$ describing the unpolarized DIS. The calculation of  the
non-singlet component of $F_1$ and the non-singlet $g_1$ is quite similar, so in
the present paper we consider $g_1^{NS}$ in detail and give
the results for $F^{NS}_1$ in Appendix \textbf{A}.
In order to  account for the logarithmic contributions it is convenient to use the asymptotic
SW transform for amplitudes $T^{(\pm)}$ in the following form:
\begin{equation}
\label{mellintf} T^{(\pm)} = \int_{-\imath \infty +
\delta}^{\imath \infty + \delta}\frac{d\omega}{2\pi \imath}
\Big(\frac{s}{\mu^2}\Big)^{\omega} \xi^{(\pm)}(\omega)
\,F^{(\pm)}(\omega, y)
\end{equation}
where  $y = \ln(Q^2/\mu^2)$ and $\xi$ are the signature factors:
\begin{equation}\label{signfact}
\xi^{(\pm)} = -[e^{-\imath \pi \omega} \pm 1]/2 \approx [1 \pm 1 +
\imath \pi \omega] /2~.
\end{equation}
The integration line in Eq.~(\ref{mellintf}) runs parallel to $\Im \omega$ and $\delta$
should be larger than the rightmost singularity of $F^{(\pm)}(\omega, y)$. Quite often
in the literature $\delta$ in Eq.~(\ref{mellintf}) is dropped.
In  the phenomenological Regge theory, the mass scale
$\mu$ in Eq.~(\ref{mellintf}) should obey $\mu^2 \ll s$, otherwise it is arbitrary.
 We are going to specify it later in the context of $g_1$.
The integration contour in Eq.~(\ref{mellintf}), which includes the line parallel to the imaginary
 $\omega$-axis, as stated above,  must be closed up to the left.Then  the
 contour includes all $\omega$-singularities of $F^{(\pm)}(\omega, y)$.
 As Eq.~(\ref{mellintf}) partly coincides with the
standard Mellin transform, it is often addressed as the Mellin
transform and we will do the same through the paper.
Nevertheless, we will use the inverse transform to
Eq.~(\ref{mellintf}) in its proper form:
\begin{equation}\label{invmellin}
F^{(\pm)}(\omega, y) =  \frac{2}{\pi \omega} \int_{0}^{\infty} d
\rho e^{-\omega \rho}\, \Im  T^{(\pm)}(s/\mu^2,y)
\end{equation}
where we have denoted $\rho  = \ln(s/\mu^2)$~.
Eqs.~(\ref{invmellin}) and ~(\ref{mellin}) are supposed to be used
at large $s$ ($s \gg \mu^2$) where the bulk of the integrals comes
from the region of small $\omega$ ($\omega \ll 1$). Obviously,
Eq.~(\ref{invmellin}) does not coincide with  the standard Mellin
transform. Finally Eqs.~(\ref{wt},\ref{tmunu},\ref{tsign}) lead to
\begin{equation}
\label{gf} g_1= \frac{1}{2}\int_{-\imath \infty }^{\imath
\infty}\frac{d\omega}{2\pi \imath}
\Big(\frac{s}{\mu^2}\Big)^{\omega} \omega F^{(-)}(\omega, y)~.
\end{equation}

\section{Treatment of $\alpha_s$  at large and small $x$}

The rigorous knowledge on $\alpha_s$ is provided by the renormalization group equation (RGE).
According to it, the total resummation of the leading radiative corrections
to the Born value of $\alpha_s$ leads to the well-known expression
\begin{equation}\label{alphas}
\alpha_s = \frac{1}{b \ln(-s/\Lambda^2)}
\end{equation}
where
$\Lambda \equiv \Lambda_{QCD}$
and $b= (11N-2n_f)/12 \pi$, with $N=3$ and $n_f$
being the number of involved flavors. Eq.~(\ref{alphas}) is the asymptotic
expression valid at $|s| \gg \Lambda^2$. The value of $s$ in Eq.~(\ref{alphas}) is negative.
Eq.~(\ref{alphas}) is often addressed as the
leading order expression for $\alpha_s$ and is obtained with the total resummation
of the leading, single-logarithmic contributions. Corrections to  Eq.~(\ref{alphas})
are also available
in the literature\footnote{For   recent progress
in RGE see e.g. the review \cite{prosp}.} but we will not use them in the present paper
because in practice the accuracy of the total resummations of the other
radiative corrections, usually accounted for
with various evolution equations, never exceeds the single-logarithmic accuracy.
The minus sign at $s$ in Eq.~(\ref{alphas})is related to  the analyticity: $\alpha_s(s)$
should be real at negative $s$, but when $s$ is positive, $\alpha_s(s)$ acquires an imaginary
part. Conventionally, $\alpha_s (s)$ at positive $s$ is understood as the value of $\alpha_s$
on the upper side of the $s$ -cut. Therefore, $-s = s \exp (- \imath \pi)$ and
\begin{equation}\label{alphaborn}
\alpha_s (s) = \frac{1}{b} \frac{1}{[\ln (s/\Lambda^2) - \imath \pi]} =
\frac{1}{b} \Big(\frac{\ln (s/\Lambda^2) + \imath \pi}{\ln^2 (s/\Lambda^2) + \pi^2}\Big).
\end{equation}
Expressions (\ref{alphas},\ref{alphaborn}) are perturbative and asymptotic. In order to
be consistent with the applicability  of the perturbative QCD, $\mu$ defined
in Eqs.~(\ref{dglapord},\ref{dlord}) should be large enough:
\begin{equation}\label{mulambda}
\mu \gg \Lambda.
\end{equation}
An alternative way is to modify Eq.~(\ref{alphas}) in order to be able to
investigate $\alpha_s$ at $s \lesssim \Lambda^2$.
For example, there is the so called Analytic Perturbation Theory
(APT) suggested in Ref.~\cite{apt1}. It is based on subtracting from Eq.~(\ref{alphas})
its pole contribution at $s = - \Lambda^2$. In the vicinity of the pole
\begin{equation}\label{alphapole}
\alpha_s (s) = \frac{1}{b \ln\Big((\Lambda^2 +|s|- \Lambda^2)/\Lambda^2\Big)}
\approx \frac{1}{b} \Big[\frac{\Lambda^2}{|s| - \Lambda^2} + \frac{1}{2}\Big]
+ \emph{O}\big(|s| - \Lambda^2\big).
\end{equation}
The result of the subtraction is called the effective
coupling and  is used
instead of $\alpha_s$. Such a coupling can be used at any value of $s$.
The recent results in this approach can be found in Refs.~\cite{apt2}.
However, APT does not allow one to get rid of the cut-off $\mu$
when the Sudakov contributions
of the higher-loop Feynman graphs  are involved. So, in the present paper we do not follow this approach.

\begin{figure}[h]
\begin{center}
\begin{picture}(160,120)
\put(0,0){ \epsfbox{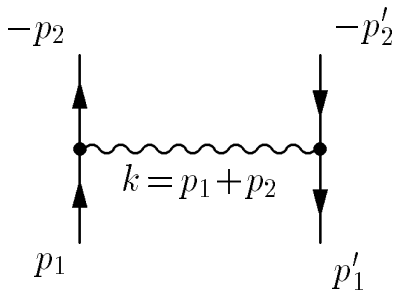} }
\end{picture}
\end{center}
\caption{\label{g1sumfig1} The Born amplitude $A_{Born}$.}
\end{figure}

Now let us discuss how  $\alpha_s$ is incorporated into the expressions for
the amplitude $A$ of the forward annihilation of the quark-antiquark pair into another pair.
The generalization to the scattering of gluons can be
 obtained easily. We assume that the  external quarks are almost on-shell, with virtualities
 $\sim \mu^2$, keeping $\mu^2 \ll s$.
 In the Born approximation, the amplitude $A_{Born}$ is given by the following expression (see Fig.~\ref{g1sumfig1}):
 \begin{equation}\label{aborn}
 A_{Born} = - 4\pi\alpha_s C^{(col)} \frac{\bar{u}(-p_2) \gamma_{\mu}u(p_1)
 \bar{u'}(p_1) \gamma_{\mu}u'(-p_2)}{s + \imath \epsilon} \equiv
\frac{ \bar{u}(-p_2) \gamma_{\mu}u(p_1)
 \bar{u'}(p_1) \gamma_{\mu}u'(-p_2)}{s} M_f^{Born}(s)
 \end{equation}
where $s= (p_1+p_2)^2$. In Eq.~(\ref{aborn}) and through the paper we use the Feynman gauge
 for intermediate gluons.
According to Appendix A, the quark color factor $C^{(col)} = C_F =
(N^2 - 1)/2N$ for the $t$ -channel color singlet and $C^{(col)}  =
-1/2N$ for the vector (octet) representation. Through the paper we
mostly discuss the color singlet amplitude. We address
$M^{Born}(s)$ as the invariant amplitude for this process in the
Born approximation:
\begin{equation}\label{mborn}
M^{Born}(s) = - 4\pi \alpha_s C^{(col)} \frac{s}{s + \imath \epsilon}~.
\end{equation}

\begin{figure}[h]
\begin{center}
\begin{picture}(125,160)
\put(0,0){ \epsfbox{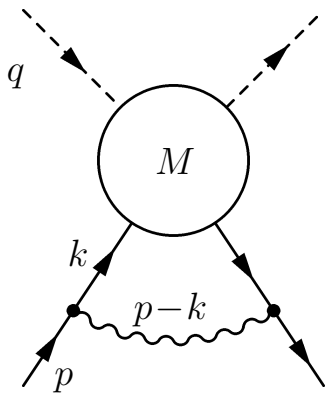} }
\end{picture}
\end{center}
\caption{\label{g1sumfig2} The integral contribution in
Eq.~(\ref{bseq}). The $w$ -cut is implied, though is not shown
explicitly.}
\end{figure}

By definition, $\alpha_s$ in the Born approximation is a constant.
The radiative correction to $M_f^{Born}$ can be divided into two groups: \\
 \textbf{(i)} The corrections contributing to $\alpha_s$. \\
 \textbf{(ii)} The other corrections. \\
 Leaving the corrections to \textbf{(ii)} for the next Sects. we consider now
 the effect of   \textbf{(i)}. They transform the fixed $\alpha_s$
  in Eqs.~(\ref{aborn},\ref{mborn}) into the running coupling. It is possible
  to fix the argument of $\alpha_s$, using the arguments of Ref.~\cite{blm}. Incorporating the
 radiative corrections from \textbf{(i)} to $A_{Born}$ leads, in
 particular, to insert  the quark bubbles into the (horizontal) propagator of the
 intermediate gluon.
 In the logarithmic approximation, each quark bubble brings the
 contribution $\sim n_f \ln s$.
The gluon logarithmic contributions, each $\sim N$,  come from more involved graphs but eventually
all contributions lead to the factor $b = (11 N - 2n_f)/(12 \pi)$ which multiplies  the overall logarithm. Obviously,
the argument of this logarithm  coincides with the argument of the logarithm from the
fermion bubble contribution and  it is $s$. The total resummation of the
leading radiative corrections from group \textbf{(i)} converts the fixed $\alpha_s$ of Eq.~(\ref{aborn})
into the well-known expression of Eq.~(\ref{alphas}) and therefore converts the Born invariant
amplitude $M^{Born}$ of Eq.~(\ref{mborn}) into $M^{(0)}$:
\begin{equation}\label{m0}
M^{(0)}(s) = - 4\pi \alpha_s(s) C^{(col)} \frac{s}{s + \imath \epsilon}~.
\end{equation}
In order to apply the Mellin transform  to $M^{(0)}$, we allow for the shift
$s \to s - \mu^2$, with $s \gg \mu^2$, in Eq.~(\ref{m0}). Then we can write
\begin{equation}\label{m0f}
M^{(0)}(s) =
\int_{-\imath \infty }^{\imath
\infty}\frac{d\omega}{2\pi \imath}
\Big(\frac{s}{\mu^2}\Big)^{\omega}  F^{(0)} (\omega),
\end{equation}
with
\begin{equation}\label{f0}
F^{(0)} (\omega) = 4 \pi C^{(col)} \frac{A(\omega)}{\omega}
\end{equation}
where $A(\omega)$ corresponds to $\alpha_s(s)$ in the $\omega$ -space
:
\begin{equation}
\label{a} A(\omega) = \frac{1}{b} \Big[\frac{\eta}{\eta^2 + \pi^2}
- \int_0^{\infty} \frac{d \rho e^{-\omega \rho}}{(\rho + \eta)^2 +
\pi^2} \Big].
\end{equation}
In Eq.~(\ref{a}) we have denoted $\eta = \ln(\mu^2/\Lambda^2_{QCD})$.
 The first term in Eq.~(\ref{a})
corresponds to the cut of the bare gluon propagator while the second term comes
from the cut of $\alpha_s (s)$. They have opposite signs because of the famous
anti-screening in QCD, which is the basis of the asymptotic freedom for $\alpha_s$.
In the literature $M_f^{(0)}$ and $F_f^{(0)}$ are
often called  Born amplitudes (and we also follow this tradition ) in spite of the
fact that they include the leading radiative correction from group \textbf{(i)}.
 The radiative
corrections to $M_f^{(0)}$, i.e. the corrections from the group (\textbf{ii}),
are often included  by using evolution equations. Such
equations involve convolutions of $M_f^{(0)}$, so they look simpler in
the $\omega$ -space. We discuss this in detail in the next Section and
focus now on the parametrization of $\alpha_s$ in the parton ladders. As the
treatment of $\alpha_s$ for the color singlet $M_0$ and octet $M_V$ amplitudes is the same,
we will not specify the channel below. \\
 First  we remind that the well-known  result
$\alpha_s = \alpha_s(Q^2)$ in the DGLAP equations
follows from the parametrization
\begin{equation}\label{alphakt}
\alpha_s = \alpha_s (k^2_{\perp})
\end{equation}
in every rung of the
ladder Feynman graphs, where the ladder (vertical) partons  can
be either quarks or gluons.
The notation  $k_{\perp}$ in Eq.~(\ref{alphakt})
stands for the transverse components of momenta $k$ of the vertical
partons (quarks and gluons).
The theoretical grounds for this parametrization
 can be found in refs. ~\cite{cg,ddt,adglap}.
The analysis of the parametrization of $\alpha_s$ directly for the DGLAP
equations was discussed in details in Ref.~\cite{ds}. In Ref.~\cite{egtalpha} we had  shown that the
arguments of Ref.~\cite{ds} in favor of using
the parametrization (\ref{alphakt}) in DGLAP can be used at large $x$ only.
Later, in Ref.~\cite{etalfa} we made a more detailed investigation on this issue and
 showed that the parametrization (\ref{alphakt}) is always an approximation regardless
of value of $x$. As the matter of fact,
$\alpha_s(k^2_{\perp})$ should be replaced by the effective coupling
$\alpha_s^{eff}$ given by the following expression:
\begin{eqnarray}\label{aeff}
\alpha_s^{eff} =  \alpha_s(\mu^2) + \frac{1}{\pi b} \Big[\arctan
\Big(\frac{\pi}{\ln (k^2_{\perp}/\beta \Lambda^2)}\Big) -\arctan
\Big(\frac{\pi}{\ln (\mu^2/\Lambda^2)}\Big) \Big]
\\ \nonumber
= \alpha_s(\mu^2) + \frac{1}{\pi b} \Big[\arctan \Big(\pi b
\alpha_s(k^2_{\perp}/\beta)\Big) -\arctan \Big(\pi b \alpha_s
(\mu^2)\Big) \Big] ,
\end{eqnarray}
where the longitudinal Sudakov variable $\beta$ is defined in Eq.~(\ref{sud}). However when
the starting point $\mu^2$ of the $Q^2$
-evolution obeys the strong inequality
\begin{equation}\label{mulambdapi}
\mu^2 \gg \Lambda^2 e^{\pi} \approx 23 \Lambda^2,
\end{equation}
$\alpha_s^{eff}$ can be approximated by the much simpler expression:
\begin{equation}\label{abigmu}
\alpha_s^{eff} \approx \alpha_s(k^2_{\perp}/\beta).
\end{equation}
If additionally
$x$ is large, $\alpha_s^{eff} \approx \alpha_s(k^2_{\perp})$. For practical use, the
inequality in
Eq.~(\ref{mulambdapi}) can be expressed in terms of the discrepancy $R(\mu)$ defined as
\begin{equation}\label{aeffakt}
R(\mu) = \frac{|(1/\pi b) \arctan (\pi/ \ln(\mu^2/\Lambda^2)) - \alpha_s(\mu^2)|}{\alpha_s(\mu^2)} .
\end{equation}
A simple calculation shows that $R(\mu)$ rapidly grows when $\mu$
decreases, ranging, for example,  from $R(\mu) = 5 \%$ at $\mu^2 = 2800 \Lambda^2$ to
$R(\mu) = 10 \%$ at $\mu^2 = 250 \Lambda^2$
and $R(\mu) = 50 \%$ at $\mu^2 = 8.7 \Lambda^2$.
As the DGLAP starting point of the $Q^2$
-evolution is typically chosen close to $1~$GeV$^2$, the latter
example shows that the DGLAP parametrization Eq.~(\ref{alphakt})
has an error of 50 $\%$ at such low scale . This statement is true for all DIS
structure functions.  In order to derive
Eq.~(\ref{aeff}) we consider now the
parametrization of $\alpha_s$ in the integral expressions for the
DIS structure functions. Here we partly follow the approach of
Ref.~\cite{ds}. To this aim we consider the forward Compton
amplitude $T(x,Q^2)$ related to the structure functions by
Eq.~(\ref{tg}). Obviously, this equation is true for all DIS
structure functions, so we drop here the signature superscript in
$T$ as unessential. One can show (and in this paper we will do
it in the context of DGLAP and our Infrared Evolution Equations)
that $T$ obeys the following Bethe-Salpeter equation:
\begin{equation}
\label{bseq}
T(x, Q^2) = T^{Born} + \imath
\int \frac{d^4 k}{(2\pi)^4}\frac{2 w k^2_{\perp}}{(k^2+ \imath \epsilon)^2 }
M((q + k)^2,k^2,Q^2)~
4\pi \frac{\alpha_s((p - k)^2)}{(p - k)^2 + \imath\epsilon}.
\end{equation}
 The integral term in Eq.~(\ref{bseq})
is shown  in Fig.~\ref{g1sumfig2}.
 The notation  $M((q + k)^2,k^2,Q^2)$ in Eq.~(\ref{bseq}) corresponds to the blob in
Fig.~\ref{g1sumfig2}. Besides the amplitude $T$, it can also include a
kernel (splitting functions).
The inhomogeneous term $T^{Born}$ is $T$ in the Born
approximation. To be specific, we consider the case when the
horizontal parton in Fig.~\ref{g1sumfig2} is the virtual gluon
with momentum $p-k$ whereas the vertical partons, with momentum
$k$, can be either quarks or gluons. When they are quarks, $k^2 +
\imath \epsilon$ should be replaced by $k^2 - m^2_q + \imath
\epsilon$, with $m_q$ being the quark mass, but this shift does
not play any role for our consideration below. The factor $2w
k^2_{\perp}$ in Eq.~(\ref{bseq}) appears as a result of the
simplification of the spin structure of the ladder Feynman graph
in Fig.~\ref{g1sumfig2}. We use the Sudakov parametrization
(\ref{sud}) for momentum $k$ of the vertical partons. In terms of
the Sudakov variables $\alpha,~\beta,~k_{\perp}$,
\begin{equation}\label{ksud}
k^2 = -w \alpha\beta -k^2_{\perp},~~2qk =  w x \alpha  + w \beta,~~2pk = -w \alpha.
\end{equation}
where  $w= 2pq$.
It is convenient to introduce a new variable $m^2 = (p-k)^2$ instead of $\alpha$.
Therefore,
\begin{equation}\label{alfam}
\alpha = \frac{m^2 + k^2_{\perp}}{w(1- \beta)},~~~ k^2 = -\frac{\beta m^2 + k^2_{\perp}}{1-\beta}~.
\end{equation}
Using Eqs. (\ref{ksud},\ref{alfam}), we can rewrite
Eq.~(\ref{bseq}) in a simpler way:
\begin{equation}
\label{bseqsud} T(x, Q^2) = T^{Born} + \frac{\imath}{4
\pi^2}\int_{\mu^2}^{w} d k^2_{\perp} \int^1_{\beta_0}d \beta \int_{-
\imath \infty}^{\imath \infty} d m^2 \frac{w(1-
\beta)k^2_{\perp}}{[m^2 \beta + k^2_{\perp}- \imath \epsilon]^2} M ((q +
k)^2,(\beta m^2 + k^2_{\perp} ), Q^2)~ \frac{\alpha_s(m^2)}{m^2  +
\imath\epsilon}~.
\end{equation}
The integration over $\beta$ and $k^2_{\perp}$ in Eq.~(\ref{bseqsud}) runs over the region
\begin{equation}\label{betaint}
\beta_0 < \beta < 1,~~~\mu^2 < k^2_{\perp} < w.
\end{equation}
The value of $\beta_0$ follows from the requirement of positivity
of the invariant energy $(q+k)^2$ of the blob in
Fig.~\ref{g1sumfig2}:
\begin{equation}\label{beta0}
\beta_0 \approx x + \frac{k^2_{\perp}}{w - m^2}.
\end{equation}
Let us notice that  the $m^2$ -dependence in
Eq.~(\ref{beta0}) can be neglected to the leading logarithmic accuracy
that we keep through this paper.

\subsection{Integration in Eq.~(\ref{bseqsud}) at fixed $\alpha_s$}

Let us consider first the calculation of  Eq.~(\ref{bseqsud}) under the approximation of
fixed $\alpha_s$. From the analysis of the ladder Feynman graphs (see e.g. the review ~\cite{g})
one can see that DL contribution comes from the region of large $k^2_{\perp}$ where
\begin{equation} \label{kkperp}
-k^2 \approx k^2_{\perp} \gg k^2_{||} = \beta m^2.
\end{equation}
It allows to neglect the dependence of $M$ on $\beta m^2$ in Eq.~(\ref{bseqsud}).
It is convenient to integrate Eq.~(\ref{bseqsud}) over $m^2$, using
the Cauchy theorem. The singularities of the integrand are:
the double pole from the vertical propagators
\begin{equation}\label{kpole}
\beta m^2 + k^2_{\perp} - \imath \epsilon =0
\end{equation}
and the simple pole from the horizontal gluon propagator
\begin{equation}\label{mpole}
m^2 +  \imath \epsilon =0.
\end{equation}
The integration contour can be equally closed up or down.
Traditionally (see e.g. Ref.~\cite{g}) the integration contour is closed down
which involves taking the residue at the
simple pole (\ref{mpole}), so we arrive at the following result:
\begin{equation}\label{tafix}
T(x, Q^2) = T^{Born} + \frac{\alpha_s}{2 \pi}\int_{\mu^2}^{w} \frac{d
k^2_{\perp}} {k^2_{\perp}} \int^1_{\beta_0} d \beta (1-\beta) M
(\beta, Q^2, k^2_{\perp}).
\end{equation}

Obviously, when $x$ is large enough, one can change the upper
limit of the integration over $k^2_{\perp}$ for $Q^2$.
Similarly, $\beta_0 \approx x$. Extracting the Born factor $1/\beta$ from
$M$ we can write
\begin{equation}\label{fipt}
M (\beta, k^2_{\perp}) = (1/\beta) P T
\end{equation}
where $P$ is a kernel. To specify it, let us  provide $T$ with the quark and
gluon subscripts through the replacement $T$ by $T_r$
(with $r = q,g$). It leads to specifying $P T$ in Eq.~(\ref{fipt}):
$P T= P_{rr'} T_{r'}$. At last, assuming that we have used the planar
gauge allows to identify $(1-\beta) P_{rr'}$ with the standard LO DGLAP splitting functions.
Differentiation of Eq.~(\ref{tafix}) with respect to the upper limit of
the  $k^2_{\perp})$ -integration (which is $Q^2$ at large $x$) leads to the
standard integro- differential DGLAP
equations for the Compton amplitudes $T_q,T_g$.

\subsection{Integration in Eq.~(\ref{bseqsud}) with running $\alpha_s$}

When $\alpha_s$ is running, it is also convenient to use the Cauchy theorem
for integrating Eq.~(\ref{bseqsud}) over $m^2$. The integration contour can
again be closed down.
However, the spectrum of singularities in the lower semi-plane now includes
the pole (\ref{mpole}) and the cut of $\alpha_s$ running along the real axis:
\begin{equation}\label{mcut}
\mu^2  < m^2- \imath \epsilon  < + \infty.
\end{equation}
Therefore,  instead of Eq.~(\ref{tafix}) we arrive at the more complicated
expression:
\begin{eqnarray}\label{tarun}
T(x, Q^2) = T^{Born}\; + && \frac{1}{2 \pi}\int_{\mu^2}^{w} d
k^2_{\perp} \int^1_{\beta_0} d \beta\; (1-\beta)\; \Big[
\frac{\alpha_s (\mu^2)}{k^2_{\perp}} M (\beta, Q^2, k^2_{\perp})\;
+ \\ \nonumber &&\int_{\mu^2}^{\infty} \frac{d m^2}{m^2} M (\beta,
Q^2,~ \beta m^2 + k^2_{\perp}) \Im \alpha_s (m^2)
\frac{k^2_{\perp}}{(\beta m^2 + k^2_{\perp})^2}\Big].
\end{eqnarray}
where we have not used the assumption of Eq.~(\ref{kkperp}). The second term in
the rhs of Eq.~(\ref{tarun}) is the result of taking the residue in the pole (\ref{mpole})
and the third terms corresponds to accounting for the cut (\ref{mcut}).
Eq.~(\ref{tarun}) demonstrates explicitly that it is impossible to factorize
$\alpha_s$, i.e. to integrate $\alpha_s(m^2)$ over $m^2$, without making the
approximation of Eq.~(\ref{kkperp}). When this approximation has been made,
we immediately obtain
\begin{equation}\label{tarun1}
T(x, Q^2) = T^{Born} + \frac{1}{2 \pi}\int_{\mu^2}^{w} \frac{d
k^2_{\perp}} {k^2_{\perp}}\int^1_{\beta_0} d \beta (1-\beta) M
(\beta, Q^2, k^2_{\perp})\; \alpha_s^{eff},
\end{equation}
with $\alpha_s^{eff}$ given by Eq.~(\ref{aeff}). Indeed, in this
case the integral over $m^2$ in Eq.~(\ref{tarun1}) is
\begin{equation}\label{intm}
I = \frac{1}{b} \int^{k^2_{\perp}/\beta}_{\mu^2} \frac{d
m^2}{m^2}\; \frac{1}{\ln^2(m^2) + \pi^2} =  \frac{1}{\pi b}
\Big[\arctan \Big(
\frac{\pi}{\ln(\mu^2/\Lambda^2)}\Big) - \arctan \Big(\frac{\pi} {\ln(k^2_{\perp}/\beta)} \Big)
\Big]~.
\end{equation}
Obviously, the term $\pi^2$ in the integrand in Eq.~(\ref{intm}) can be
neglected when $\mu$
obeys Eq.~(\ref{mulambdapi}). It leads to the approximative expression
of Eq.~(\ref{abigmu}) for $\alpha_s^{eff}$.
The approximation $\beta m^2 \ll k^2_{\perp}$ was also made in Ref.~\cite{ds} for
the integration Eq.~(\ref{bseqsud}) over $m^2$ in the
case of running $\alpha_s$. However, the integration contour in that paper
was closed up in order to take
the residue of the double pole at $k^2 = \beta m^2 + k^2_{\perp} = 0$,
which contradicts  the assumption of Eq.~(\ref{kkperp}) made in Ref.~\cite{ds}.
Taking this residue automatically led to the wrong conclusion that
$\alpha_s^{eff} = \alpha_s (-k^2_{\perp}/\beta)$ regardless of the value
of $\mu$. This error was found and corrected in
Ref.~\cite{etalfa}.

\section{Description of $g_1$ in the Region B: Total resummation of the leading logarithms}

The region \textbf{B} is defined in Eq.~(\ref{regb}). As it includes small $x$ and
large $Q^2$, both logs of $1/x$ and $Q^2$ are equally important in this
region and should be summed up. The most important logarithmic contributions
to $g_1$ in region \textbf{B} are the double-logarithmic (DL) ones, i.e. the terms
\begin{equation}\label{dl}
\sim \alpha_s^n \ln^{2n -k} (1/x) \ln^k (Q^2/\mu^2),
\end{equation}
with $k = 0,1,..,n$, so they should be accounted for in the first place. Then the
sub-leading, single-logarithmic (SL), contributions can also be taken into account etc.
Therefore, an appropriate evolution equation for $g_1$ in region \textbf{B} should
account for the evolution both with respect to $x$ and $Q^2$ while DGLAP controls
the $Q^2$ -evolution only and cannot sum up the logarithms of $x$. In addition,
the running coupling effects in Eq.~(\ref{dl}) should be taken into account. To this
end,
a special attention should be given to the parameterizations of $\alpha_s$ in
region \textbf{B} and we will use here the results of Sect.~IV.
In order to resum DL contributions we use the
 alternative method of the Infrared Evolution Equations
 (IREE), first suggested by L.N.~Lipatov (see Ref.~\cite{l}).
 Then it was applied to the elastic scattering of quarks in Ref.~\cite{kl}
 and in Ref.~\cite{efl}, with the  generalization
 to inelastic processes (radiative $e^+ e^-$ annihilation).
 Since then the IREE method has been applied to various problems and
 a brief review of the applications can be found in Ref.~\cite{acta}. It is
 convenient to compose IREE for the Compton amplitudes $T^{(-)}$
 related to $g_1$ by Eq.~(\ref{tsigng}).

\subsection{The essence of the method}

  As we have mentioned above, the DGLAP -ordering of Eq.~(\ref{dglapord}) makes impossible to collect
 all DL contributions, regardless of their arguments, to all
 powers in $\alpha_s$.
 In order to account for them,  the ordering of Eq.~(\ref{dglapord}) should be changed
 as in Eq.~(\ref{dlord}). This leads to the
 infrared (IR) singularities emerging from the graphs with soft gluons. In order to regulate
 them an IR cut-off $\mu$ should be introduced and therefore the result of
 such calculation becomes $\mu$ -dependent. The fermion (quark)
 ladders contributing e.g. to the non-singlet components of the DIS structure
 functions do not need an IR cut-off as long
 as the quark masses are accounted for. But in order to treat  them similarly to the
 graphs with soft gluons, one can choose $\mu \gg$ masses of involved quarks.
 After that the quark masses can be dropped and the only remaining mass scale
 is $\mu$. Generally, the value of $\mu$ is arbitrary,
 with one important exception: in order to use the perturbative QCD, $\mu$
 should obey Eq.~(\ref{mulambda}).\\
On one hand, such flexibility can be used to resum DL contributions
 through the use of  evolution equations with respect to $\mu$, which is the basis of our
 approach. \\
On the other hand, after such a resummation has been done, we arrive
to a result which depends on this indefinite parameter\footnote{We remind that
 DGLAP is free of this problem due to the different ordering
 of Eq.~(\ref{dglapord}).}. Of course, the problem of fixing
 the IR cut-off is not a new one. It has been known since long time
 ago, appearing first in QED,  where $\mu$ is replaced in the final
 expressions by a suitable mass or energy scales. However, in the  context of QCD this problem
  becomes more involved. Indeed, besides regulating the IR divergencies, $\mu$ acts also
 as a border line between Perturbative  and Non-Perturbative QCD. Of course such a border is
 totally artificial from the
 point of view of the physics of hadrons . \\
 Below we will discuss how we fix the value of $\mu$ for the non-singlet
 (where $\mu =1~$GeV approximately) and singlet ( $\mu =5.5~$GeV)
 components of the structure function $g_1$. \\
 The last  point deserving a discussion concerns the possible dependence of our results on
 the way of introducing the IR cut-off: basically, different ways can
 lead to different results as  it was shown explicitly in
 Ref.~\cite{ioffe}. However, such a
 discrepancy appears far beyond the leading logarithmic approximation (LLA),
 that we keep through this paper.  We will  discuss now the technical
 details of our approach.

\subsection{The  IREE for $T^{(-)}$ in DLA}

\begin{figure}[h]
\begin{center}
\begin{picture}(380,160)
\put(0,0){
\epsfbox{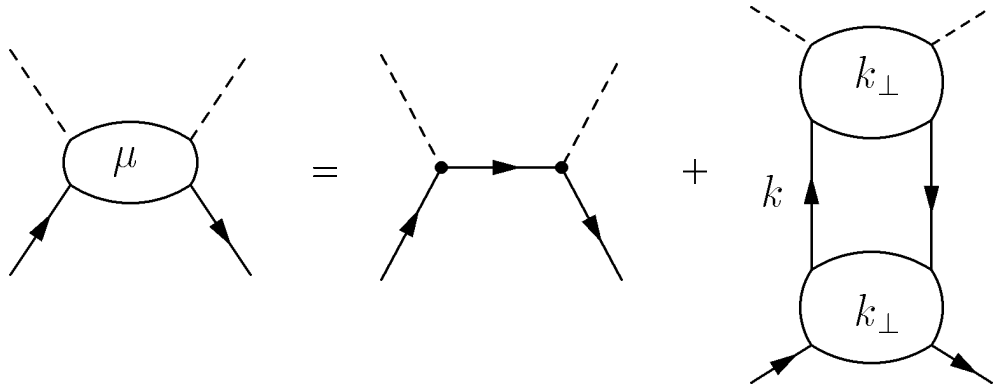} }
\end{picture}
\end{center}
\caption{\label{g1sumfig3}The IREE for the Compton amplitude
$T^{(-)}$.}
\end{figure}

We will consider, from now on,  the invariant amplitude $T^{(-)}_1$, defined in
Eqs.~(\ref{tmunu}) and (\ref{tsign}) and related to the structure function
$g_1$ by Eq.~(\ref{tsigng}). To simplify our notations, we drop both the subscript
and superscript at $T^{(-)}_1$ and denote $T \equiv T^{(-)}_1$.
When the cut-off $\mu$ is used for calculating Feynman graph contributions to
 $T$, this
amplitude acquires the additional dependence:
\begin{equation}\label{tmu}
T = T (w, Q^2, \mu^2).
\end{equation}
This amplitude is in the left-hand side of the equation in
Fig.~\ref{g1sumfig3}.
 Beyond the Born approximation, $T$ depends on its arguments
through their logarithms, so we can parameterize it in terms of
logarithms:
\begin{equation}\label{troy}
T = T\Big(\ln(w/\mu^2), \ln(Q^2/\mu^2)\Big) \equiv
T (\rho, y).
\end{equation}
Therefore,
\begin{equation}\label{lhsiree}
- \mu^2 \frac{\partial T}{\partial \mu^2} =
\frac{\partial T}{\partial \rho} + \frac{\partial T}{\partial y}.
\end{equation}
Eqs.~(\ref{troy}) and (\ref{lhsiree}) are the left-hand sides of the IREE for $T$
 (see Fig.~\ref{g1sumfig3}) in the integral and differential form respectively. The
right-hand side of the IREE includes, in the first place, the Born
amplitude $T^{(-)}_{Born}$ given by Eq.~(\ref{tborn}). It
corresponds to the second term in Fig.~\ref{g1sumfig3}. In
general, the other terms of IREE are obtained by factorizing the DL
contributions of the softest partons. It is well-known that the DL
contributions are technically obtained from the integration over both
longitudinal and transverse momenta, each integration bringing a
logarithm. Logarithmic contributions from the  integration over each
transverse momentum come from the kinematic regions where those
momenta differ from each other: $ k_{i~\perp} \gg k_{j~\perp}$. This
implies that the set of virtual partons (quarks and gluons) always
contains a parton with minimal transverse momentum. In other
words, the transverse phase space can be represented as the sum of
sub-regions $\emph{D}_i$, each of them contains the parton with a
minimal transverse momentum. We address such a parton as the
softest parton even if its energy is not small and denote $k_{\perp}$
its transverse momentum. The IR divergence arising by integrating
$dk_{\perp}/k_{\perp}$ can be regulated with the IR cut-off $\mu$:
\begin{equation}\label{kmu}
k_{\perp} > \mu .
\end{equation}
Of course, $\mu$ should obey Eq.~(\ref{mulambda}) to guarantee applicability of the Pert. QCD.
It was shown in Ref.~\cite{ggfl} that in QED the DL contributions
can be of two different kinds: \\
\textbf{(A)}: the Sudakov DL contributions
coming from soft non-ladder partons; \\
\textbf{(B)}: non-Sudakov DL
contributions calculated first in Ref.~\cite{ggfl}. They arise from ladder Feynman
graphs.
This classifications stands also for QCD. The
DL contributions of the softest partons from groups \textbf{(A)} and \textbf{(B)}
are  factorized
differently.\\

\textbf{Factorization of the softest gluon from group A:} \\
The DL contributions from the softest non-ladder gluons can be factorized by
using the QCD -generalization\cite{efl,e,ce} of the Gribov factorization theorem
(often called the Gribov bremsstrahlung theorem) of the
soft photons obtained in Ref.~\cite{gr}. According to it, the non-ladder
gluon having the minimal $k_{\perp}$ and being polarized in the plane
formed by the external momenta can be factorized, i.e. its propagator
is attached to the external lines only. When the Feynman gauge is used,
the softest gluon propagator connects all available pairs of the external lines.
The integration over other transverse momenta have $k_{\perp}$ as the lowest
integration limit. Such a factorization deals with $k_{\perp}$ only and does not involve
longitudinal momenta.  Obviously, there is no way to attach the softest gluon
propagator to the external lines of amplitude $T^{(-)}$  with the DL
accuracy\footnote {We will use this kind of factorization in the next Sect. for
calculating the lower blob in Fig.~\ref{g1sumfig3}}.\\

\textbf{Factorization of the softest partons from the group B:} \\
Both the DGLAP-ordering in Eq.~(\ref{dglapord}) and the ordering in
Eq.~(\ref{dlord}) imply that one can always find a ladder
(vertical) parton (quark or gluon) with minimal transverse
momentum. However, there is a difference between the two cases:
 the softest parton in (\ref{dglapord}) is
always the lowest parton at the ladder whereas in
(\ref{dlord})  the softest gluon can be anywhere in the
ladder, from the bottom to the top. Therefore, it corresponds to the
factorization of the lowest ladder rung in the DGLAP ordering
(\ref{dglapord}), and the factorization of an arbitrary ladder rung
under (\ref{dlord}). The latter option corresponds to the last
term in Fig.~\ref{g1sumfig3}. By definition, in  both cases the
integration over $k_{\perp}$ involves $\mu$ as the lowest limit
whereas integrations over
other $k_{i~\perp}$ are $\mu$ -independent. \\

Now we can compose the IREE for $T$ in the integral form.
The lhs is just $T$ while the rhs consists of the Born contribution and the
term obtained by using the factorization \textbf{B}, therefore we
arrive at the following IREE

\begin{equation}\label{ireetint}
T_r(\rho, y) = T_{r~Born} + \imath \int \frac{d^4 k}{(2 \pi)^4}
\frac{2 w k^2_{\perp}}{(k^2+ \imath \epsilon)^2 }
T_{r'}((q+k)^2, Q^2, k^2) M_{r'r}((p-k)^2, k^2)
\end{equation}
where any of $r,r'$ denotes $q$ or $g$; the factor $2 w
k^2_{\perp}$comes by simplifying the spin
structure; The negative signature amplitudes $M_{r'r}$ of the $2
\to 2$ -forward scattering of partons correspond to the lower
blobs in the rhs of Fig.~\ref{g1sumfig3}. This will account for the
total resummation of the leading logarithms as we'll see in detail in
the next Sect. When the DGLAP-ordering (\ref{dglapord}) is used
instead of (\ref{dlord}), only the lowest ladder rung can be
factorized and therefore $M^{(-)}_{r'r}$ are in this case given
by the DL part of the LO DGLAP splitting functions. In other
words, we arrive in this case to Eq.~(\ref{bseq}). Applying the
operator $-\mu^2 \partial/\partial \mu^2$ to Eq.~(\ref{ireetint}) it
converts the lhs into Eq.~(\ref{lhsiree}). On the other hand Eq.~(\ref{tborn}) shows
that the Born contribution in the rhs of Eq.~(\ref{ireetint}) vanishes
under the differentiation.because it does not depend on  $\mu$.
Using the SW transform as in Eq.~(\ref{mellintf}) we rewrite
Eq.~(\ref{ireetint}) in terms of the amplitudes $F_r$, related to $T_r$
through Eq.~(\ref{mellintf}):

\begin{equation}\label{ireef}
\omega F_r (\omega, y)+ \frac{\partial F_r (\omega,y)}{\partial y}
= \frac{1}{8 \pi^2} F_{r'} (\omega,y) L_{r'r}(\omega)
\end{equation}
where $L_{r'r}$ is related to $M_{r'r}$ by the transform
Eq.~(\ref{mellintf}). The derivation of Eq.~(\ref{ireef}) from Eq.~(\ref{ireetint}) is given
in detail in Ref.~\cite{egtsns}. The general technique of simplifying the convolution
in Eq.~(\ref{ireetint}) is given in Appendix C. It is useful to rewrite Eq.~(\ref{ireef}) in
terms of the flavor singlet,
$T^S$, and non-singlet, $T^{NS}$, components of the Compton amplitude $T^{(-)}_r$:

\begin{equation}\label{eqfns}
 \frac{\partial F^{NS} (\omega,y)}{\partial y}= \big(- \omega +
\frac{1}{8 \pi^2} L_{qq}(\omega) \big) F^{NS} (\omega,y),
\end{equation}
and
\begin{eqnarray}\label{eqfs}
 \frac{\partial F^S_q (\omega,y)}{\partial y}= \Big( -\omega +
\frac{1}{8 \pi^2} L_{qq}(\omega)\Big) F^S_q (\omega, y) +
\frac{1}{8 \pi^2} F^S_g (\omega,y) L_{gq}(\omega),  \\ \nonumber
 \frac{\partial F^S_g (\omega,y)}{\partial y}= \Big(- \omega +
 \frac{1}{8 \pi^2} F^S_q (\omega,y) L_{qg}(\omega)\Big)F^S_g (\omega, y) +
\frac{1}{8 \pi^2} L_{gg}(\omega)F^S_g (\omega,y).
\end{eqnarray}
It is also convenient to introduce the amplitudes $H_{ik}$ related to $L_{ik}$ as follows:
\begin{equation}\label{fl}
H_{ik} = \frac{1}{8 \pi^2}\; L_{ik}.
\end{equation}

$F^S$, and  $F^{NS}$ are related to the singlet and non-singlet components of $g_1$
with Eq.~(\ref{gf}). Eqs.~(\ref{eqfns},\ref{eqfs}) are written in the DGLAP-like form,
with the derivative with respect to $Q^2$, but actually they combine the evolution
with respect to $Q^2$ and $w$.

Let us consider
how to incorporate the single-logarithmic corrections from group (\textbf{ii}) of Sect.~IV
into Eqs.~(\ref{eqfns},\ref{eqfs}).

\subsection{Inclusion of single-logarithmic contributions into Eq.~(\ref{ireef})}

Technically, the DL contributions appear from the integrals over the loop
momenta $k_i$ of the following form:
\begin{equation}\label{dlint}
\sim \int \frac{d k^2_{i~\perp}}{k^2_{i~\perp}} \frac{d
\beta_i}{\beta_i} \;\varphi (p_r, k_j),
\end{equation}
where we have used notations $p_r$ for external momenta and presumed that $j\neq i$.
The  function $\varphi (p_r, k_j)$  is independent of $k_i$, which follows from
imposing the strong inequalities giving rise to  the DL integration region:
\begin{equation}\label{dlineq}
k_{i~\perp} \ll k_{j~\perp},~~\beta_i \ll \beta_j.
\end{equation}

When, for example, linear terms in $k_i$   are present  in $\varphi$,
one of the integrations in Eq.~(\ref{dlint}) does not give rise to  a logarithm and
as a consequence a single-logarithmic (SL) contributions appears. This  takes place
in the integration region where the strong inequalities (\ref{dlineq})
do not apply. In particular,when the inequality $k_{i~\perp} \ll k_{j~\perp}$
is not fulfilled there is not a single
soft parton in this region and therefore the method we use cannot account for
such contributions. On the other hand, replacing the DL inequality  $\beta_i \ll \beta_j$
by the single-logarithmic one, $\beta_i < \beta_j$ is not essential for
the method and these SL contributions can be taken into account.
  This replacement converts Eqs.~(\ref{eqfns},\ref{eqfs}) into
\begin{equation}\label{eqfnssl}
 \omega F^{NS}(\omega,y) + \frac{\partial F^{NS} (\omega,y)}{\partial y}=
\frac{1}{8 \pi^2}(1 + \lambda_{qq} \omega) h_{qq}(\omega) F^{NS} (\omega,y) ,
\end{equation}
and
\begin{eqnarray}\label{eqfssl}
 \omega F^S_q (\omega,y) + \frac{\partial F^S_q (\omega,y)}{\partial y}= (1 + \lambda_{qq} \omega)
 h_{qq}(\omega) ~F^S_q (\omega, y) +
 (1 + \lambda_{gq} \omega) h_{gq}(\omega) ~F^S_g (\omega,y) ,  \\ \nonumber
 \omega F^S_g (\omega,y) + \frac{\partial ~F^S_g (\omega,y)}{\partial y}=
 (1 + \lambda_{qg} \omega)h_{qg}(\omega) ~F^S_q (\omega,y) +
 (1 + \lambda_{gg} \omega) h_{gg}(\omega) ~F^S_g (\omega, y)
\end{eqnarray}
with $\lambda_{rr'}$ given by the following expressions:
\begin{equation}\label{lambdaik}
\lambda_{qq} = 1/2~,\qquad\lambda_{qg} = -1/2~,\qquad\lambda_{gq}
= -2~,\qquad\lambda_{gg}= -13/24 + n_f/(12 N),
\end{equation}
obtained with one-loop calculations in the planar gauge.
The new amplitudes $h_{ik}$ are defined similarly to $H_{ik}$ and  account not only for the
total resummation of the DL contributions but also contain the  resummation of
the SL contributions. Therefore  Eqs.~(\ref{eqfnssl},\ref{eqfssl}) account for
the resummation of the leading (DL) together with sub-leading (SL) contributions to
the $w$ -evolution and for the resummation of the
leading (DL) contributions to the $Q^2$ -evolution. For this reason they
can be considered as the generalization of the DGLAP equations to the region (\textbf{B}).
In contrast to the well-developed technology of resummation of double logarithms,
no regular methods for the total resummation of SL
contributions are presently  available in the literature.

\section{IREE for the amplitudes $H_{ik}$ and $h_{ik}$}

The expressions for the parton amplitudes $H_{ik}$ and $h_{ik}$
in Eqs.~(\ref{eqfns}-\ref{eqfssl}) can be found by explicitly
 solving the IREE for them. Such equations were obtained
and discussed in detail first in Ref.~\cite{ber} where $\alpha_s$
was kept fixed, while the running coupling effects were implemented in
Ref.~\cite{egtsns}. The technique  for solving the IREE for the
anomalous dimensions is similar to the one for amplitudes $T^S$,
so we give below just a short comment on it.  As it was done
before for the Compton amplitudes, we begin with the  IREE
for amplitudes $M_{ik}(\rho)$ related to the Mellin amplitudes
$H_{ik}(\omega)$ through Eq.~(\ref{mellintf}). The amplitudes
$L_{ik}(\rho)$ do not depend on $Q^2$, so the left-hand side of
the IREE for $H_{ik}(\omega)$ does not involve derivatives and is
equal to $\omega H_{ik}(\omega)$.
  The invariant amplitudes $M$ in the
Born approximation were
introduced in Eq.~(\ref{m0}). It was shown that in the $\omega$ -space all of them are $\sim 1/\omega$, so
we can write
\begin{equation}\label{ha}
H^{Born}_{ik} = a_{ik}/\omega,
\end{equation}
with
\begin{equation}\label{aik}
a_{qq} = \frac{A(\omega) C_F}{2 \pi}~,\quad a_{qg} = \frac{
A'(\omega) C_F}{\pi}~,\quad a_{gq} = - \frac{n_f A'(\omega)}{2
\pi}~,\quad a_{gg} = \frac{4 N A(\omega)}{2 \pi}~,
\end{equation}
where we have used the standard notations $C_F = (N^2 - 1)/2N = 4/3,~n_f$ is the number of the
quark flavors, $A$ is defined in Eq.~(\ref{a}) and
\begin{equation}\label{aprime}
A'(\omega) = \frac{1}{b} \Big[ \frac{1}{\eta} - \int_{0}^{\infty}
 \frac{d \rho e^{- \omega \rho}}{(\rho + \eta)^2} \Big]~.
\end{equation}
The reason for the replacement of $A(\omega)$ by $A'(\omega)$ is that
the argument of $\alpha_s$ in $L_{qg}$ and $L_{gq}$ is space-like,
so the coupling does not lead to  $\pi$ -terms.
The next term in the rhs of the IREE are the convolutions of the
anomalous dimensions appearing  as
the convolution in Eqs.~(\ref{eqfns},\ref{eqfs}). At last, the rhs
contains the contribution from the factorization of the soft (Sudakov)
gluons. We account for this contribution approximately (see
Ref.~\cite{egtsns}) for more details. Eventually we arrive at the system
of IREE for the singlet anomalous dimensions $H_{ik}$  represented in
Fig.~\ref{g1sumfig4}. In the case of the non-singlet anomalous
dimension, all gluon contributions in Eq.~(\ref{eqhik}) should be
dropped. Rewriting Fig.~\ref{g1sumfig4} in a detailed form, we
arrive at the system of the algebraic non-linear equations for
$H_{ik}$~:
\begin{figure}[h]
\begin{center}
\begin{picture}(400,200)
\put(0,0){
\epsfbox{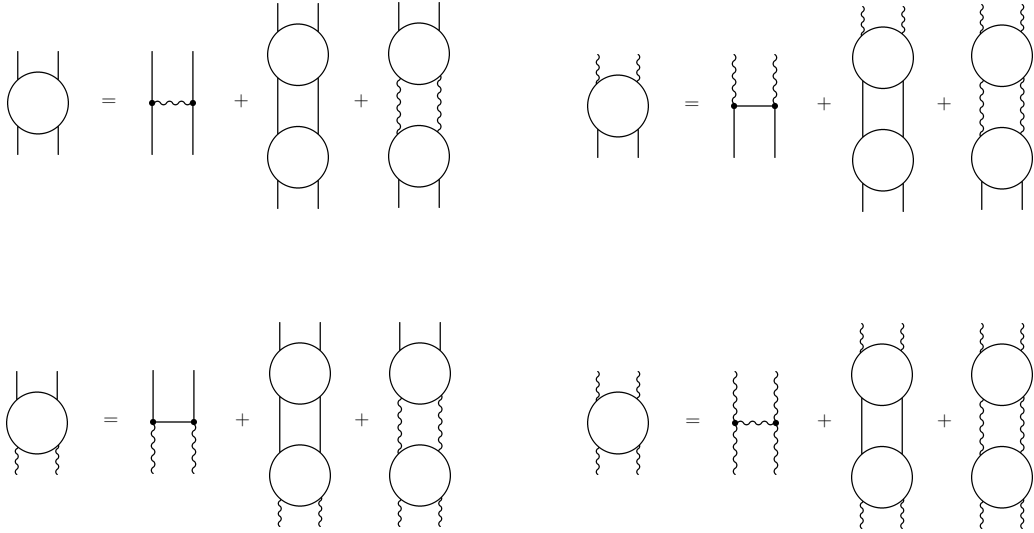} }
\end{picture}
\end{center}
\caption{\label{g1sumfig4} The IREE for singlet amplitudes
$H_{ik}$ . The solid lines correspond to quarks and the wavy lines denote gluons.}
\end{figure}
\begin{eqnarray}\label{eqhik}
\omega H_{qq} = b_{qq} + H_{qq}H_{qq} + H_{qg} H_{gq},~~
\omega H_{qg} = b_{qg} + H_{qq}H_{qg} + H_{qg} H_{gg},
\\ \nonumber
\omega H_{gq} = b_{gq} + H_{gq}H_{qq} + H_{gg} H_{gq},~~
\omega H_{gg} = b_{gg} + H_{gq}H_{qg} + H_{gg} H_{gg}.
\end{eqnarray}
where we have denoted
\begin{equation}\label{bik}
b_{ik} = a_{ik} + V_{ik},
\end{equation}
with $a_{ik}$ given by Eq.~(\ref{aik}). Then
\begin{equation}
\label{vik} V_{ik} = \frac{m_{ik}}{\pi^2} D(\omega)~,
\end{equation}
\begin{equation}
\label{mik} m_{qq} = \frac{C_F}{2 N}~,\quad m_{gg} = - 2N^2~,\quad
m_{gq} = n_f \frac{N}{2}~,\quad m_{qg} = - N C_F~,
\end{equation}
and
\begin{equation}
\label{d} D(\omega) = \frac{1}{2 b^2} \int_{0}^{\infty} d \rho
e^{- \omega \rho} \ln \big( (\rho + \eta)/\eta \big) \Big[
\frac{\rho + \eta}{(\rho + \eta)^2 + \pi^2} + \frac{1}{\rho +
\eta}\Big]~.
\end{equation}
Let us add a comment on the terms $V_{ik}$. They appear from those
graphs in Fig.~\ref{g1sumfig4} where the softest virtual
gluon is factorized, i.e. its propagator is attached to the
external lines in all possible ways. The blob obtained after the
factorization is not a color singlet but an octet in the $t$
(vertical) channel because the factorized gluon bears the color
and belongs to an octet representation of the color group
$SU(3)$. So, these new amplitudes $M^{(8)}_{ik}$ should be
calculated independently. It is not a big deal for the non-singlet
$g_1$ which involves  $M^{(8)}_{qq}$ only, but becomes a serious
technical problem when all $M^{(8)}_{ik}$ are involved (see
Ref.~\cite{ber}). Fortunately, all $M^{(8)}_{ik}$ rapidly decrease
with energy, and then  is possible to approximate them by their
Born values with a few per cent accuracy as was suggested in
Ref.~\cite{etoctet}.

Eqs.~(\ref{eqhik}), similarly to Eqs.~(\ref{eqfns},\ref{eqfs}),
combine  the total resummation of DL contributions
and the running coupling effects but do not include other SL
contributions. The part of SL contributions accounted through
Eq.~(\ref{lambdaik}) can be easily incorporated into
Eqs.~(\ref{eqhik}), leading to the following equations:
\begin{eqnarray}\label{eqhiksl}
&&\omega h_{qq} = b_{qq} + (1 + \lambda_{qq} \omega) h_{qq}h_{qq}
+ (1 + \lambda_{qg} \omega) h_{qg} h_{gq},~~
\\ \nonumber
&&\omega h_{qg} = b_{qg} + (1 + \lambda_{qg} \omega)h_{qq} h_{qg}
+ (1 + \lambda_{qg} \omega) h_{qg} h_{gg},
\\ \nonumber
&&\omega h_{gq} = b_{gq} + (1 + \lambda_{qq} \omega)h_{gq}h_{qq} +
(1 + \lambda_{qq} \omega) h_{gg} h_{gq},~~
\\ \nonumber
&&\omega h_{gg} = b_{gg} + (1 + \lambda_{qq} \omega) h_{gq}h_{qg}
+ (1 + \lambda_{gg} \omega) h_{gg} h_{gg}.
\end{eqnarray}

Now we  present the expressions for
 the non-singlet case. Dropping all gluon contributions in Eq.~(\ref{eqhiksl}),
we immediately arrive to the equation for the non-singlet amplitude $\widetilde{h}_{qq}$:
\begin{equation}\label{eqhns}
\omega \widetilde{h}_{qq} = b_{qq} + (1 + \lambda_{qq} \omega) (\widetilde{h}_{qq})^2
\end{equation}
with the obvious solution
\begin{equation}\label{fins}
\widetilde{h}_{qq} = \frac{[\omega - \sqrt{\omega^2 - B_{NS}}]}
{2(1 + \lambda_{qq} \omega)}
\end{equation}
where
\begin{equation}\label{bns}
B_{NS} = 4(1 + \lambda_{qq} \omega) b_{qq}.
\end{equation}

Unfortunately, the system of non-linear algebraic equations in Eq.~(\ref{eqhiksl}) can be solved
analytically only if all $\lambda_{ik}$
are dropped. In this case the solution to the system is
\begin{eqnarray}\label{hik}
&& H_{qq} = \frac{1}{2} \Big[ \omega - Z + \frac{b_{qq} -
b_{gg}}{Z}\Big],\qquad H_{qg} = \frac{b_{qg}}{Z}~, \\ \nonumber &&
H_{gg} = \frac{1}{2} \Big[ \omega - Z - \frac{b_{qq} -
b_{gg}}{Z}\Big],\qquad H_{gq} =\frac{b_{gq}}{Z}~
\end{eqnarray}
where
\begin{equation}
\label{z}
 Z = \frac{1}{\sqrt{2}}\sqrt{(\omega^2 - 2(b_{qq} + b_{gg})) +
\sqrt{(\omega^2 - 2(b_{qq} + b_{gg}))^2 - 4 (b_{qq} - b_{gg})^2 -
16b_{gq} b_{qg} }}~.
\end{equation}
The non-linear algebraic equations in Eq.~(\ref{eqhns}) and Eq.~(\ref{eqhik}) have more than one solution, however the
solution chosen in Eqs.~(\ref{fins},\ref{hik}) obeys the matching condition
\begin{equation}\label{matchhikborn}
\widetilde{h}_{qq} \to a_{qq}/\omega,~~~ H_{ik} \to H_{ik}^{Born} = a_{ik}/\omega
\end{equation}
at $\omega \to \infty$, with $a_{ik}$ given by Eq.~(\ref{aik})~.
In other words, this matching condition in Eq.~(\ref{matchhikborn}) implies that
these amplitudes are represented at low energies by their Born values.
 Indeed, from Eq.~(\ref{mellintf})  high energies correspond to small
$\omega$ and vice versa.

\section{Solution to the IREE~(\ref{eqfnssl}) for $g_1$ non-singlet}

As soon as the expressions for  $h_{qq}$ are obtained,
one can easily find the general solution to the linear differential
equation (\ref{eqfnssl}) for the Mellin amplitude $F^{NS}$.
As the  procedure between the non-singlet and singlet cases has a purely technical difference,
we consider in detail  the former case and proceed to the singlet
case in a much shorter way.

 \subsection{General solution to the non-singlet equation (\ref{eqfnssl})}

Obviously, the general solution to Eq.~(\ref{eqfnssl}) is
\begin{equation}\label{qensolfns}
F^{NS} =  \widetilde{F}^{NS}(\omega) e^{-\omega y + y (1+\lambda
\omega)\widetilde{h}_{qq}}
\end{equation}
and therefore
\begin{equation}\label{gentns}
T^{NS}(x,Q^2) = \int_{- \imath \infty}^{\imath \infty} \frac{d
\omega}{2 \imath \pi} \Big(w/\mu^2\Big)^{\omega}
\widetilde{F}^{NS}(\omega) e^{-\omega y + y (1 +
\lambda_{qq}\omega) \widetilde{h}_{qq}} = \int_{- \imath
\infty}^{\imath \infty} \frac{d \omega}{2 \imath \pi}
(1/x)^{\omega} \widetilde{F}^{NS}(\omega) e^{y (1 +
\lambda_{qq}\omega) \widetilde{h}_{qq}}
\end{equation}
with $\widetilde{F}^{NS}(\omega)$ being arbitrary. In order to
specify it, we use the matching condition
\begin{equation}\label{mathns}
T^{NS}(x,Q^2) = \widetilde{T}^{NS}(w/\mu^2)
\end{equation}
when $y = 0$. The new amplitude $\widetilde{T}^{NS}(w/\mu^2)$ describes again the forward Compton
scattering off the same quark, however the virtual photon has now a virtuality $\approx \mu^2$.
 The IREE for $\widetilde{T}^{NS}(w/\mu^2)$ should be obtained independently.

\subsection{Composing the IREE for $\widetilde{T}^{NS}(w/\mu^2)$.}

The IREE for $\widetilde{T}^{NS}(w/\mu^2)$ is similar to the IREE for $T^{NS}(x, Q^2)$, Eq.~(\ref{ireetint}).
 It has the same structure and involves the same amplitude $M_{qq}$.
Still, it differs from Eq.~(\ref{ireetint})
 because of two following points: \\
 \textbf{(a)} $\widetilde{T}^{NS}(w/\mu^2)$ does not depend on
 $Q^2$, so the differential IREE for it does not involve $\partial/\partial y$;\\
 \textbf{(b)} the
 Born amplitude $\widetilde{T}^{NS}_{Born}(w/\mu^2)$ can be obtained from $T^{NS}(x, Q^2)$,
 putting $Q^2 = \mu$, so its contribution
 does not vanish under differentiation with respect to $\mu$.  Then introducing
 the Mellin amplitude $\widetilde{F}^{NS}(\omega)$ related to $\widetilde{T}^{NS}(w/\mu^2)$
 through the transform (\ref{mellintf}), we arrive at the following IREE:
 \begin{equation}\label{eqfnstilde}
 \omega \widetilde{F}^{NS}(\omega) = (e^2_q/2) +  (1 + \lambda_{qq}\omega)
 \widetilde{h}_{qq} \widetilde{F}^{NS}(\omega).
 \end{equation}
and therefore
\begin{equation}\label{fnstilde}
 \widetilde{F}^{NS}(\omega) = \frac{(e^2_q/2)}{\omega - (1 + \lambda_{qq}\omega)
 \widetilde{h}_{qq} (\omega)}
 \end{equation}

\subsection{Expression for $g_1^{NS}$ in the Region B}

Combining Eqs.~(\ref{fnstilde}, \ref{qensolfns}) and (\ref{gf}) and convoluting the
perturbative expression with the initial quark density immediately leads to

\begin{eqnarray}\label{gnsb1}
g_1^{NS}  = \frac{e^2_q}{2} \int_{- \imath \infty}^{\imath \infty} \frac{d \omega}{2 \pi \imath}
\Big(\frac{w}{\mu^2}\Big)^{\omega} \frac{\omega \delta q (\omega)}{\omega - (1 + \lambda_{qq}\omega)
 \widetilde{h}_{qq} (\omega)}e^{-\omega y + y \widetilde{h}_{qq}} \\ \nonumber
= \frac{e^2_q}{2} \int_{- \imath \infty}^{\imath \infty} \frac{d \omega}{2 \pi \imath}
\Big(\frac{1}{x}\Big)^{\omega} \frac{\omega \delta q (\omega)}{\omega - (1 + \lambda_{qq}\omega)
 \widetilde{h}_{qq} (\omega)}e^{y \widetilde{h}_{qq}}
\end{eqnarray}
where $\delta q (\omega)$ is the initial quark density in  $\omega$ -space.
Confronting Eq.~(\ref{gnsb1}) to Eq.~(\ref{g1nsdglap}) it is  clear that
\begin{equation}\label{hns}
h_{NS}(\omega) = (1 + \lambda_{qq}\omega)\widetilde{h}_{qq}
(\omega) = (1/2) \Big[\omega - \sqrt{\omega^2 -
B_{NS}(\omega)}\Big]
\end{equation}
 is the new non-singlet
anomalous dimension. It contains the
total resummation of DL contributions together with the running $\alpha_s$
effects and a part of SL contributions as explained in the previous Sect. Similarly,
\begin{equation}\label{cns}
C_{NS} = \frac{\omega}{\omega - (1 + \lambda_{qq}\omega)
 \widetilde{h}_{qq} (\omega)} = \frac{\omega}{\omega - h_{NS} (\omega)}
 = \frac{2 \omega}{\omega + \sqrt{\omega^2 - B_{NS}(\omega)}}
\end{equation}
is the new non-singlet coefficient function. It is expressed through the anomalous dimension and
therefore incorporates the same kind of logarithmic contributions. Eventually we arrive at the
final expression for $g_1^{NS}$ in the region \textbf{B} of large $Q^2$ and small $x$:
\begin{equation}\label{gnsb}
g_1^{NS} (x, Q^2) = \frac{e^2_q}{2} \int_{- \imath \infty}^{\imath
\infty} \frac{d \omega}{2 \pi \imath} x^{- \omega}\; C_{NS}
(\omega)\; \delta q (\omega)\; e^{h_{NS}(\omega) \ln(Q^2/\mu^2)}
.
\end{equation}

\section{Solution to the IREE (\ref{eqfs}) for the singlet $g_1$}

Eq.~(\ref{eqfs}) for $g_1^S$ can be solved in a similar  way as the IREE for $g_1^{NS}$.  First
a general solution should be obtained and then constrained with a boundary condition.
The general solution is easy to obtain:

\begin{eqnarray}\label{gengs}
F_q &=&   e^{- \omega y}\Big[C^{(+)} e^{\Omega_{(+)} y} + C^{(-)}
e^{\Omega_{(-)} y} \Big]~, \\ \nonumber F_g &=&   e^{- \omega
y}\Big[ C^{(+)}\frac{X+ \sqrt{R}}{2 H_{qg}} e^{\Omega_{(+)} y} +
C^{(-)} \frac{X- \sqrt{R}}{2 H_{qg}} e^{\Omega_{(-)} y}\Big]
\end{eqnarray}
where
\begin{equation}\label{x}
X = H_{gg}- H_{qq}
\end{equation}
and
\begin{equation}\label{r}
R = (H_{gg}- H_{qq})^2 + 4 H_{qg} H_{gq}~.
\end{equation}
We remind that the anomalous dimensions $H_{ik}$ are found in
Eq.~(\ref{hik}). The exponents $\Omega_{(\pm)}$ are also expressed in terms of
$H_{ik}$~:
\begin{equation}\label{omegapm}
\Omega_{(\pm)} = \frac{1}{2} \Big[H_{qq} + H_{gg} \pm
\sqrt{(H_{gg} - H_{qq})^2 + 4 H_{qg}H_{gq}}\Big]~.
\end{equation}
Finally the quantities $C^{(+)}$ and $C^{(-)}$ have to be specified.

In order to constrain Eq.~(\ref{gengs}) we use the matching condition at $Q^2 = \mu^2$:
\begin{eqnarray}\label{matchfs}
&&F_q (\omega,Q^2= \mu^2)= C^{(+)} + C^{(-)} =
\widetilde{F}_q(\omega)~, \\ \nonumber &&F_g(\omega,Q^2 = \mu^2) =
C^{(+)}\frac{X+ \sqrt{R}}{2 H_{qg}} + C^{(-)}\frac{X - \sqrt{R}}{2
H_{qg}} =\widetilde{F}_g(\omega)~.
\end{eqnarray}
 The new Mellin amplitudes $\widetilde{F}_{q,g}(\omega)$ correspond to the forward Compton
scattering, when the photon virtuality is $\mu^2$. They should be found independently. We again proceed by using new IREE for them. They have a structure similar to Eq.~(\ref{eqfs}). The
difference is that the new equations do not
contain derivatives with respect to $y$ and account for the Born contributions
\begin{equation}\label{fstildeborn}
\widetilde{F}_q^{Born} = \frac{<e^2_q>}{\omega} ,~~~
\widetilde{F}_g^{Born} = 0~,
\end{equation}
where $<e^2_q>$ is the standard notation for the averaged $e^2_q$. So, the IREE for amplitudes
$\widetilde{F}_{q,g}$ are

\begin{eqnarray}\label{eqfstilde}
 &&\omega \widetilde{F}_q^S= <e^2_q> h_{qq}(\omega) ~\widetilde{F}^S_q (\omega) +
 h_{gq}(\omega) ~\widetilde{F}^S_g (\omega,y)~,  \\ \nonumber
 &&\omega \widetilde{F}_g^S= h_{qg}(\omega) ~\widetilde{F}^S_q (\omega) +
 h_{gg}(\omega) ~\widetilde{F}^S_g (\omega)
\end{eqnarray}
and the solution to Eq.~(\ref{eqfstilde}) is
\begin{eqnarray}\label{fstilde}
  \widetilde{F}_q^S &=&  <e^2_q> \frac{\omega - H_{gg}}
  {\omega^2 - \omega (H_{qq} + H_{gg}) + (H_{qq}H_{gg} - H_{qg}H_{gq})}~,
  \\ \nonumber
  \widetilde{F}_g^S &=& <e^2_q> \frac{H_{gq}}
  {\omega^2 - \omega (H_{qq} + H_{gg}) + (H_{qq}H_{gg} - H_{qg}H_{gq})}~.
\end{eqnarray}
Combining Eqs.~(\ref{fstilde}) and (\ref{matchfs}), we obtain the
explicit expressions for $C^{(\pm)}$:
\begin{eqnarray}\label{cpm}
C^{(+)} &=& <e^2_q> \frac{2 H_{gq}H_{qg} - (X - \sqrt{R})(\omega -
H_{gg})} {2 \sqrt{R}[\omega^2 - \omega (H_{qq} + H_{gg}) +
(H_{qq}H_{gg} - H_{qg}H_{gq})]}~, \\ \nonumber C^{(-)} &=& <e^2_q>
\frac{-2 H_{gq}H_{qg} + (X + \sqrt{R})(\omega - H_{gg})} {2
\sqrt{R}[\omega^2 - \omega (H_{qq} + H_{gg}) + (H_{qq}H_{gg} -
H_{qg}H_{gq})]}~.
\end{eqnarray}
Introducing the initial quark and gluon densities $\delta q$ and   $\delta g$ respectively
and using Eq.~(\ref{gf}), we finally arrive at the following expression for $g_1^S$ in region \textbf{B}:
\begin{eqnarray}
\label{gsb} g_1^S(x, Q^2) = \frac{1}{2} \int_{-
\imath \infty}^{\imath \infty} \frac{d \omega}{2 \pi \imath}
 \Big(\frac{1}{x}\Big)^{\omega} \Big[\Big(C^{(+)} e^{\Omega_{(+)}y} + C^{(-)} e^{\Omega_{(-)}y}\Big) \omega \delta q(\omega) +
 \\ \nonumber
 \Big(C^{(+)} \frac{(X + \sqrt{R})}{2 H_{qg}} e^{\Omega_{(+)}y} +
 C^{(-)} \frac{(X - \sqrt{R})}{2 H_{qg}} e^{\Omega_{(-)}y}\Big)\omega \delta g(\omega)\Big]~.
\end{eqnarray}
where $\delta q (\omega)$ and $\delta g (\omega)$ are the initial quark and gluon densities in the $\omega$
-space. $C_q^{(\pm)}(\omega)$ and $C_g^{(\pm)}(\omega)$
are the singlet coefficient
functions calculated in LLA. The exponents $\Omega_{(\pm)}$ are expressed in terms of $H_{ik}$ in the same way as the DGLAP exponents
are related to the DGLAP anomalous dimensions. So, we conclude that $H_{ik}$ are the anomalous dimensions
for $g_1^S$ in LLA. The total resummation of DL contributions to the singlet $g_1$ under the approximation of
fixed $\alpha_s$ was done in Ref.~\cite{ber}. This result was used in Ref.~\cite{blums} where $\alpha_s$ was
running: $\alpha_s =\alpha_s (Q^2)$ according to the renorm group concept. However, in Sect.~IV we showed that
this parametrization does not stand at the small-$x$ region and should be changed by the parameterizations
 of Eqs.~(\ref{a},\ref{aprime}).  The application of the expressions for $g_1$ in Eqs.~(\ref{gnsb},\ref{gsb}) and
the study of  their impact on the Bjorken sum rule can be found in Ref.~\cite{kotl1}.

\section{Asymptotics of the non-singlet $g_1$ in the region B}

The expressions Eqs.~(\ref{gnsb},\ref{gsb}) represent $g_1$ in the region \textbf{B}. Before discussing them in detail,  we  consider first  their asymptotics at fixed $Q^2 \gg \mu^2$ and $x \to 0$. Strictly speaking, such asymptotics can be obtained by applying
the saddle-point method. When the small-$x$ behaviour  is proved to be of the Regge type (power-like),
one can use a short cut by finding the position of the leading (rightmost) singularity in the $\omega$ -plane.
Of course, such singularities can be different for $g_1^{NS}$ and $g_1^S$ and should be found independently.
In the present Sect. we consider the small -$x$ asymptotics of the non-singlet $g_1$.

\subsection{Asymptotic scaling}

Let us assume that the initial quark density $\delta q$ in Eq.~(\ref{gnsb}) is non-singular in $x$ at $x \to 0$,
so it does not contribute to the small- $x$ asymptotics. Then by applying the saddle-point method to Eq.~(\ref{gnsb})
one deals with the non-singlet coefficient function and the anomalous dimension only. In this case the stationary point
is (see Appendix E for details)
\begin{equation}
\label{saddle} \omega_0 = \sqrt{B_{NS}} \Big[ 1 + (1 -
\kappa)^2 (y/2 + 1/\sqrt{B_{NS}})^2/(2\ln^2 \xi) \Big],
\end{equation}
with
\begin{equation}
\label{kappa}
 \kappa = d \sqrt{B}/ d\omega |_{\omega = \omega_0}
\end{equation}
and $y = \ln(Q^2/\mu^2)$, $\xi = \sqrt{Q^2/(x^2 \mu^2)}$.

It immediately leads to the Regge asymptotics for the
non-singlets:

\begin{equation}
\label{as} g_1^{NS} \sim \frac{e_q^2}{2} \delta q (\omega_0)
\Pi_{NS} \xi^{\omega_0/2},
\end{equation}
with
\begin{equation}
\label{pi} \Pi_{NS} = \frac{\Big[2(1 -
\kappa)\sqrt{B_{NS}}\Big]^{1/2} \big(y/2 +
1/\sqrt{B_{NS}} \big)}{\pi^{1/2}\ln^{3/2}\xi} .
\end{equation}

When in Eq.~(\ref{saddle}) $y \ll 2/\sqrt{B_{NS}(\omega)}$
(let us notice in advance that at $\omega = \Delta_{NS}$
it means that $y \ll 150 ~\mu^2$),
the value of
$\omega_0$ does not depend on $y$ at all.
 Therefore Eq.~(\ref{as}) can be rewritten as:

\begin{equation}
\label{asympt}
g_1^{NS}(x,Q^2) \sim (e_q^2/2) \delta q(\omega_0)
c_{NS} T_{NS}(\xi)
\end{equation}
with
\begin{equation}
\label{tns}
 T_{NS}(\xi) = \xi^{\omega_0/2}/\ln^{3/2}\xi.
\end{equation}
The factor $c_{NS}$ is
\begin{equation}
\label{c}
 c_{NS} = \Big[2(1 - \kappa)/(\pi \sqrt{B_{NS}}) \Big]^{1/2}
\end{equation}
and does not depend on $y$. Eq.~(\ref{asympt}) predicts the
scaling behavior for the non-singlet structure functions: in the region
 $Q^2\ll 150\, \mu^2$, with $T^{(\pm)}$ depending on one argument $\xi$
instead of $x$ and $Q^2$ independently. Therefore in this region
\begin{equation}\label{asscal}
g_1^{NS} \sim \widetilde{g}_1^{NS}  \equiv \Pi_{NS}(\omega_0)
 \delta q (\omega_0)\big(Q^2/x^2\mu^2
\big)^{\omega_0/2}
\end{equation}
at $x \to 0$, with $\omega_0$ being the largest root of Eq.~(\ref{eqbrns1}):

\begin{equation}\label{eqbrns1}
\omega^2 - B_{NS} =0 .
\end{equation}
We call the result of  Eq.~(\ref{asscal}) \textbf{asymptotic scaling}:
$g_1^{NS}$ asymptotically depends on one variable $Q^2/x^2$ only,
instead of two variables $x$ and $Q^2$. The
DGLAP prediction for the asymptotics of $g_1^{NS}$ in Eq.~(\ref{asdglap}) is quite
different. Below we compare these results in detail \\
 According to the results of Ref.~\cite{egtinp} (see also Appendix E), in the opposite case, when
 $Q^2 \gtrsim ~150~\mu^2$, the rightmost and non-vanishing at $w \to \infty$ stationary point is again given by
Eq.~(\ref{eqbrns1}) but the pre-exponential factor $\Pi_{NS}$ essentially depends on $y$,
so the asymptotic scaling in this region holds for $g_1^{NS}/y$. \\
Let us notice that the sign of the non-singlet asymptotic behaviour is
positive when $\delta q(\omega_0)$ is positive and coincides with
the sign of $g_1^{NS}$ in the Born approximation. In other words,
both the $x$ and $Q^2$ -evolutions do not affect the sign of
$g_1^{NS}$.
 Now we focus on solving Eq.~(\ref{eqbrns1}).

\subsection{Estimate for the non-singlet intercept}

\begin{figure}[h]
\begin{center}
\begin{picture}(320,200)
\put(0,0){
\epsfbox{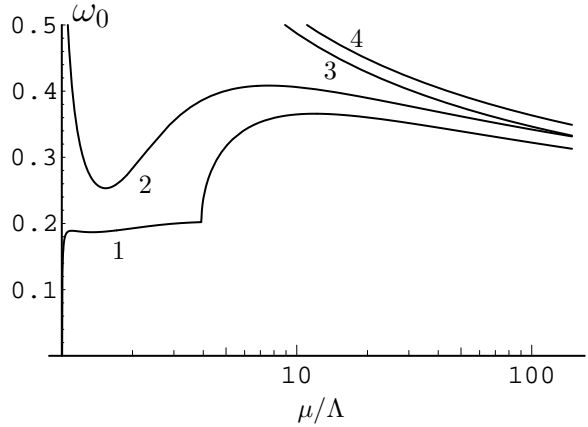} }
\end{picture}
\end{center}
\caption{\label{g1sumfig5}Dependence of the intercept $\omega_0$
on infrared cutoff $\mu$~: 1-- for $F_1^{NS}$; 2-- for $g_1^{NS}$;
3-- and 4-- for $F_1^{NS}$ and $g_1^{NS}$ respectively without
account of $\pi^2$-terms. The structure function $F_1^{NS}$ is
discussed in the Appendix B.}
\end{figure}

In a more detailed form, equation Eq.~(\ref{eqbrns1}) is
\begin{eqnarray}\label{eqbrns}
\omega^2 - (1 + \lambda_{qq}\omega) \Big[\Big(\frac{2 C_F}{\pi b}\Big)
\Big(\frac{\eta}{\eta^2 + \pi^2} - \int_0^{\infty} \frac{d \rho e^{- \omega \rho}}
{\rho^2 + \pi^2}\Big) + \\ \nonumber
\Big(\frac{2 C_F}{\pi b}\Big)^2 \frac{1}{4N C_F}
\int_0^{\infty} d \rho e^{- \omega \rho} \ln ((\rho + \eta)/\eta)
\Big(\frac{\rho + \eta}{(\rho + \eta)^2 + \pi^2}\ + \frac{1}{\rho + \eta}\Big)\Big] =0
\end{eqnarray}
where we have used the notation $\eta = \ln (\mu^2/\Lambda^2)$. Let us remind that in the
case of fixed $\alpha_s$ Eq.~(\ref{eqbrns}) is much simpler and can be solved analytically:
\begin{equation}\label{eqbrnsafix}
\omega^2 - \frac{2 \alpha_s C_F}{\pi} - \Big(\frac{2 \alpha_s C_F}{\pi}\Big)^2
\frac{1}{\omega^2 4N C_F} = 0,
\end{equation}
with the obvious solution $\omega_0^{DL}$ given by the following expression\cite{ber}:
\begin{equation}\label{brnsafix}
\omega_0^{DL} = (2 \alpha_s C_F/\pi)^{1/2} \sqrt{\frac{1}{2}\Big[
1 + \Big(1 + \frac{4}{N^2 - 1}\Big)^{1/2}\Big]}.
\end{equation}
In contrast, Eq.~(\ref{eqbrns}) cannot be solved analytically.
Besides, there is a big qualitative difference between the cases
of  fixed and running $\alpha_s$. Although the IR cut-off $\mu$
is used for regulating the IR divergencies in both cases,
Eq.~(\ref{eqbrnsafix}) is  free of any $\mu$
-dependence whereas Eq.~(\ref{eqbrns}) is obviously $\mu$
-dependent and therefore the solution $\omega_0$ also depends on
$\mu$: $\omega_0 = \omega_0 (\mu)$. The value of $\mu$ is
restricted by Eq.~(\ref{mulambda}) only. As a result, we arrive at the solution to
Eq.~(\ref{eqbrns}) in the form of the curve plotted in
Fig.~\ref{g1sumfig5}.
Eq.~(\ref{eqbrns}) shows that $\omega_0$ depends on $\mu$ through
$\eta$, therefore $\omega_0$ depends on the ratio
$\mu / \Lambda$ and on $n_f$.  Besides the $\eta$ -dependence, $\omega_0$
is not sensitive to the value of $\Lambda$.
The plot
in Fig.~\ref{g1sumfig5} shows that the curve $\omega_0 =
\omega_0 (\mu)$ rapidly grows at  $\mu \lesssim \Lambda$, however this
region contradicts Eq.~(\ref{mulambda}), so the
perturbative expression (\ref{alphas}) for $\alpha_s$ cannot be used at so small
$\mu$. Both Eq.~(\ref{eqbrns}) and the plot in Fig.~\ref{g1sumfig5} are consistent in the region
(\ref{mulambda}) only and should not be considered out of this region. Eq.~(\ref{eqbrns})
has one maximum in region (\ref{mulambda}):
\begin{equation}\label{intns}
\Delta_{NS} \equiv \max \big[\omega_0 \big]= \omega_0 (\mu_{NS}) = 0.42
\end{equation}
at
\begin{equation}\label{muns}
\mu =\mu_{NS} \equiv \Lambda e^{2.3} \approx 10 \Lambda.
\end{equation}
For the sake of simplicity we chose in Ref.~\cite{egtsns} $n_f = 3$ and
$\Lambda = 0.1$~GeV. It would have been more realistic to choose $\Lambda = 0.5$~GeV,
and  Eq.~(\ref{muns}) shows that such a change of $\Lambda$ leads to multiply by a
factor of $5$ the values of $\mu$ for the singlet and non-singlet obtained in Ref.~\cite{egtsns}.
Comparison of the curves 1 and 2 in Fig.~\ref{g1sumfig5} to the curves 3 and 4 shows that the
important role played by the $\pi^2$ -terms in $\alpha_s$ (i.e. respecting the analyticity) for
producing a maximum in the curves and 2.
Furthermore in the vicinity of this maximum, the power expansion
\begin{equation}\label{max}
\omega_0 (\mu) = \omega_0 (\mu_{NS}) +  \frac{d\omega_0 (\mu_{NS})}{d \mu} (\mu - \mu_{NS})
+ \frac{1}{2} \frac{d^2\omega_0 (\mu_{NS})}{d \mu^2} (\mu - \mu_{NS})^2 + ...
\end{equation}
does not contain the linear term,  so $\omega_0 (\mu_{NS}) \equiv \Delta_{NS}$ is much less dependent on $\mu$
than all other points on the curves 1 and 2.
This remarkable feature allows us to identify $\Delta_{NS}$ as the
best candidate\footnote{We are grateful to P.~Castorina for this
very useful observation.} for the perturbative estimate of
the  genuine intercept of the non-singlet $g_1$. According to the
prediction of the Regge approach, the genuine intercept should
be a constant, with no other dependence. However,
Eq.~(\ref{eqbrns}) and its solution (\ref{intns}) account for the
leading logarithmic contributions only and leave aside sub-leading
perturbative contributions and possible non-perturbative ones, so
it is hardly possible to identify (\ref{intns}) with the genuine
intercept. Nevertheless, it turned out that our estimate
(\ref{intns}) is in a good agreement with the results of
Ref.~\cite{kat} obtained by fitting all available experimental data.
This leads to a very interesting conclusion: by some unknown reason
all sub-leading and non-perturbative contributions to the
non-singlet intercept happen to be either small or irrelevant
at $\mu = \mu_{NS}$, so that the LLA prediction (\ref{intns}) proves to be a
good estimate for the non-singlet intercept. Motivated by this result, we call
$\mu_{NS}$ the Optimal non-singlet mass scale. However, it is worth stressing that this scale is an
artefact of our approach and should disappear when non-perturbative contributions (also dependent
on the same scale) would be accounted for. To conclude, let us notice that the Regge form of the
small-$x$ asymptotics of $g_1$ is the direct consequence of the total
 resummation of logarithms of $x$ and cannot appear at fixed orders in $\alpha_s$.
 This small-$x$ asymptotics depends on $Q^2$ only through
 the factor $(Q^2/\mu^2)^{\Delta_{NS}/2}$ (see Eq.~(\ref{asscal})).
In particular it means that the intercept $\Delta_{NS}$
has no dependence on $Q^2$. On the other hand, the well-known DGLAP small- $x$ asymptotics

\begin{equation}\label{asdglap}
g_1^{NS~DGLAP} \sim \exp \Big[\sqrt{\frac{2 C_F}{\pi b}\ln (1/x)\ln
\Big(\frac{\ln (Q^2/\Lambda^2)}{\ln(\mu^2/\Lambda^2)}\Big)}\Big] .
\end{equation}

can also be obtained (see Appendix F for detail)
with the saddle-point method providing the initial parton densities are
not singular
at $x \to 0$ (In Sect.~XII we consider the alternative case presently used in the Standard Approach for the analysis of
experimental data at small $x$).
The DGLAP -asymptotics (\ref{asdglap}) clearly
does not exhibit the Regge behavior. The same is true for the case where the anomalous dimensions and
coefficient functions are calculated in high but fixed orders in $\alpha_s$ which would
correspond to the NN..NLO DGLAP accuracy (see Appendix G for detail).
In principle, one might think that a
generalization of Eq.~(\ref{asdglap}) could lead to the Regge asymptotics, however with the intercept depending on
$Q^2$. We show now that there are no theoretical grounds for such a scenario.
Indeed, it follows from Eq.~(\ref{dglapstatpoint}) that the $Q^2$ -dependence in Eq.~(\ref{asdglap})
is the consequence of the use of the DGLAP -parametrization $\alpha_s = \alpha_s (k^2_{\perp})$
and the DGLAP -ordering (\ref{dglapord}). As explained in detail in Appendix F, at small $x$ this ordering should be changed
by the
ordering (\ref{dlord}). Then the upper limit $Q^2$ in Eq.~(\ref{dglapord})
in the small-$x$ region should be modified to $w$ (see Ref.~\cite{etalfa} for detail).
After that the DGLAP asymptotics will not depend on $Q^2$ but at the same time will
not have a Regge-type form. The Regge asymptotics is achieved
by accounting for the resummation of the
leading logarithms of $x$. It exhibits an asymptotic behavior
much steeper than the DGLAP result (\ref{asdglap}),
not only with respect to $x$ but also with respect to $Q^2$.  The comparison of Eq.~(\ref{asdglap})
to Eq.~(\ref{asscal}) shows that
$g_1^{NS~DGLAP}/g_1^{NS} \to 0$ when $x \to 0$. The question however arises: how small should
$x$ be in order to allow our asymptotic expression Eq.~(\ref{asscal}) to represent
$g_1^{NS}$ reliably? We answer this question below.

\subsection{Applicability region of the small-$x$ asymptotics}

\begin{figure}[h]
\begin{center}
\begin{picture}(240,160)
\put(0,0){
\epsfbox{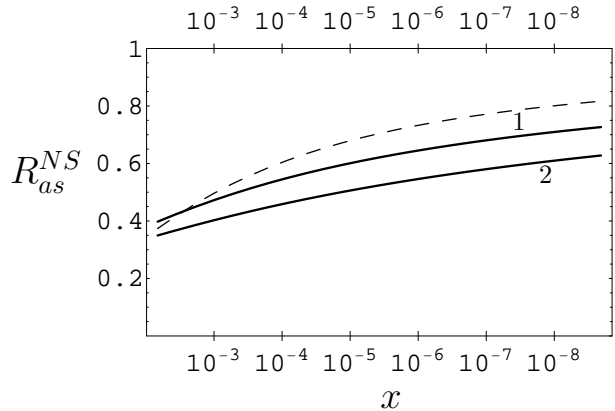} }
\end{picture}
\end{center}
\caption{\label{g1sumfig6} Rate of  the $g_1^{NS}$ approach to
asymptotics for different $Q^2$: solid curve 1 for $Q = 10\mu$~,
solid curve 2 for $Q = 100\mu$~, dashed curve for $Q = \mu$.}
\end{figure}

The asymptotic expression (\ref{asscal}) for $g_1^{NS}$ is
obviously much simpler than the integral representation
(\ref{gnsb}) and also much easier to work with. However, it is valid
for very small $x$ only. In order to determine when
Eq.~(\ref{asscal}) reliably represents Eq.~(\ref{gnsb}), let us
study numerically the ratio
\begin{equation}\label{ras}
R^{NS}_{as} (x, Q^2) = \frac{g_1^{NS}(x,
Q^2)}{\widetilde{g}_1^{NS}(x, Q^2)}
\end{equation}
at fixed $Q^2$ and different values of $x$. The result is plotted
in Fig.~\ref{g1sumfig6}.
According to it, $g_1^{NS}$ is reliably represented by its
asymptotic expression  $\widetilde{g}_1^{NS}(x, Q^2)$ at $x \lesssim 10^{-6}$
only. So, strictly speaking, Eq.~(\ref{asscal}) should not be used
at available values of $x$. However, in the literature one can
find that Regge type ($\sim x^{-a}$) fits of the experimental
data are used at much larger values of $x$, and such fits are reported to
work well. We suggest a simple explanation to this:
the phenomenological parameterizations including the Regge type fits
 have nothing in common with the expression in
Eq.~(\ref{asscal}) obtained with the saddle-point method. In order to use
such parameterizations at relatively large values of $x$ (at $x \gg 10^{-6}$),
one can choose the exponents $a$ in the fits greater than the genuine intercepts.
An analysis of such Regge parameterizations can be found in
Ref.~\cite{kotl2}. To conclude, we also notice also that sometimes
 Regge parameterizations are used with intercepts depending on $Q^2$: they  have no
theoretical ground and contradict  Eq.~(\ref{asscal}).

\section{Small-$x$ asymptotics of the singlet $g_1$ in the region B}

The small- $x$ asymptotics of $g_1^S$ in the region\textbf{ B} can be obtained quite similarly to the non-singlet case,
 by applying the saddle-point method to Eq.~(\ref{gsb}).
The singlet asymptotics also exhibits the Regge behavior:
\begin{equation}\label{sas}
g_1^S (x,Q^2) \sim (1/x)^{\omega_0} (Q^2/\mu^2)^{\omega_0/2}
[A(\omega_0) \delta q (\omega_0)+ B (\omega_0) \delta g
(\omega_o)]
\end{equation}
where $\omega_0$ is the stationary point and $A,~~B$ include the asymptotic form
 of the coefficient functions.
The position of the leading singularity corresponds the largest root of the equation
\begin{equation}\label{eqbrs}
(\omega^2 - 2(b_{qq} + b_{gg}))^2 - 4 (b_{qq} - b_{gg})^2 -
16b_{gq} b_{qg}  = 0.
\end{equation}
Similarly to the non-singlet case, the position of the singlet leading singularity depends on $\mu$, with
one maximum
\begin{equation}\label{ints}
\omega_0^S \equiv \Delta_S = 0.86
\end{equation}
achieved at (we choose again $n_f = 3$)
\begin{equation}\label{mus}
\mu = \mu_S \approx \Lambda e^4 \approx 55 \Lambda
\end{equation}
which gives $\mu_S/\Lambda \approx 55$~GeV when $\Lambda = 0.1~$GeV. By repeating the arguments given  also in the
previous Sect., we call $\Delta_S$ the singlet intercept and
 call $\mu_S$ the Optimal singlet mass scale. The Optimal singlet and non-singlet
mass scales are quite different.
Our perturbative estimate (\ref{ints}) is  also in a very
good agreement with the result obtained in Ref.~ \cite{koch} by fitting the experimental data.
Eq.~(\ref{sas}) shows that the asymptotic scaling is also valid
for the singlet $g_1$: asymptotically $g_1^S$ depends on one
argument $Q^2/x^2$ only.
 In contrast to the case of the non-singlet
asymptotics (\ref{asscal}), the interplay between $\delta q$ and
$\delta g$ can affect the sign of $g_1^S$. Indeed, in the Born
approximation $g_1^S > 0$ but it can be negative (positive)
asymptotically depending on the sign of $A \delta q + B \delta g$ in
Eq.~(\ref{sas}).

\section{Applicability region of the IREE method}
In this Sect. we discuss the region of applicability of our approach and also answer a claim on a
possible contradiction in our method that we have got in the past: in the IREE
technology that we use to sum up the leading logarithms we work in the $\omega$ -space and
systematically keep $\omega$ small. On the other hand when we calculate the non-singlet and especially
the singlet
intercepts, they are found not so small. Therefore we should guarantee the validity of our
method not only at small but also at large $\omega$. We start from the conventional analysis of the
double-logarithmic QCD power series
(\ref{dlser}), so we would like to stress at once that the use of (\ref{dlser}) for analysis of QCD processes
and all estimates (e.g. the one in Eq.~(\ref{xmin})) based on it originate from the QED results
(see e.g. Ref.~\cite{g}) where the running coupling
effects can be neglected. They can become unreliable in QCD and therefore these conventional estimates
should be
replaced by more accurate estimates which we present in this Sect.

Now let us remind the basic principle of the Leading Logarithmic Approximation, and DLA in particular.
The straightforward calculation of the Feynman graphs contributing to a certain quantity (for instance, to the
non-singlet coefficient) yields the double-logarithmic contributions. In the $n$-th order of the perturbation theory
they are  $\sim \alpha_s^n \ln^{2n}(1/x)$. For the sake of simplicity we keep here $\alpha_s$ fixed and leave out
other numerical parameters  like $1/\pi$, color factors, etc. Accounting for the running coupling
effects and sub-leading logarithmic contributions does not change the essence of the problem.
The series of such contributions, for example
\begin{equation}\label{dlser}
 c_1 \alpha_s \ln (1/x) + c_2 \alpha_s^2 \ln^3 (1/x) + c_3 \alpha_s^4 \ln^5 (1/x) + ...
\end{equation}
converges when $\alpha_s \ln^2 (1/x) < 1$
only.
It brings us to a  rough estimate for the lowest limit for $x$:
\begin{equation}\label{xmin}
x > x_{min} = \exp [-1/\sqrt{\alpha_s}].
\end{equation}
 On the other hand, it is interesting to know the result of the
total resummation of the DL terms at really small $x < x_{min}$
and even at $x \to 0$. The reason is that the DL terms are not so
large compared to other contributions at $x \gtrsim x_{min}$ but
dominate at small $x$. However, the series in
Eq.~(\ref{dlser}) diverges at $x < x_{min}$ and cannot be summed
up in this region. The solution to this problem is well-known: in the first place the series
Eq.~(\ref{dlser}) should be summed up at $x > x_{min}$ and then the
result of the resummation can be analytically continued into
the region $x < x_{min}$. Let us notice that the series Eq.~(\ref{dlser}) becomes
divergent in region $x < x_{min}$ and is called an asymptotic
series. In the $\omega$ -space the series Eq.~(\ref{dlser}),
according to the relation
\begin{equation}\label{mellinfix}
\int_{- \imath \infty}^{\imath \infty} \frac{d \omega}{2 \pi \imath}~
e^{\omega \ln(1/x)}\frac{1}{\omega^{1+2n}} = \frac{1}{(2n)!}\ln^{2n}(1/x) ,
\end{equation}
is given by:
\begin{equation}\label{dlsermel}
\widetilde{c}_1 \frac{\alpha}{\omega^2} + \widetilde{c}_2 \frac{\alpha^2}{\omega^4} +
\widetilde{c}_3 \frac{\alpha^3}{\omega^6} + ...
\end{equation}
This series converges when, roughly,
\begin{equation}\label{dlconvmel}
\omega > \omega_{min} = \sqrt{\alpha_s}
\end{equation}
whereas the DL terms becomes large at  $\omega < \omega_{min}$. To be specific, let us notice
that the expressions for the coefficient
functions in Eqs.~(\ref{cns},\ref{cpm}) represent the total sum
of the DL contributions. Strictly speaking, the coefficient functions should first be
calculated for large $\omega$: $\omega > \omega_{min}$ (where the DL contributions are small) and then continued
to the region of small $\omega$.
However,  anticipating  the analytical continuation into the small -$\omega$ region, quite often a
short cut is taken and we follow this way: we treat $\omega$  as small since the beginning.  It
gives us the reason to neglect
 all non-logarithmic corrections regardless of their relatively large
values in the region (\ref{dlconvmel}). After the total resummation of the
DL terms has been done, the  formulae obtained are insensitive to the value of $\omega$ and therefore
 can be used at any $x$. The resummed expressions (\ref{cns},\ref{cpm}) contain new singularities $\omega_0$
(branching points in our case but, generally, they can also be poles),
which are absent in the series (\ref{dlsermel}). The rightmost singularities, i.e. the intercepts
($\Delta_{NS}$ and $\Delta_S$ in our case), determine the range of convergence of the series Eq.~(\ref{dlser})
instead of $\sqrt{\alpha_s}$: the series (\ref{dlsermel}) converges only if
\begin{equation}\label{dlconvmeldelta}
\omega > \Delta ,
\end{equation}
with $\Delta$ being the intercept. However, after the total resummation in (\ref{dlsermel})
has been performed, the result of the resummation can be used at arbitrary
values of $\omega$.
To conclude, we note that all  equations for the resummation of the  leading logarithms, and
in particular the IREE we have used, are not the equations for finding the intercepts. Indeed,
the intercepts are the singularities and the values of $\omega$ in the IREE should be kept
pretty far away from
them by definition. The intercepts appear in the asymptotic expressions and therefore they
should be found independently of the resummation methods,
usually by applying the saddle-point method.

\section{Comparison of $g_1$ to  $g_1^{DGLAP}$ in the region B}

In this Sect. we compare our results (\ref{gnsb},\ref{gsb})  to
the DGLAP expressions for $g_1$ at small $x$.  We are not going
to use here the asymptotic expressions in Eqs.~(\ref{asscal},\ref{sas}) or (\ref{asdglap}),
but  we compare the two
approaches at small but finite $x$.  First  we will compare the basic ingredients of
the expressions for $g_1$: the anomalous dimensions and the coefficient functions.
Whenever it is possible, we will consider in  detail, for the sake of simplicity,
the non-singlet $g_1$ and more briefly generalize our results to the case of $g_1^S$ .

\subsection{Comparison of the coefficient functions and the anomalous dimensions}

Eqs.~(\ref{gnsb}) and (\ref{g1nsdglap}) have a similar
structure: each integrand contains the initial parton density, the
coefficient function and the exponent with the non-singlet
anomalous dimension to govern the $Q^2$ -evolution. However,
$C_{NS}$ and $h_{NS}$ in Eq.~(\ref{gnsb}) contain the total
resummation of the leading logarithms of $x$ whereas in
Eq.~(\ref{g1nsdglap}) the coefficient function and the
anomalous dimension are considered to LO and NLO accuracy,
namely  they are given in
Eqs.~(\ref{cnsdglap},\ref{gnsdglap}). Originally DGLAP was
suggested for studying the region \textbf{A} of large $x$ and
large $Q^2$. Due to the oscillating  factor $x^{- \omega}$ in
the Mellin integrals, the main contribution to $g_1$ in the
region \textbf{B} comes from small $\omega$. On the contrary, the
main contribution in region \textbf{A} comes from large $\omega$.
At large $\omega$, the expressions for $C_{NS}$ and $h_{NS}$  in
Eq.~(\ref{gnsb}) can be expanded into a converging series in
$1/\omega$:
\begin{equation}\label{cnsser}
C_{NS} = 1 + \frac{A(\omega) C_F}{2 \pi} \Big[\frac{1}{\omega^2} + \frac{1}{2 \omega}\Big] + ...,
\end{equation}
\begin{equation}\label{gnsser}
h_{NS} = \frac{A(\omega) C_F}{2 \pi} \Big[\frac{1}{\omega} + \frac{1}{2}\Big] + ...
\end{equation}
Obviously we observe a large discrepancy between
Eqs.~(\ref{cnsser},\ref{gnsser}) and the LO DGLAP expressions in
Eqs.~(\ref{cnsdglap},\ref{gnsdglap}). However, this discrepancy
almost disappears when we come back to the region \textbf{B} where
$\omega$ is small and therefore regular terms $\sim \omega^k$ in
Eqs.~(\ref{cnsdglap},\ref{gnsdglap}) can be dropped. The remaining
discrepancy is due to  the different   treatment of the QCD
coupling. When the starting point of the $Q^2$
-evolution obeys Eq.~(\ref{mulambda}),  then $A(\omega)$  with very good
approximation can be replaced by $\alpha_s (k^2_{\perp}/x)$, but
definitely  not by $\alpha_s (k^2_{\perp})$. Taking into account  more terms
in the series and adding them to Eqs.~(\ref{cnsser},\ref{gnsser}) does not change
the situation. So, we conclude that in region \textbf{B}  the first and
second
terms of the $1/\omega$ -expansion of Eqs.~(\ref{hns},\ref{cns})
reproduce the most important LO and NLO DGLAP results in the non-singlet
anomalous dimension and coefficient function, with the exception of  the different treatment
of the QCD coupling. Expanding Eqs.~(\ref{hik},\ref{cpm}) into a
series in $1/\omega$ and comparing the result to the singlet DGLAP anomalous dimensions
and coefficient functions, we arrive at the same conclusion.

\subsection{Numerical comparison of the $x$ -evolutions in Eqs.~(\ref{g1nsdglap}) and (\ref{gnsb})}

The integrands in Eqs.~(\ref{g1nsdglap}) and (\ref{gnsb}) for the
non-singlet $g_1$ contains also a  phenomenological ingredient: the
initial quark densities $\delta q$.
 Let us introduce the ratio

\begin{equation}\label{rns}
R_{NS} = \frac{g_1^{NS}}{g_1^{NS~DGLAP}}
\end{equation}
and study its $x$ -dependence at fixed $Q^2$, for example at $Q^2 = 10~$GeV$^2$. Obviously, this
cannot be done until $\delta q$ is fixed. The choice
\begin{equation}\label{qarkfit}
\delta q (x)= N_q \delta (1-x)
\end{equation}
corresponds to approximate the initial hadron by a quark and to neglect all influence of the hadron structure. Of course, such a choice cannot be used for phenomenological applications, but it
makes possible to compare the $x$ -evolutions in Eqs.~(\ref{g1nsdglap}) and (\ref{gnsb}). The substitution of
the bare quark input into Eqs.~(\ref{g1nsdglap}) and (\ref{gnsb}) leads to the $x$ -dependence of $R_{NS}$
plotted in Fig.~7.
\begin{figure}[h]
\begin{center}
\begin{picture}(240,160)
\put(0,0){
\epsfbox{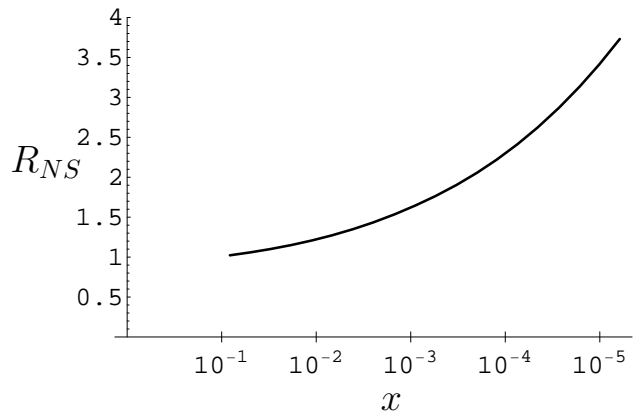} }
\end{picture}
\end{center}
\caption{\label{g1sumfig7}Rise of $R_{NS}$ of Eq.~(\ref{rns}) at
small $x$ in case of bare quark input and for $Q^2 = 10$~GeV$^2$.}
\end{figure}
This shows that the impact of the leading logarithms becomes quite sizable at
$x_0 \approx 10^{-2}$. So, we arrive to a sort of puzzle: \\

According to the different behaviour of  the $x$ -evolution in Eqs.~(\ref{g1nsdglap}) and (\ref{gnsb}),
the DGLAP -description of $g_1^{NS}$ should have failed for $x < 10^{-2}$, but phenomenologically  it is well-known that DGLAP works well at $x < 10^{-2}$.
The solution to this puzzle is given below.

\subsection{The role of the initial parton densities}

In order to clarify the problem, let us consider in more detail a standard fit to the initial quark density, as in Eq.~(\ref{fitsb}):
\begin{equation}\label{fita1}
\delta q (x) = N_q x^{-\alpha}(1-x)^{\beta} (1 + \gamma x^{\delta}) \equiv
N_q x^{-\alpha} \varphi(x). \nonumber
\end{equation}
with all parameters $N_q,~\alpha,~\beta,~\gamma,~\delta$ being positive. As the fit is defined at certain
fixed values of
$x= x_0$ and $Q^2 = Q^2_0$, all its parameters depend on $x_0,~Q^2_0$.
We define $N_q$ as the normalization. As the term $x^{- \alpha} \to \infty$  when $x \to 0$, we call it  the singular term,
although  the fit is introduced at large $x$. We call $\varphi$ the regular part of the fit because
$\varphi \to 1$ when $x \to 0$.
Once transformed into the $\omega$ -space, the fit becomes a sum of  pole contributions:
\begin{equation}\label{fitamel}
\delta q (\omega) = N_q \Big[(\omega - \alpha)^{-1} +
\sum_{k=1}^{\infty} \Big(m_k (\omega + k - \alpha) + \gamma
(\omega + k +\delta - \alpha)^{-1}\Big)\Big]
\end{equation}
where $m_k = \beta (\beta-1)...(\beta-k+1)/k!$. The first pole in
Eq.~(\ref{fitamel}) corresponds to the singular term $x^{-a}$ in
Eq.~(\ref{fita}). We call it the leading pole. The other,
non-leading poles in Eq.~(\ref{fitamel}) originate from the
interference between $\varphi(x)$ and $x^{-a}$. Substituting
Eq.~(\ref{fitamel}) into the DGLAP expression (\ref{g1nsdglap}),
we see that the contribution of the leading pole,
$\widetilde{g}_1^{NS~DGLAP}$ to $g_1^{NS~DGLAP}$ is (we drop the
NLO contribution here)
\begin{equation}\label{leadpol}
\widetilde{g}_1^{NS~DGLAP} (x,Q^2) = \frac{e^2_q}{2} N_q
\Big(\frac{1}{x}\Big)^{\alpha} C^{NS~DGLAP}(\alpha)
\Big(\frac{\ln(Q^2/\Lambda^2)}{\ln(\mu^2/\Lambda^2)}\Big)^{\gamma_{(0)}(\alpha)/(2\pi
b)}.
\end{equation}
Substituting the other terms of Eq.~(\ref{fitamel}) into Eq.~(\ref{g1nsdglap}) it leads to a
contribution quite similar to that in Eq.~(\ref{leadpol}), however with
$\alpha \to \alpha_k = \alpha - k$ and $\alpha - k - 1$. Obviously, $\alpha > \alpha_k$.
Therefore  $\widetilde{g}_1^{NS~DGLAP}$ in Eq.~(\ref{leadpol}) is really the leading
contribution to $g_1^{NS~DGLAP}$ at small $x$ and actually it represents the
small- $x$ asymptotics of $g_1^{NS~DGLAP}$. Confronting Eq.~(\ref{leadpol}) to the very well-known
expression (\ref{asdglap}) for the DGLAP asymptotics, we see that they are totally different.
The singular terms are also included into the DGLAP parametrization of
the  singlet parton densities. It leads to the steep growth of $g_1$ at small $x$ and provides
the reason for the  agrement between the DGLAP -description of the structure functions and the experimental data.
Therefore the DGLAP success st small $x$ is related to the use of  singular fits for  the
initial parton densities.This is the solution to the puzzle.
Now let us discuss the most important consequences of this result.

\begin{figure}[h]
\begin{center}
\begin{picture}(320,200)
\put(0,0){
\epsfbox{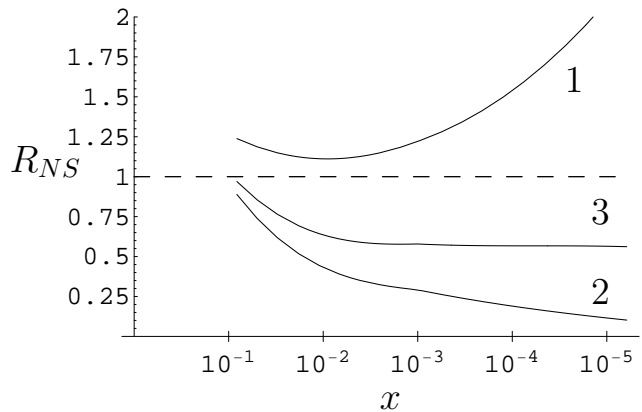} }
\end{picture}
\end{center}
\caption{\label{g1sumfig8}Examples of the small $x$ -behaviour of
$R_{NS}$ for different singular terms in the initial quark
distribution fits of Eq.~(\ref{fita}): $\alpha = 0$ (curve 1),
$\alpha = 0.576$ (curve 2), $\alpha= 0.36$ (curve 3). All curves
correspond to $Q^2 = 10$~GeV$^2$, and involve only a regular part,
$\varphi$, in the parametrization  of the initial quark density used for $g_1$ of Eq.~(\ref{gnsb}). }
\end{figure}

 First, let us confront the asymptotics of
Eqs.~(\ref{leadpol}) and (\ref{asscal}). We see that the $x$
-dependence in these expressions is identical: both formulae exhibit the Regge
(power-like) behavior. It allows us to conclude that the singular
term $x^{-\alpha}$  in the standard DGLAP fits mimics the total
resummation of the leading logarithms of $x$. Therefore, the
singular factors can be dropped when the total resummation of the
leading logarithms of $x$ is accounted for. In order to show it
explicitly, let us study numerically $R_{NS}$, using the
standard DGLAP fit of Eq.~(\ref{fita}). The results are plotted in
Fig.~\ref{g1sumfig8}.
 We can observe that $R_{NS}$
is pretty close to unity only when the fit (\ref{fita}) is used in
the expression for $g_1^{NS~DGLAP}$, whereas only the regular part,
$\varphi$, of the fit is used in the resummed expression
(\ref{gnsb}). All other options drive $R_{NS}$ far away from unity
at small $x$. So, the resummation of  the leading logarithms
leads to simplify the standard fits. Fig.~\ref{g1sumfig8} explicitly demonstrates that
the singular fit (\ref{fita}) and the total resummation of
the logarithms lead to close values of $g_1$ in region \textbf{B}. From a
practical point of view, the use of the
resummation is preferable because it allows one to construct new fits with a reduced
number of parameters. From the theoretical point of view, the resummation is even more preferable. Indeed,
the DGLAP intercept $\alpha$ in Eq.~(\ref{leadpol}) depends on the starting point $x_0,~Q^2_0$, where $\alpha$ is fixed,
and such a
dependence can hardly be deduced from theoretical considerations. On the contrary, the intercept $\Delta_{NS}$ in
Eq.~(\ref{asscal}) is independent of the initial parton densities.

Now we would like to comment on an apparent puzzle arising first from the results of
Refs.~\cite{bv,fitsa}  and then also in
other subsequent publications (see e.g. the recent review \cite{lead}): \\
On one hand, the direct comparison in Ref.~\cite{bv} of the NLO DGLAP result,
$g_1^{NS~DGLAP}$ to the expression for $g_1^{NS}$ obtained in Ref.~\cite{ber}
in the limit of fixed $\alpha_s$
showed
that
\begin{equation}\label{bluml}
g_1^{NS}/g_1^{NS~DGLAP} \ll 1
\end{equation}
 at the small values of $x$
available in experiment and therefore the conclusion was made that
the resummation of $\ln^k(1/x)$ can yield a small impact.
On the other hand, it is clear that the small-$x$ asymptotics (\ref{asscal}) is much steeper than
the well-known DGLAP
asymptotics (\ref{asdglap}), which proves that asymptotically
\begin{equation}\label{antibluml}
g_1^{NS}/g_1^{NS~DGLAP} \gg 1
\end{equation}
 and therefore the total resummation of
$\ln^k(1/x)$ should be essential. So, Eqs.~(\ref{bluml}) and (\ref{antibluml}) obviously contradict
each other, which is puzzling.

Eq.~(\ref{bluml})
was interpreted in the literature as follows: the resummation of leading logarithms
at values of $x$ available in experiment is much less important
than the impact of the sub- leading (compared to the double-logarithmic contributions)
terms in the DGLAP coefficient functions and anomalous dimensions.  \\
However, a close inspection of
Eqs.~(\ref{fita},\ref{fita1}) for the standard fit suggests another
solution to this puzzle. Fig.~8 clearly demonstrates that the main impact on the small-$x$ behavior of
$g_1^{NS~DGLAP}$ comes not from the NLO DGLAP coefficient functions but
from the singular term $x^{-\alpha}$ in Eq.~(\ref{fita}). Indeed, when this factor is removed from
Eq.~(\ref{fita}), we arrive at curve~1 despite the sub-leading contributions are accounted in $g_1^{NS~DGLAP}$.
Therefore, their impact leads to Eq.~(\ref{antibluml}) instead of Eq.~(\ref{bluml}).
On the contrary, when the singular term is accounted for, we arrive at curve~2 where $R_{NS} \sim 1$.
It proves that the conclusion of the extreme importance of the sub-leading contributions on the small-$x$ behavior of $g_1$
advocated in Refs.~\cite{bv, blums}  is groundless.
Now let us compare the small-$x$ asymptotics of $g_1^{NS}$ and $g_1^{NS~DGLAP}$.
Parameters $\alpha$ in the DGLAP fits obey

\begin{equation}\label{aint}
\alpha > \Delta ,
\end{equation}
with $\Delta$ in Eq.~(\ref{aint}) being either the non-singlet or singlet intercept, depending on
 the case. Eq.~(\ref{aint}) naturally leads to Eq.~(\ref{bluml}).
When the singular factor $x^{-\alpha}$ in Eq.~(\ref{fita}) is dropped, the small-$x$ behavior of $g_1^{DGLAP}$
is given by Eq.~(\ref{asdglap})  and
the strong inequality sign in Eq.~(\ref{bluml}) should be reversed.
The reason why the exponents $\alpha$ in the singular factors of the DGLAP fits should obey Eq.~(\ref{aint})
is clear: indeed, we have just shown above that the asymptotic
regime is actually achieved at very small $x$, so in order to reproduce it at values of
$x$ accessible at  present experiments, the parameter $\alpha$, playing the role of the
intercept, should be larger than the intercepts $\Delta$.
In this connection we remind that our predictions agree very
 well with results of Refs~\cite{kat, koch}, whereas the phenomenological
 value of the intercept $\alpha$ in Eq.~(\ref{fita}) contradicts  those results.
It is clear that combining
 singular fits with the total resummation of logarithms  also implies a double counting
 of the same logarithmic contributions: explicitly in the first case and implicitly
 in the latter, through the
 singular factors $x^{- \alpha}$.
Furthermore, Eqs.~(\ref{asscal}) and (\ref{leadpol}) explicitly show that neither the DGLAP intercept $\alpha$
nor our intercept $\Delta_{NS}$  depend on $Q^2$. Such a dependence, sometimes appearing in the
literature as a possible generalization of Eq.~(\ref{asdglap}) is an ad hoc assumption and
never appears as a result of QCD calculations.\\
Finally, let is notice that
it is commonly believed that the  expression for
the fit in Eq.~(\ref{fita}) mimics the effect of the hadron
structure, including basically unknown non-perturbative
contributions. On the other hand, when the leading logarithms are
accounted for and  the initial parton densities are
fitted at not too large $x$,  the $x$
-dependent terms in $\varphi$ can be almost dropped, so the fit can be
simplified down to $N_q$. It means that the impact of the
non-perturbative contributions is greater at large $x$ whereas in
the small-$x$ region it is reduced to a simple normalization. \\

\section{Reggeon structure of $g_1$}

According to the Regge theory (see e.g. Ref.~\cite{col}), any
forward scattering amplitude, including the invariant Compton
amplitude $T$ related to $g_1$ through Eq.~(\ref{tsigng}),
asymptotically exhibits the Regge (power-like) behavior and can be
written as a sum of such power-like terms called Reggeons. The
same should be true for $g_1$. In this Sect. we show that
both the standard approach (SA) and our description of $g_1$ agree with such a
representation. However, the Reggeons in these two approaches are
different and the reasons for this Reggeon representation are also
quite different. As usually we begin with considering in detail
the non-singlet $g_1$.

\subsection{Reggeon structure of $g_1$ in the SA description}

Eq.~(\ref{fitamel})  with the standard DGLAP fit in the $\omega$
-representation can be re-written as:

\begin{equation}\label{inpregge}
\delta q (\omega) = \frac{r}{\omega - j} + \sum_{k =
1}^{\infty}\frac{r_k}{\omega - j_k} + \sum_{k =
1}^{\infty}\frac{\widetilde{r}_k}{\omega - \widetilde{j}_k}
\end{equation}
where $r,~r_k,~\widetilde{r}_k $ and $j,~j_k,~\widetilde{j}_k$ are
expressed through the parameters of the fit as follows:
\begin{equation}\label{dglapj}
j = \alpha,~~j_k = \alpha - k,~~ \widetilde{j}_k = \alpha - k -
\delta ,
\end{equation}
\begin{equation}\label{dglapr}
 r = N_q,~~r_k = (1 + \gamma) N_q \beta (\beta
-1)..(\beta - k + 1)/(k!),~~ \widetilde{r}_k = \gamma N_q \beta
(\beta -1)..(\beta - k + 1)/(k!) .
\end{equation}
By inserting Eq.~(\ref{inpregge}) into Eq.~(\ref{g1nsdglap}),
integrating over $\omega$, and  taking the residues of the poles of
Eq.~(\ref{inpregge}), allows us to write  $g_1^{NS~DGLAP}$ in
the region \textbf{B } as the following series:
\begin{equation}\label{gnsdglapr}
g_1^{NS~DGLAP} (x,Q^2) = \frac{e^2_q}{2} \Big[S(x,Q^2) +
\sum_{k=1}^{\infty} \Big(S_k (x,Q^2)+
\widetilde{S}_k(x,Q^2)\Big)\Big]
\end{equation}
where, to the LO accuracy, the terms $S,~S_k,~\widetilde{S}_k$
are
\begin{eqnarray}\label{dglapreggeons}
S(x,Q^2) &=&  \Big(\frac{1}{x}\Big)^j C^{NS~DGLAP}(j) ~r~
\Big(\frac{\ln(Q^2/\Lambda^2)}{\ln(\mu^2/\Lambda^2)}\Big)^{\gamma_{DGLAP}(j)/(2\pi
b)}, \\ \nonumber S_k(x,Q^2) &=& \Big(\frac{1}{x}\Big)^{j_k}
C^{NS~DGLAP}(j_k) ~r_k
\Big(\frac{\ln(Q^2/\Lambda^2)}{\ln(\mu^2/\Lambda^2)}\Big)^{\gamma_{DGLAP}(j_k)/(2\pi
b)}, \\ \nonumber \nonumber \tilde{S}_k(x,Q^2) &=&
\Big(\frac{1}{x}\Big)^{\tilde{j}_k} C^{NS~DGLAP}(j_k) ~\tilde{r}_k
\Big(\frac{\ln(Q^2/\Lambda^2)}{\ln(\mu^2/\Lambda^2)}\Big)^{\gamma_{DGLAP}(\tilde{j}_k)/(2\pi
b)}.
\end{eqnarray}
It is clear that the $x$ -dependence of each of the terms
$S,~S_k,~\widetilde{S}_k$ is Regge-like, so we call them the DGLAP
Reggeons contributing to the non-singlet $g_1$. The intercept $j$ of the
Reggeon $S$ is the largest, and we call $S$ the
leading Reggeon and address $S_k$ and $\widetilde{S}_k$ as the
sub-leading Reggeons. Only the leading Reggeon has the positive intercept. All other
intercepts are negative. We remind that all features of the DGLAP
Reggeons are due to the assumed form of the initial quark density  and are related to the phenomenological parameters of the fit
(\ref{fita}).  Obviously,  one can decompose the DGLAP expression for
the singlet $g_1$ quite similarly into a set of  Reggeons.

\subsection{Reggeon structure of Eq.~(\ref{gnsb})}

Let us consider once more the limit of $g_1$ at $x \to 0$. In Sects.~X, XI
we have shown that the use of the
saddle-point method to Eqs.~(\ref{gnsb},\ref{gsb}) led  to the
Regge asymptotics (\ref{asscal},\ref{sas}). The intercepts $\Delta_{NS}$ and
$\Delta_{S}$ were
determined in Eqs.~(\ref{intns},\ref{ints})
as the largest roots of Eqs.~(\ref{eqbrns}) and (\ref{eqbrs}) respectively.
They are not simple poles in the $\omega$ -plane but
the rightmost square-root branching points. They were found
by solving numerically Eqs.~(\ref{eqbrns},\ref{eqbrs}).  However,
each of this equations can have more than one root. Applying the same argument
we are able to find the additional non-singlet and singlet
intercepts
$\Delta^{(k)}_{NS}$ and $\Delta^{(k)}_{S}$, with $k = 1,2,..$. Accounting for these
contributions allows
us to represent the non-singlet and singlet $g_1$ in a form similar to
Eq.~(\ref{gnsdglapr}):

\begin{equation}\label{gnsr}
g_1 (x,Q^2) \sim \frac{e^2}{2} \Big[B (x,Q^2) + \sum_k B_k
(x,Q^2)\Big]
\end{equation}
where the leading contribution $B(x,Q^2)$ is given by Eq.~(\ref{asscal}) for $g_1^{NS}$  and
by Eq.~(\ref{sas}) for $g_1^S$.   The other Reggeons $B_k(x,Q^2)$
look quite similarly. Namely,
they can be obtained from Eqs.~(\ref{asscal}) and (\ref{sas})
with the replacement $\Delta_{NS} \to \Delta^{(k)}_{NS}$ and
$\Delta_{S} \to \Delta^{(k)}_{S}$ respectively. In particular, for $g_1^{NS}$
we have
\begin{equation}\label{reggeonsbk}
B_k =\Pi(\Delta^{(k)}_{NS}) \delta q
(\Delta^{(k)}_{NS})\big(Q^2/x^2\mu^2 \big)^{\Delta^{(k)}_{NS}/2}
\end{equation}
and Reggeons for $g_1^S$ have the structure of Eq.~(\ref{sas}).
We call  $B,¬B_k$ QCD Reggeons because they are obtained from the total
resummation of the leading logarithms in the QCD perturbation series.
The Reggeon $B$ has the maximal intercept compared to $B_k$, so we call it
the leading QCD Reggeon and Reggeons $B_k$ are the sub-leading
(secondary) QCD Reggeons. It turns out that only the leading non-singlet
Reggeon has the positive intercept (\ref{intns}) whereas the next non-singlet
intercept is $\Delta_{NS}^{(1)}\approx 0$.
On the contrary, there are three singlet Reggeons
with positive intercepts:
$\Delta^{(1)}_{S} = 0.55,~\Delta^{(2)}_{S} =0.35,~\Delta^{(3)}_{S} =0.21$.

\subsection{Comparison between the DGLAP and the QCD Reggeons}

The Regge theory, in the DIS context, states that the Regge (power-like) form of
$g_1$ should be achieved at $x \to 0$ only, while $g_1$ looks quite differently
at large $x$. It perfectly agrees
with the features of the QCD
Reggeons $B,~B_k$ obtained with the saddle-point method from the expressions
for $g_1$  due to the QCD radiative  corrections. They appear as a result
of the total resummation of the QCD perturbation series and are never present to
any fixed order of the perturbative expansions, including, of course, the Born term.
  Also they are not simple
poles but square-root branching points,  Their intercepts are found in terms of
the basic QCD constants as the number of the colors $N$, the number
of the flavors $n_f$, and  $\Lambda_{QCD}$.

On the
contrary, the SA Reggeons are produced by the poles present
in any fixed order in $\alpha_s$, including the Born approximation.
They exist at any $x$, even at $x \sim 1$,
because they are generated by the structure of the fit for $\delta
q$ instead of QCD radiative corrections.
The intercepts of the SA Reggeons are expressed in terms of the
phenomenological parameters of the
fit (\ref{fita}) and have nothing to do with QCD calculations,
so we call them  input Reggeons in contrast
to the QCD Reggeons $B,~B_k$.
On one hand, the existence of such Reggeons contradicts  the concepts of the
Regge theory,  On the other hand, we have shown that the phenomenological success
of DGLAP  at small $x$ is due to
the singular factors $x^{- \alpha}$ in the fits for the
initial parton densities which mimic the total resummation of the QCD
radiative corrections.
Obviously, these parameters  are chosen to match the experimental data.
 So one should not be surprised that a truncated set of
 input Reggeons $S,~S_k,~\widetilde{S}_k$ could be close
to the experiments with a good accuracy, the agreement being
 entirely due to the choice of their
phenomenological parameters. So any theoretical interpretation of
such Reggeons in the QCD context would be groundless.

\section{Description of $g_1$ in the unified region A$\bigoplus$B}

In this Sect. we construct a description of $g_1$ valid in both
 regions \textbf{A} and \textbf{B}. Again, we focus on
the non-singlet $g_1$ in the first place. To begin with, let us
remind that in region \textbf{A }, where $x$ is large, the
non-singlet $g_1$ is described by the DGLAP expression
Eq.~(\ref{g1nsdglap}) where both the coefficient function and the
anomalous dimension are known to the NLO (two-loop) accuracy. In
order to describe $g_1^{NS}$ in the small- $x$ region \textbf{B},
we took into account the leading logarithms of $x$ and arrived at
Eq.~(\ref{gnsb}). When Eq.~(\ref{gnsb}) is
considered in the region \textbf{A }, the LL contributions become
small. On the other hand, non-logarithmic contributions accounted
for in Eqs.~(\ref{cnsdglap},\ref{gnsdglap}) are quite important in
this region.  So, a possible option is to create an
interpolation formula for $g_1^{NS}$ which would coincide with
Eq.~(\ref{g1nsdglap}) and Eq.~(\ref{gnsb}) in regions \textbf{A }
and \textbf{B} respectively. To this aim, let us define   new
coefficient function $\widetilde{C}^{NS}$ and anomalous dimension
$\widetilde{C}^{NS}$ by  combining directly the DGLAP results of
Eqs.~(\ref{cnsdglap},\ref{gnsdglap}) and the LL results of
Eqs.~(\ref{cns},\ref{hns}):
\begin{eqnarray}\label{chtilde}
&&\widetilde{C}^{NS} = C^{NS} +  C^{NS~DGLAP}_{LO} +
\frac{A(\omega)}{2 \pi} C^{NS~DGLAP}_{LO}~, \\ \nonumber
&&\widetilde{h}^{NS} = h^{NS} + \frac{A(\omega)}{2 \pi}
\gamma^{(0)}(\omega) + \Big(\frac{A(\omega)}{2 \pi}\Big)^2
\gamma^{(1)}(\omega)~.
\end{eqnarray}
Because of the obvious double counting in Eq.~(\ref{chtilde}),
 let us make the necessary subtractions and define
$C^{NS}_{comb}$ and $h^{NS}_{comb}$, which we call the combined
coefficient function and anomalous dimension:
\begin{eqnarray}\label{chcomb}
&&C^{NS}_{comb} = \widetilde{C}^{NS} - \Delta C^{NS}~, \\
\nonumber &&h^{NS}_{comb} = \widetilde{h}^{NS} - \Delta h^{NS}
\end{eqnarray}
where $\Delta C^{NS}$ and $\Delta h^{NS}$ are the first- and
second- loop terms of the expansion of $C^{NS}$ and $h^{NS}$ into
the series (see Eqs.~(\ref{cnsser},\ref{gnsser})):
\begin{eqnarray}\label{deltach}
&&\Delta C^{NS} = 1 + \frac{A(\omega)C_F}{2 \pi} \Big[
\frac{1}{\omega^2} + \frac{1}{2 \omega} \Big]~, \\
\nonumber &&\Delta h^{NS} = \frac{A(\omega)C_F}{2 \pi} \Big[
\frac{1}{\omega} + \frac{1}{2} \Big]~.
\end{eqnarray}

By inserting $C^{NS}_{comb}$ and $h^{NS}_{comb}$ in
Eq.~(\ref{gnsb}), we arrive at the final expression for $g_1^{NS}$ valid
in the  region \textbf{A}$\bigoplus$\textbf{B} :

\begin{equation}\label{gnsab}
g_1^{NS} (x,Q^2) = \frac{e^2_q}{2} \int_{- \imath \infty}^{\imath
\infty} \frac{d \omega}{2 \pi \imath}
\Big(\frac{1}{x}\Big)^{\omega} C^{NS}_{comb}(\omega)\; \delta q
(\omega)\; e^{y h^{NS}_{comb}(\omega)} .
\end{equation}

Quite similarly we obtain the combined coefficient functions
$C^{(\pm)}_{comb}(\omega)$ and anomalous dimensions
$h_{ik}^{comb}$ for the singlet $g_1$:

\begin{equation}\label{chscomb}
C^{(\pm)}_{comb} = C^{(\pm)} + C^{(\pm)}_{DGLAP} - \Delta
C^{(\pm)},~~h^{comb}_{ik} = h_{ik} + h^{DGLAP}_{ik}- \Delta h_{ik}
\end{equation}

where $C^{(\pm)}_{DGLAP}$ correspond to the DGLAP coefficient
functions with the replacement $\alpha_s \to A(\omega)$ and
$h^{DGLAP}_{ik}$ are the DGLAP anomalous dimensions with the same
replacement. The subtraction terms in Eq.~(\ref{chscomb}) to the LO
accuracy are:
\begin{equation}\label{deltachs}
\Delta h_{ik} = \frac{a_{ik}}{\omega}~,\qquad \Delta C^{(\pm)} =
 \frac{<e^2_q>}{ 2\omega} \Big[1 \mp \frac{a_{gg} -
a_{qq}}{\sqrt{(a_{qq} - a_{gg})^2 + 4 a_{qg}a_{gq}}}\Big].
\end{equation}

Replacing $C^{(\pm)}$ and $h_{ik}$ in Eq.~(\ref{gsb}) by
$C^{(\pm)}_{comb}$ and $h^{comb}_{ik}$, we finally obtain the
expression for $g_1^S$  valid in both  regions\textbf{ A}
and\textbf{ B}:

\begin{eqnarray}
\label{gsab} g_1^S(x, Q^2) = \frac{1}{2} \int_{- \imath
\infty}^{\imath \infty} \frac{d \omega}{2 \pi \imath}
 \Big(\frac{1}{x}\Big)^{\omega} \Big[\Big(C^{(+)}_{comb} e^{\Omega_{(+)}y} +
 C^{(-)}_{comb} e^{\Omega_{(-)}y}\Big) \omega \delta q(\omega) +
 \\ \nonumber
 \Big(C^{(+)}_{comb} \frac{(X + \sqrt{R})}{2 h_{qg}^{comb}} e^{\Omega_{(+)}y} +
 C^{(-)}_{comb} \frac{(X - \sqrt{R})}{2 h_{qg}^{comb} \Omega_{(-)}}
 e^{\Omega_{(-)}y}\Big)\omega \delta g(\omega)\Big]
\end{eqnarray}
where $\Omega_{(\pm)},~X$ and $R$ are also expressed in terms of
$h_{ik}^{comb}$.
In the sub-region $\textbf{A}$ of  $\textbf{ A}\bigoplus\textbf{B}$ the main contribution in
Eqs.~(\ref{gnsab},\ref{gsab}) comes from the DGLAP terms in the coefficient
functions and anomalous dimensions while the logarithmic terms are small,
so that
Eqs.~(\ref{gnsab},\ref{gsab}) almost coincide with the DGLAP expressions.
On the contrary, in the sub-region $\textbf{B}$ the main role is  played by the LL terms and
therefore  Eqs.~(\ref{gnsab},\ref{gsab}) are
pretty close  to the expressions of Eqs.~(\ref{gnsb},\ref{gsb}).
Therefore Eqs.~(\ref{gnsab},\ref{gsab}) really represent the interpolation expressions for $g_1$
in region $\textbf{A}\bigoplus\textbf{B}$.

\section{Description of $g_1$ in the region C}

\begin{figure}[h]
\begin{center}
\begin{picture}(360,200)
\put(0,0){
\epsfbox{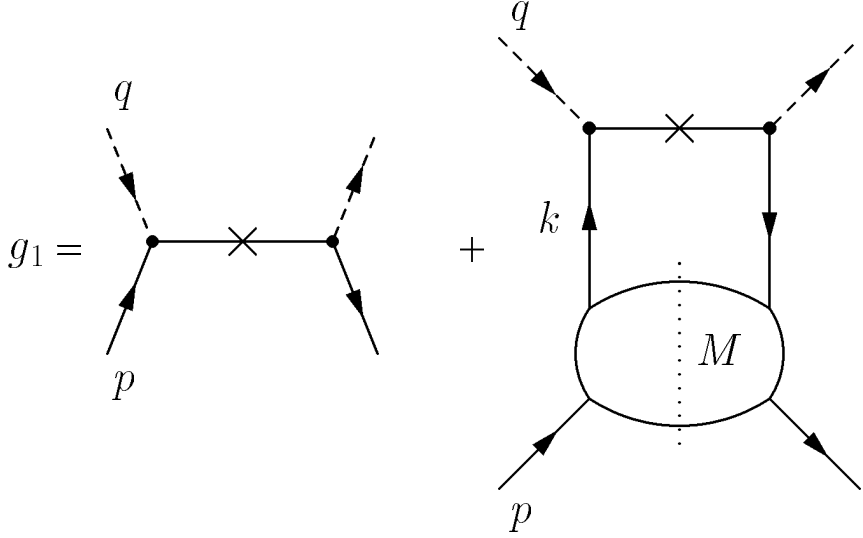} }
\end{picture}
\end{center}
\caption{\label{g1sumfig9} The Bethe-Salpeter equation for $g_1$
in the region \textbf{C}~.}
\end{figure}

The small $Q^2$ -region \textbf{C} is defined in Eq.~\ref{regc}.
Contrary to the regions \textbf{A} and \textbf{B}, the SA cannot be used
in region \textbf{C} at all. Indeed, the basic ingredient of SA, the DGLAP
evolution equations, control the evolution
with respect to $\ln(Q^2/\mu^2)$
in the regions \textbf{A}, \textbf{B} and do not apply at small $Q^2$.
In Ref.\cite{egtsmallq} we have proposed a method to describe $g_1$ at small
$Q^2$, which is a kinematic region studied experimentally.  It turned out that our results for $g_1$ in region B can be
generalized into the  region C by introducing the shift

\begin{equation}\label{shift}
Q^2 \to \bar{Q}^2  \equiv Q^2 + \mu^2
\end{equation}
where $\mu$ is the infrared cut-off. Numerically, we have suggested to use the
Optimal mass scales $\mu_{NS} = 1~$GeV and $\mu_S= 5.5~$GeV for the non-singlet and singlet case,
respectively. The reasons for introducing those scales
 were given in Sect.~X. Other shifts in $Q^2$ similar to
Eq.~(\ref{shift}) were suggested in various papers, see e.g. Refs.~\cite{nacht, bad}.
 In the literature, such shifts were introduced from phenomenological
considerations whereas we suggest it from the analysis of the Feynman graphs contributing
to $g_1$. Let us notice that introducing this shift we go beyond the logarithmic
approximation we have kept so far, so in this sense we consider our description of $g_1$
in the region \textbf{C} model-dependent. To begin with,
let us notice that both the singlet and non-singlet component of $g_1$
obey the Bethe-Salpeter equation shown in Fig.~\ref{g1sumfig9}.
In the analytical form this equation is written

\begin{equation}\label{bsregc}
g_1 = g_1^{Born}  + \imath \kappa \int \frac{d^4 k}{(2\pi)^4} (-2 \pi
\imath)\delta((q+k)^2 - m^2_q) \frac{2w k_{\perp}^2}{(k^2 -
m^2_q)^2} \frac{E (2pk, k^2)}{2pk},
\end{equation}
where $\kappa$ stands for the numerical factors $e^2/2$  and  $<e^2/2>$ for the non-singlet
and the singlet respectively.
We have skipped in Eq.~(\ref{bsregc}) the convolution with the initial parton densities
in order  to prove the shift in  Eq.~(\ref{shift}), that the densities
cannot affect. The $\delta$ -function (together with the factor $-2\pi
\imath$) corresponds to the cut propagator of the upper quark
with momentum $k$ and mass $m_q$ coupled to the virtual photon
lines and the factor  $2k_{\perp}^2$ appears after simplifying the spin
structure of the equation. Similarly to Eq.~(\ref{tsigng}), $E (2pk, k^2)$ in
Eq.~(\ref{bsregc}) is related to the invariant amplitude $M(2pk,k^2)$:
\begin{equation}\label{em}
E (2pk, k^2) \equiv (1/2 \pi) \Im M(2pk,k^2).
\end{equation}
The invariant amplitude $M(2pk,k^2)$ describes the forward
scattering of partons, with the upper partons being quarks.
In other words, $M(2pk,k^2)$ can be any of $M^{NS}(2pk,k^2)$ (for $g_1^{NS}$) or
$M_{qg}(2pk,k^2),~M_{qq}(2pk,k^2)$ (for the singlet $g_1$).
These amplitudes incorporate the total resummation of the leading logarithms.
Obviously, $E$ in Eq.~(\ref{bsregc}) does not depend on $Q^2$.
Contrary to the parton amplitudes $M_{ik}$ entering
in Eq.~(\ref{ireetint}), where $k$ stands for the softest momenta, the amplitudes
$E$ in Eq.~(\ref{bsregc}) are
essentially off-shell, and depend on two arguments: the invariant total
energy $(p+k)^2 \approx 2pk$ and the virtuality $k^2$ which is not small now.
Therefore they should be calculated independently.
The amplitudes $M^{NS}(2pk,k^2),~M_{qg}(2pk,k^2),~M_{qq}(2pk,k^2)$ are
considered in detail in Appendix C.

\subsection{Infrared regularization in Eq.~(\ref{bsregc})}

First, we remind that in order to account for the LL contributions
in the small- $x$ region \textbf{C}, one
should use the ordering (\ref{dlord}).  This  leads to the IR
singularities  of soft gluons and therefore an IR cut-off in the
divergent propagators must be introduced. In order to treat the quarks and gluon
ladders similarly, we drop the quark masses and introduce the IR cut-off in the
quark and gluon ladders, providing both the ladder (vertical) partons and the
soft non-ladder gluons with the fictitious mass $\mu$,
assuming  that $\mu > m_{quark}$. In the previous Sects. we introduced
$\mu$ in somewhat different way: according to Eq.~(\ref{kmu}), $\mu$ is the lowest limit
in the integrations over $k_{\perp}$. Although it was noticed in
Ref.~\cite{ioffe} that different ways of introducing the IR cut-off lead to different
results, this goes well beyond the accuracy we keep. Let us also notice that
there is no need to introduce $\mu$ into the horizontal propagators of the ladder
 because they are IR stable. Then Introducing $\mu$  Eq.~(\ref{bsregc}) is modified into

\begin{equation}\label{bsmu}
g_1 = g_1^{Born}  + \imath \kappa \int \frac{d^4 k}{(2\pi)^4} (-2 \pi
\imath)\delta((q+k)^2) \frac{2w k_{\perp}^2}{(k^2 - \mu^2)^2} \frac{E(2pk,
 k^2+\mu^2)}{2pk} .
\end{equation}

\subsection{Solving the Bethe-Salpeter equation (\ref{bsmu})}

It is convenient to write Eq.~(\ref{bsmu}) in terms of the Sudakov variables
defined in Eq.~(\ref{sud}).
Eq.~(\ref{e}) shows that the $2pk$ and $k^2$
-dependence for any of
$E^{NS},~E_{qq},~E_{qg}$ looks much alike, so below we consider the Bethe-Salpeter
equation for $g_1^{NS}$ only. Substituting $E^{NS}$ into Eq.~(\ref{bsmu}) and
changing the order of the integrations, we arrive at

\begin{eqnarray}\label{bssud}
g_1 = g_1^{Born} +
\kappa \int_{- \imath \infty}^{ \imath \infty}
\frac{d \omega}{2 \pi \imath} \Big(\frac{2pk}{k^2}\Big)^{\omega}
\omega h^{NS} (\omega) \int \frac{d \alpha}{\alpha} d \beta d k^2_{\perp}
\frac{w k^2_{\perp}}{(w \alpha \beta + k^2_{\perp}+ \mu^2)^2}  \\ \nonumber
\delta (w \beta + wx\alpha - w \alpha \beta - k^2_{\perp} - Q^2)
\Big(\frac{2pk}{k^2}\Big)^{\omega}
\Big(\frac{w \alpha \beta + k^2_{\perp} + \mu^2}{\mu^2} \Big)^{h^{NS}}~.
\end{eqnarray}

\begin{figure}[h]
\begin{center}
\begin{picture}(300,200)
\put(0,0){
\epsfbox{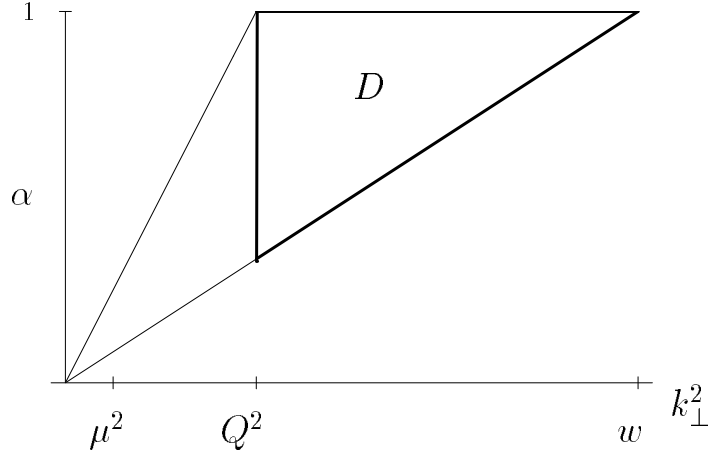} }
\end{picture}
\end{center}
\caption{\label{g1sumfig10} The integration region in
Eq.~(\ref{bsak})~.}
\end{figure}

In the region \textbf{C}  $x$ is small, so we drop the second term in the
argument of the $\delta$ -function, which is used  for the
integration over $\beta$. We obtain
\begin{equation}\label{bsak}
g_1 = g_1^{Born} +
\kappa \int_{- \imath \infty}^{ \imath \infty}
\frac{d \omega}{2 \pi \imath}
\omega h^{NS} (\omega) \int \frac{d \alpha}{\alpha} \frac{d k^2_{\perp}}{(\alpha Q^2 + k^2_{\perp} + \mu^2)}
\Big(\frac{w \alpha}{\alpha Q^2 + k^2_{\perp} + \mu^2}\Big)^{\omega}
\Big(\frac{\alpha Q^2 + k^2_{\perp} + \mu^2}{\mu^2}\Big)^{h^{NS}}
\end{equation}
The region of integration in Eq.~(\ref{bsak}) is shown in
Fig.~\ref{g1sumfig10}.
It is restricted by the following limits: \\
\textbf{(a)}: $w > k^2_{\perp} + \mu^2 > \alpha Q^2 > 0;~~$
\textbf{(b)}: $w \alpha > \alpha Q^2 + k^2_{\perp} + \mu^2$. \\
The result of the integration over this region depends on the
relations between $Q^2$  and $k^2_{\perp}/\alpha$. The leading
contribution comes from the sub-region \emph{D} in
Fig.~\ref{g1sumfig10}. Integrating over $\alpha$ in \emph{D} leads
to
\begin{equation}\label{bsk}
g_1 = g_1^{Born} +
\kappa \int_{- \imath \infty}^{ \imath \infty}
\frac{d \omega}{2 \pi \imath} \omega h^{NS} (\omega) \frac{1}{\omega}
\int^{w}_{Q^2} \frac{d k^2_{\perp}}{k^2_{\perp} + \mu^2}
\Big(\frac{w}{k^2_{\perp} + \mu^2}\Big)^{\omega}\Big(\frac{k^2_{\perp}  + \mu^2}{\mu^2}\Big)^{h^{NS}} .
\end{equation}
Replacing $k^2_{\perp} + \mu^2$ by $t$ in Eq.~(\ref{bsk}), we get
\begin{equation}\label{bst}
g_1 = g_1^{Born} +
\kappa \int_{- \imath \infty}^{ \imath \infty}
\frac{d \omega}{2 \pi \imath} h^{NS} (\omega) \int_{Q^2 + \mu^2}^{w + \mu^2}
\frac{d t}{t} (w/t)^{\omega} (t/\mu^2)^{h^{NS}}.
\end{equation}
In the region \textbf{C} $w \gg \mu^2$, so the upper limit of integration in Eq.~(\ref{bst})
can be approximated by $w$. The lowest limit is definitely $Q^2 + \mu^2$ which proves the validity of the shift we have suggested in Eq.~(\ref{shift}). Performing the integration over $t$ in
Eq.~(\ref{bst}) leads to the expression of Eq.~(\ref{gnsb}) but with the shifted value of  $Q^2$.
Indeed, the integration over $t$ yields
\begin{equation}\label{bst1}
g_1 = g_1^{Born} +
\kappa \int_{- \imath \infty}^{ \imath \infty} \frac{d \omega}{2
\pi \imath} \frac{h^{NS}}{(\omega - h^{NS})}
\Big[\Big(\frac{w}{\bar{Q}^2}\Big)^{\omega}
\Big(\frac{\bar{Q}^2}{\mu^2}\Big)^{h^{NS}}-
\Big(\frac{w}{\mu^2}\Big)^{h^{NS}}\Big] .
\end{equation}
The integration of the second term in the squared bracket yields zero,  by closing the integration contour  to the right of the singularity of
$h^{NS} /(\omega - h^{NS})$.  Using the identity
\begin{equation}\label{omegah}
\frac{h^{NS}}{(\omega - h^{NS})} = -1 +\frac{\omega}{(\omega - h^{NS})}
\end{equation}
and noticing that the first term in Eq.~(\ref{omegah}) cancels
the term $ g_1^{Born}$ in Eqs.~(\ref{bsregc}- \ref{bst1}), we arrive at the following expression:

\begin{equation}\label{gnsc}
g_1^{NS} =
\kappa \int_{- \imath \infty}^{ \imath \infty}
\frac{d \omega}{2 \pi \imath} \frac{\omega}{(\omega - h^{NS})}
\Big(\frac{1}{\bar{x}}\Big)^{\omega}
\Big(\frac{\bar{Q}^2}{\mu^2}\Big)^{h^{NS}}
\end{equation}
which coincides with $g_1^{NS}$ from Eq.~(\ref{gnsb}) with the replacement
 $Q^2 \to \bar{Q}^2$.
We have used here the shifted variable
\begin{equation}\label{shiftx}
\bar{x} = \bar{Q}^2/w = x + \mu^2/w \equiv x + z .
\end{equation}
So, we have proved that our result for $g_1^{NS}$ in region \textbf{B} can be extended to region
\textbf{C} with the shift $Q^2 \to \bar{Q}^2 = Q^2 + \mu^2$.

It is not difficult to repeat the above
 calculations for the singlet $g_1$. Eventually we conclude that our expressions
 (\ref{gnsb},\ref{gsb}) for $g_1$ in region \textbf{B}
 ($\equiv g_1^{(\textbf{B})} (x,Q^2)$)  can represent $g_1$ in the region \textbf{C} with the shifts
 $Q^2 \to \bar{Q}^2 = Q^2 + \mu^2,~~x \to \bar{x} = x +z$:
\begin{equation}\label{gc}
g_1^{(\textbf{C})} (x,z,Q^2,\mu^2) = g_1^{(\textbf{B})}(\bar{x},\bar{Q}^2).
\end{equation}
Obviously, Eq.~(\ref{gc}) is valid in the unified region $\textbf{B} \oplus \textbf{C}$.

\section{Description of $\textbf{g}_1$ in the full region
$\textbf{A} \oplus \textbf{B} \oplus \textbf{C} \oplus \textbf{D}$}

In this Sect we will show that combining the shift of $Q^2$ introduced in Eq.~(\ref{shift})
and the interpolation expressions of
Eqs.~(\ref{gnsab} \ref{gsab}) for $g_1$ in region $\textbf{A}$, it
allows us to generalize  expressions for $g_1$ which can be used in the full
region $\textbf{A} \oplus \textbf{B} \oplus \textbf{C} \oplus \textbf{D}$.
 Let us first discuss the perturbative description of $g_1$ in the
region $\textbf{D}$.


The interpolation expression for $g_1^{NS}$ valid at large $x$ and small $Q^2$ is
\begin{equation}
\label{gnsd} g^{(\textbf{D})}_{1~NS}(\bar{x}, \bar{Q^2}) = (e^2_q/2) \int_{-\imath
\infty}^{\imath \infty} \frac{d \omega}{2\pi\imath }\Big(
\frac{1}{\bar{x}} \Big)^{\omega} C_{comb}^{NS}(\omega) \delta q(\omega)
e^{\bar{y} h_{comb}^{NS}(\omega)}~,
\end{equation}
where $\bar{x} = x + z,~~\bar{y} = \ln [(Q^2 + \mu^2)/\mu^2]$ and
the combined coefficient function $C_{comb}^{NS}$ and anomalous dimension
$h_{comb}^{NS}$ are
given in Eq.~(\ref{chcomb}).

Similarly, combining the shift and Eq.~(\ref{gsab}),  the expression
for the singlet $g_1$ in region $\textbf{D}$ is:
\begin{eqnarray}
\label{gsd}
g_{1~S}^{(\textbf{D})} (x, Q^2) = \frac{1}{2} \int_{- \imath
\infty}^{\imath \infty} \frac{d \omega}{2 \pi \imath}
 \Big(\frac{1}{\bar{x}}\Big)^{\omega} \Big[\Big(C^{(+)}_{comb}(\omega) e^{\Omega_{(+)}\bar{y}} +
 C^{(-)}_{comb}(\omega) e^{\Omega_{(-)}\bar{y}}\Big) \omega \delta q(\omega) + \\ \nonumber
 \Big(C^{(+)}_{comb} (\omega) \frac{(X + \sqrt{R})}{2 h_{qg}^{comb}} e^{\Omega_{(+)}\bar{y}} +
 C^{(-)}_{comb} (\omega) \frac{(X - \sqrt{R})}{2 h_{qg}^{comb}}
 e^{\Omega_{(-)}\bar{y}}\Big)\omega \delta g(\omega)\Big] ,
\end{eqnarray}


Actually, Eqs.~(\ref{gnsd},\ref{gsd}) represent $g_1$ not only in region $\textbf{D}$ but also in the full region  $\textbf{A} \oplus \textbf{B} \oplus \textbf{C} \oplus \textbf{D}$. Indeed,
they are expressed in terms  of the shifted variables $\bar{x},~\bar{y}$ and therefore can be used at any values of
$Q^2$. Also they include the total resummation of  the leading logarithms, so  can be used in the
regions $\textbf{B}$ and $\textbf{C}$. Finally, they contain non-logarithmic one-loop contributions
\footnote{The second-loop contributions can be included similarly.} to
the coefficient functions and anomalous dimensions obtained by  the DGLAP- expressions, so
this makes it possible to use them in the large-$x$ regions $\textbf{A}$ and $\textbf{D}$.
 Let us remind that the use of the shift in $Q^2$ drives us out of the logarithmic accuracy and
also recall our suggestion is to use different values of $\mu$ for the
singlet and non-singlet components of $g_1$, given in Eqs.~(\ref{muns},\ref{mus}).
We  proceed now to the applications of the results on $g_1$ obtained so far..

\section{Perturbative $Q^2$ -power corrections}

In this Section we discuss the power-$1/(Q^2)^k$ corrections to $g_1$. Basically,
there are various sources of such corrections but we focus only on those arising
 when the experimental results of $g_1^{exp}$ are
confronted to the theoretical predictions $g_1^{theor}$. The numerical analysis of
 the discrepancy between the
non-singlet component of $g_1^{exp}$
and $g_1^{theor}$ shows (see for example Ref.~\cite{sid} and Refs. therein) that

\begin{equation}\label{htw}
\big(g_{1}^{NS}\big)^{exp}- \big(g_{1}^{NS}\big)^{theor} \sim \sum_{k = 1,2,..} \frac{T_k}{(Q^2)^k}
\end{equation}
and the reason of the discrepancy is attributed to the impact of higher twists.
 Conventionally, the DGLAP expression of Eq.~(\ref{gnsdglap}) is used for describing
$\big(g_{1}^{NS}\big)^{theor}$ and this is called the leading twist contribution.  Naively
one could expect from Eq.~(\ref{htw}) that the impact of the power corrections should
increase when $Q^2$ decreases, especially in the limit  $Q^2 \to 0$. On the contrary,
the power corrections become negligible
 when $Q^2$ decreases down to values $\sim 1$~GeV$^2$. \\
We are going now to explain this behavior and give an alternative description of the
 power corrections. To this aim,  first  let us notice that the kinematic region
of  $g_1^{NS}$  studied in Ref.~\cite{sid} mainly coincides with the region
$\textbf{B} \oplus \textbf{C}$ and therefore the total resummation of the leading logarithms
together with the shift of $Q^2$ should be included into expressions for $g_1^{theor}$.
Eq.~(\ref{gc})  contains  both terms and therefore  in the
region $\textbf{B} \oplus \textbf{C}$
\begin{equation}\label{gnsbc}
g_1^{NS} = \frac{e^2_q}{2} \int_{- \imath \infty}^{\imath \infty}
\frac{d \omega}{2 \pi \imath}
\Big(\frac{w}{Q^2 + \mu^2_{NS}}\Big)^{\omega}
\Big(\frac{Q^2 + \mu^2_{NS}}{\mu^2_{NS}}\Big)^{h^{NS}}
C^{NS}(\omega) \delta q (\omega)
\end{equation}
where $w = 2pq$ and $\mu_{NS}$ defined in Eq.~(\ref{muns}).
The terms with $Q^2 + \mu^2_{NS}$ in
Eq.~(\ref{gnsbc}) can be expanded in the  region $\textbf{B}$, where by
definition $Q^2 > \mu^2_{NS}$, as follows:

\begin{equation}\label{explargeq}
\Big(\frac{w}{Q^2 + \mu^2_{NS}}\Big)^{\omega} \Big({Q^2 + \mu^2_{NS}}{\mu^2_{NS}}\Big)^{h^{NS}} =
\Big(\frac{1}{x}\Big)^{\omega} \Big(\frac{Q^2}{\mu^2_{NS}}\Big)^{h^{NS}} \Big[1 + \sum_{k = 1}
T^{NS}_k (\omega)\Big(\frac{\mu^2_{NS}}{Q^2}\Big)^k\Big]
\end{equation}
with
\begin{equation}\label{tnsk}
T_k^{NS} =\frac{(-\omega + h_{NS})(-\omega + h_{NS} - 1)..(-\omega
+ h_{NS} - k + 1)}{k!}~.
\end{equation}
Obviously, the power terms in the series of Eq.~(\ref{explargeq}) have a  perturbative
origin and have nothing to do with the
higher twists. Such terms are absent in the Standard  Approach.  Of course we are aware that
higher twists can contribute to $g_1$ but we argue that the perturbative power contributions
of Eq.~(\ref{explargeq}) should be accounted for  first,  and only after a reliable
estimate of the impact of the higher twists  can be made.
In contrast, in the region \textbf{C} where $Q^2 < \mu^2_{NS}$, the power $Q^2$ -expansion takes the
different form:

\begin{equation}\label{expsmallq}
\Big(\frac{w}{Q^2 + \mu^2_{NS}}\Big)^{\omega} \Big(\frac{Q^2 +
\mu^2_{NS}}{\mu^2_{NS}}\Big)^{h^{NS}} =
\Big(\frac{1}{z}\Big)^{\omega}  \Big[1 + \sum_{k = 1} T^{NS}_k
(\omega)\Big(\frac{Q^2}{\mu^2_{NS}}\Big)^k\Big] .
\end{equation}

The power series in
Eqs.~(\ref{explargeq},\ref{expsmallq}) for large and small
$Q^2$ are derived from the same formulae. However after the
expansion has been made, they cannot be related to each other by
simply varying $Q^2$.  Our estimate $\mu_{NS} \approx 1$~GeV
gives a  natural explanation to the observation made in
Refs.~\cite{sid} that the power $Q^2$ -corrections
die out when $Q^2$ approaches values $\sim 1$ GeV$^2$ and do not appear at smaller values of $Q^2$.
Let us remind that our estimate of $\mu_{NS}$
in Eq.~(\ref{muns}) was obtained by studying
the asymptotic properties of $g_1^{NS}$, i.e. absolutely independently of
any analysis of the power corrections. We suggest that  the new
source of the power contributions given by
Eqs.~(\ref{explargeq},\ref{expsmallq}) can sizably change the
conventional analysis of the higher twists contributions to the
Polarized DIS. Obviously, the power expansion of the singlet $g_1$ can be made quite similarly.

\section{Application to the COMPASS experiment}

Now let us discuss the application of our results  to the recent COMPASS data  on the singlet
$g_1$. We
consider here the results of Refs.~\cite{egtsmallq, egtcompass}.
The COMPASS experiment, carried out at the Super Proton Synchrotron at CERN.
has investigated  $g_1$ by measuring the asymmetries in the
scattering of a polarized 160 GeV $\mu^+$ -beam on polarized deuterons from a
fixed $^6$LiD target (see Ref.~\cite{compass1}). As there is only one target,
the COMPASS collaboration can measure the singlet $g_1$ only. Values of $Q^2$ at
the COMPASS data are basically small: events with
$Q^2 < 1$~GeV$^2$ correspond to about 90\% of the total data set.  From
Refs.~\cite{compass1, compass2} one can conclude
that the COMPASS kinematic region for measuring $g_1$, $G_{COMPASS}$, is
\begin{equation}\label{compassreg}
G_{COMPASS}:\qquad 10^{-4} \lesssim x \lesssim 10^{-1}~,\qquad
10^{-1}~GeV^2 \lesssim Q^2 \lesssim 1~GeV^2~.
\end{equation}
This makes clear that
the Standard Approach cannot be used for the analysis of the COMPASS data. On the contrary,
our expressions (\ref{gsb}, \ref{gsd})  can be used
in the COMPASS kinematic region. In  the region $G_{COMPASS}$ $Q^2 \ll \mu^2_S$ and therefore in this
region
\begin{equation}\label{expgssmallq}
\Big(\frac{w}{Q^2 + \mu^2_{S}}\Big)^{\omega} \Big(\frac{Q^2 +
\mu^2_{S}} {\mu^2_{S}}\Big)^{\Omega_{(\pm)}} =
\Big(\frac{1}{z}\Big)^{\omega}  \Big[1 + \sum_{k = 1} T^{(\pm)}_k
(\omega)\Big(\frac{Q^2}{\mu^2_{S}}\Big)^k\Big]~,
\end{equation}
with
\begin{equation}\label{tpm}
T_k^{(\pm)} = \frac{(-\omega + \Omega_{\pm})(-\omega + \Omega_{\pm}
-1)..(-\omega + \Omega_{\pm} -k + 1)}{k!}~.
\end{equation}
Substituting Eq.~(\ref{expgssmallq}) into Eq.~(\ref{gsd}) leads to the following expression:
\begin{equation}\label{gsser}
 g_1(x,z,
Q^2) \approx g_1(z) + (Q^2/\mu^2_S) \frac{\partial g_1(z,x,Q^2)}{\partial Q^2/\mu^2_S} +
O\big((Q^2/\mu^2_S)^2\big)
\end{equation}
where $z = \mu^2_S/w$. The first term in Eq.~(\ref{gsser}) is
\begin{equation}\label{gsz}
g_1(z)= \frac{<e^2_q>}{2} \int_{- \imath
\infty}^{\imath \infty} \frac{d \omega}{2 \pi \imath}
 \Big(\frac{1}{z}
\Big)^{\omega} \Big[\widetilde{C}_q(\omega) \delta q +
\widetilde{C}_g(\omega) \delta g\Big]~.
\end{equation}
As stated earlier, the combined coefficient functions $\widetilde{C}_{q,g}$ include the total resummation
of the leading logarithms of $z$ and the non-logarithmic contributions $\sim \alpha_s$.
They are defined as follows:
 \begin{equation}\label{cqgcomb}
\widetilde{C}_q = C_q + C^{DGLAP}_q - \Delta C_q~,
\qquad\widetilde{C}_g = C_g + C^{DGLAP}_g - \Delta C_g~.
\end{equation}

The terms $C^{DGLAP}_q$ and $C^{DGLAP}_g$ in Eq.~(\ref{cqgcomb}) are the
NLO DGLAP coefficient functions and
\begin{eqnarray}\label{cqcg}
C_q &=& \frac{\omega (\omega - H_{gg})} {\omega^2 - \omega (H_{gg}
+ H_{qq}) + H_{qq}H_{gg} - H_{qg}H_{gq}}~, \qquad \Delta C_q = 1 +
\frac{a_{qq}}{\omega^2}~, \\ \nonumber C_g &=& \frac{\omega
H_{gq}}{\omega^2 - \omega (H_{gg} + H_{qq}) + H_{qq}H_{gg} -
H_{qg}H_{gq}}~, \qquad \Delta C_g =  \frac{a_{gq}}{\omega^2}~.
\end{eqnarray}

The presence of the terms $C^{DGLAP}_{q,g}$ in
Eqs.~(\ref{cqgcomb},\ref{cqcg}), as the DGLAP coefficient functions,
may sound irrelevant or strange because the DGLAP description of $g_1$
cannot be used in the small-$Q^2$ kinematics we are
discussing. Nevertheless, the direct calculation of $g_1(z)$ to
 order $\sim \alpha_s$ yields a contributions coinciding with
the NLO DGLAP coefficient functions. In other words the presence of
the coefficient functions has nothing
to do with the $Q^2$-evolution.
Eq.~(\ref{gsser}) explicitly shows that the $Q^2$- dependence of
$g_1^S$ in the region $G_{COMPASS}$ should be weak, and also that
 $g_1^S$ practically does not depend on  $x$ in the kinematical region
$G_{COMPASS}$, even at very small $x$. However, its absolute value
 cannot be fixed from theoretical grounds.
Indeed, Eq.~(\ref{gsz}) implies that sign of $g_1^S(z)$ at any given
$z$ is determined by the interplay between the quark and gluon
contributions and eventually depends on $\delta q/\delta g$, which
cannot be determined theoretically. At the same time,
Eq.~(\ref{gsser}) predicts that the $z$-dependence of $g_1^S$ is
pretty far from being trivial, so the experimental investigation of
this dependence would be quite interesting: it can yield
information about the initial quark and gluon densities. The
predictions of the essential independence of $g_1^S$ on  $x$ made in
Ref.~\cite{egtsmallq}, was confirmed in Ref.~\cite{compass2} where
a flat dependence of $g_1^S$ was found. More precisely,
Ref.~\cite{compass2} reported that
\begin{equation}\label{gscompass}
g_1^S \approx 0
\end{equation}
in the region $G_{COMPASS}$ with small errors. Unfortunately, the COMPASS data
do not allow to study the
$z$-dependence of $g_1^S$ in the proper way. Nevertheless, the COMPASS result (\ref{gscompass})
was used in Ref.~\cite{egtcompass} in order to obtain some rough estimates for $\delta q/\delta g$.
Below we consider this issue in detail.

\subsection{Interpretation of the COMPASS data on $\textbf{g}_1^\textbf{S}$}

 The variable  $w = 2pq$ in the COMPASS experiment runs in the interval
\begin{equation}\label{wreg}
30~ GeV^2 \lesssim w \lesssim 270~ GeV^2
\end{equation}
and therefore the range for $z$ is
\begin{equation}\label{zreg}
 1 \lesssim z \lesssim 0.1.
\end{equation}
The variable $z$ is related to the standard variable
$\nu = w/(2M)$ measured in GeV, with $M = 1$ GeV:
\begin{equation}\label{znu}
z = \Big(\frac{\mu^2}{2M}\Big) \frac{1}{\nu} \approx \frac{15}{\nu}~,
\end{equation}
so the region (\ref{zreg})  covered in the COMPASS experiment corresponds to the
$\nu$-region (in GeV)
\begin{equation}\label{nureg}
15 \lesssim \nu \lesssim 150.
\end{equation}
We remind that only the $x$ -dependence of $g_1^S$ was studied in the COMPASS experiment.
The values of $w$ and $Q^2$ were not reported in the COMPASS data, which makes
impossible the straightforward application of Eq.~(\ref{gsz}) to the COMPASS results.
However there are several options for the interpretation of Eq.~(\ref{gscompass}) and below we
consider them in detail:  \\
\textbf{Option (i)}:

Eq.~(\ref{gscompass}) means that $g_1^S (z) = 0$ for any $z$
from the whole interval of Eq.~(\ref{zreg}).\\
In this case Eq.~(\ref{gsz}) implies  a strong correlation
between $\delta q$ and $\delta g$ at any $\omega$ :
\begin{equation}\label{correl}
C_q (\omega) \delta q (\omega) + C_g (\omega) \delta g (\omega) = 0.
\end{equation}
We don't find theoretical grounds for understanding this fact and think
that next option is more realistic. \\
\textbf{Option (ii)}:

 Eq.~(\ref{gscompass}) holds in the
 average,  namely in the region (\ref{zreg}):
\begin{equation}\label{gsav}
<g_1^S (z)> = 0~.
\end{equation}
Obviously,
in order to fulfill the Eq.~(\ref{gsav}),
$g_1(z)$ should acquire both positive and negative values in the region (\ref{zreg}).
This could be realized by an appropriate choice for
the initial parton densities $\delta q (z)$ and $\delta g (z)$.
 In ref.~\cite{egtinp} we suggested  that in region \textbf{B} one can approximate
the initial parton densities by constants. Guided by this result, we suggested
in Ref.~\cite{egtsmallq}
to approximate  $\delta q (z)$ and $\delta g (z)$ at small $z$ by simple constants
to get a rough  estimate. However,
 in the COMPASS region (\ref{zreg})  $z$ is not small enough to use such a simple approximation.
As the DGLAP-fits from Ref.~\cite{fitsa} work quite well and also other parameterizations have a similar
structure, we suggest a similar but regular fit :
\begin{equation}\label{fitsb}
\delta q (z) = N_q z (1-z)^3 (1+3z),~~\delta g(z) = N_g
(1-z)^4(1+3z)~.
\end{equation}

The main difference with Ref.~\cite{fitsa}
is in the absence of the power factors $z^a$ while the terms in the brackets in
Eq.~(\ref{fitsb}) and in Ref.~\cite{fitsa} coincide ( $x$ in
Ref.~\cite{fitsa} is replaced by $z$ in Eq.~(\ref{fitsb})). Indeed
 in Ref.~\cite{egtinp}, as also discussed  previously,  we have proved that the role
played by the singular terms $x^{-a}$ in the DGLAP fits is to
mimic the total resummation of $\ln^k(1/x)$~.   On the other hand, we would like to
keep the same ratio $\delta q/\delta g$ as in Ref.~\cite{fitsa} and therefore we also
change the power factor
for $\delta q$ in Eq.~(\ref{fitsb}).
Now it is easy to check that the fit (\ref{fitsb})
do not lead to a flat $z$-dependence for  $g_1$ and
cannot keep $g_1(z) =0$ in the whole COMPASS region (\ref{compassreg}).

\begin{figure}[h]
\begin{center}
\begin{picture}(490,570)
\put(0,0){
\epsfbox{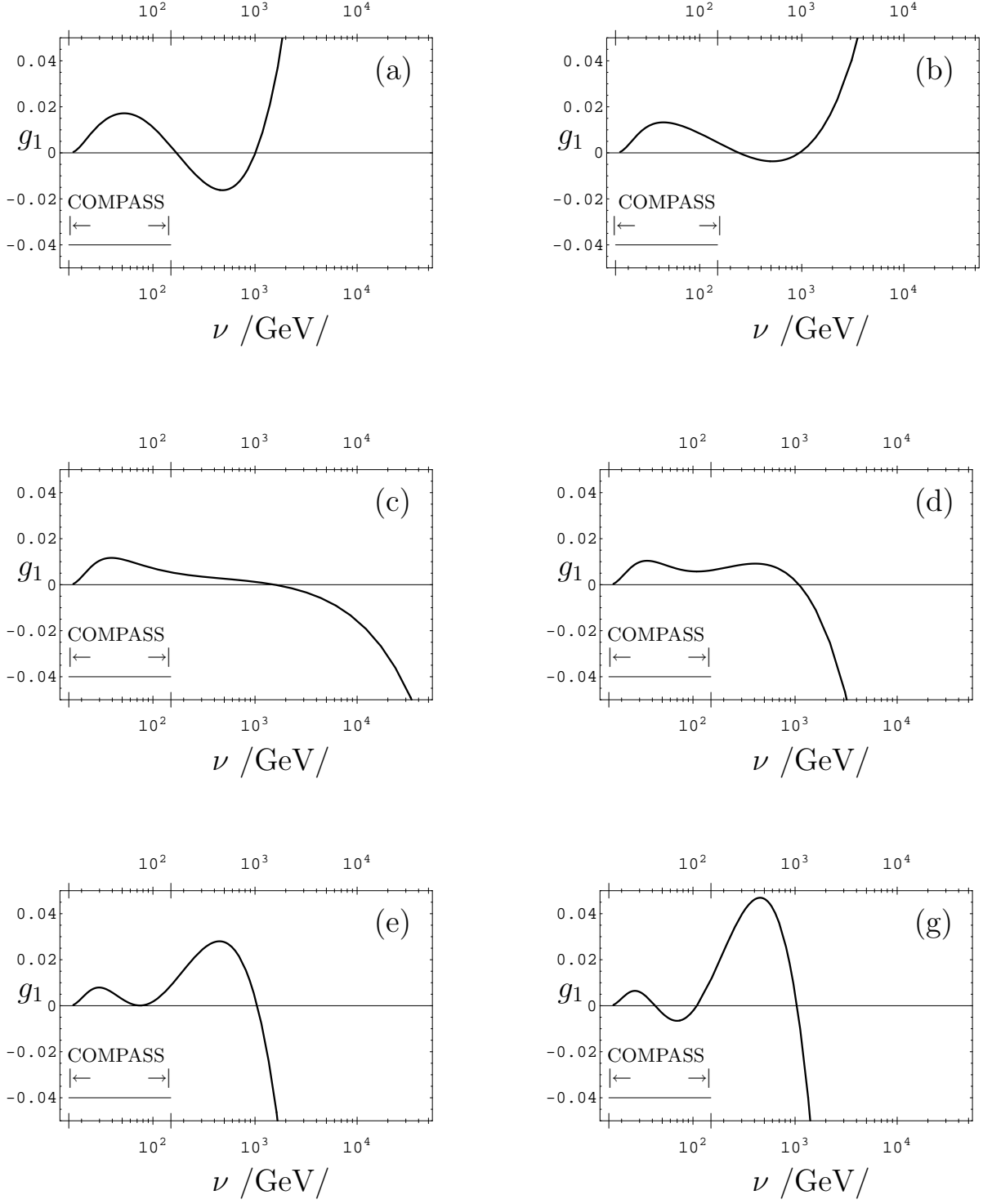}}
\end{picture}
\end{center}
\caption{\label{g1sumfig12} The $\nu$ -dependence of $g_1(\nu)$,
with $\delta q$, $\delta g$ defined in Eq.~(\ref{fitsb}), for
$N_q=0.5$ and different values of $N_g$: (a) -1.5, (b) -0.5, (c)
0, (d) 0.5, (e) 2, (g) 3.5; the COMPASS $\nu$ -region corresponds
to Eq.~(\ref{nureg})~.}
\end{figure}

In more detail by substitution of  Eq.~(\ref{fitsb}) into
Eq.~(\ref{gsd}) and performing the integration over $\omega$
numerically, with fixed and positive $N_q$~, and varying the
values of $N_g$~, we plot our results in Fig.~\ref{g1sumfig12}~.

By a close inspection of the various configurations shown, we can easily conclude
 that these fits could be compatible  with Eq.~(\ref{gsav}) only
if $N_g > 0$ and $N_g > N_q$~.

As the way of averaging $g_1$ over $z$ in the COMPASS data is
unknown to us, we can try another possibility, approximating
\begin{equation}\label{zav}
<g_1(z)> \approx g_1 (<z>) = 0~,
\end{equation}
where $<z> = 0.25$ ~(i.e. $<\nu> \approx 60$ GeV) is the mean
value of $z$ from the region (\ref{zreg}). Then using
Eqs.~(\ref{gsd},\ref{fitsb}), keeping  positive $N_q$ and varying
$N_g$~, as shown in Fig.~\ref{g1sumfig12}, we suggest again that
$N_g$ are positive and $N_g > N_q$~. \\

\subsection{Comments on the measurement of $g_1$ in different kinematic regions.}

To conclude this Sect., let us make a brief comment on the parametrization
of $g_1$.  In the Born
approximation $g_1$ is given by Eq.~(\ref{gborn}) and depends on the only
argument $x$ which corresponds to the famous scaling in the DIS.
The radiative corrections in the higher loops bring the violation
of  scaling, so $g_1$ acquires, additionally, the $Q^2$ -dependence. At this
stage one can parameterize $g_1$ by the set of variables $x,~Q^2$ or,
alternatively,
$w,~Q^2$, or $\nu,~Q^2$. As it is well-known, in the target rest frame
$w \equiv 2pq = 2M (E - E')$
where $M$ is the target (nucleon) mass and $E$ ($E'$) is the energy of the
incoming (outgoing) lepton. Then $Q^2$ in the same frame
involves the above energies and the scattering angle $\theta$. Therefore,
both $x$ and $Q^2$ depend  on $E'$ and $\theta$ and these
variables are not always independent of each other. Indeed,
there are experiments where
the $x$ -dependence of $g_1$ is measured at fixed $Q^2$; then $Q^2$ is
varied to another value and the $x$ -dependence is studied again. In this
case  $x$ and $Q^2$
are really independent variables. In the opposite case of fixed $w$ and
varying $Q^2$,
 Then $x =x(Q^2)$ and therefore these
variables are not exactly independent, so it is more convenient to use $w$
instead of $x$, as independent variable.
These examples show that using $w$ and $Q^2$
instead of the standard set $x,~Q^2$ could be more convenient
for $g_1$ . In particular, when $Q^2$ is
very small, using $x$ instead of $w$ becomes really inconvenient.
Nevertheless, the  $w$ -dependence of $g_1$ in the COMPASS
experiments predicted in Eq.~(\ref{gsser}) can be clearly extracted from
 the dependence   on $x$ and $Q^2$.
Indeed, the variable $\bar{x}$ defined in
Eq.~(\ref{shiftx}) can  be written in the following form:

\begin{equation}\label{zxq}
\bar{x} = \mu^2 x/Q^2 + x  \approx \mu^2 x/Q^2.
\end{equation}
Although $g_1$ depends on $x$ and on $Q^2$ at small $Q^2$  very
weakly,
its dependence on $x/Q^2$ is quite essential.
Indeed  Eq.~(\ref{gsser}) can be regarded as a sort of a new scaling law
where $g_1$  depends on the variable $x/Q^2$ only.

\section{Summary and outlook}

In the present paper we have presented an overview of our results on the spin structure
function $g_1$ at arbitrary $x$ and $Q^  2$.
We have  divided the whole kinematic region of $x$ and $Q^2$ into the set of four
regions \textbf{A}~\textbf{-}~\textbf{D} defined in Eqs.~(\ref{rega},~\ref{regb}-\ref{regd}) and considered $g_1$
in each of these regions. The region \textbf{A } is covered by the Standard Approach, based on the DGLAP evolution equations.
This is briefly discussed in Sect.~II. The application of the integral transforms to $g_1$ is given in Sect.~III.
In Sect.~IV we have  discussed in detail
the parametrization of $\alpha_s$ and shown  that the popular parametrization
$\alpha_s = \alpha_s (k^2_{\perp})$ is valid at  large $x$ only.
Otherwise it should be replaced by the
effective coupling $\alpha^{eff}_s$ defined in Eq.~(\ref{aeff}).  When $\mu^2$ obeys
Eq.~(\ref{mulambdapi}), $\alpha^{eff}_s$ can be approximated by $\alpha_s (k^2_{\perp}/\beta)$
(with $\beta$ being the longitudinal Sudakov variable)
and when, in addition, $x$ is large, it can be simplified down to $\alpha_s (k^2_{\perp})$.
According to Eq.~(\ref{mulambdapi}), the deviation of $\alpha^{eff}_s$
from $\alpha_s$ strongly depends on $\mu^2$. For example, when $\mu^2= 2.5$~GeV$^2$,
$\alpha_s(\mu^2)/\alpha^{eff}_s(\mu^2) \approx 0.9$ but very quickly
$\alpha_s(\mu^2)/\alpha^{eff}_s(\mu^2) \approx 0.5$ at $\mu^2 \approx 1$~GeV$^2$, which is a typical DGLAP starting point of the $Q^2$ -evolution  \\
The small-$x$ region \textbf{B},  where
the total resummation of the leading logarithms of $x$ is essential, was considered in Sect.~V.
We account there for the resummation of the leading logarithms
 by solving an Infrared Evolution Equations (IREE), so
in Sect.~V the essence of the method together with the technology of IREE was discussed and the
IREE for $g_1$ were obtained.
These IREE involve new anomalous dimensions and coefficient
functions. Explicit expressions for them were obtained in Sect.~VI.
Then the expression of Eq.~(\ref{gnsb}) for the non-singlet component
of $g_1$ in the region \textbf{B} was obtained in Sect.~VII, while in
Sect.~VIII  the result
 for the singlet $g_1$ in region \textbf{B} was given in Eq.~(\ref{gsb}).  \\
 Obviously, the impact of the total resummation of logarithms of $x$ is big at small $x$ and
becomes maximal at $x \to 0$, where the expressions in Eqs.~(\ref{gnsb},\ref{gsb})
behave  asymptotically as in Eqs.~(\ref{asscal},\ref{sas}).
The small-$x$ asymptotics of the non-singlet $g_1$ is
considered in detail in Sect.~IX while Sect.~X  contains the asymptotics of the singlet $g_1$.
Both asymptotic behaviors are of the Regge type. The estimates for their
the intercepts are given by Eqs.~(\ref{intns}) and (\ref{ints}).
The small-$x$ rise of $g_1$ predicted by Eqs.~(\ref{asscal},\ref{sas})
is much steeper than the well-known small-$x$ DGLAP prediction in Eq.~(\ref{asdglap}).
On the other hand, a numerical analysis shows that the use of the asymptotic formulae at
the presently available $x$ is not reliable, so Eqs.~(\ref{gnsb},\ref{gsb}) should not be replaced
by their asymptotic expressions of Eqs.~(\ref{asscal},\ref{sas}) in the region \textbf{B}.
The comparison of our results
 to the DGLAP -expressions for $g_1$, which  is
impossible without fixing the initial parton densities $\delta q$ and $\delta g$,  shows
that the impact of the resummation of the logarithms becomes essential for values smaller
$x \approx 10^{-2}$ 
In Sect.~XII,  by considering in detail a  standard DGLAP fit (\ref{fita}) for the initial parton densities, we
have shown that the singular factors in the fits mimic the total resummation of
the logarithms and provide the rise of  $g_1$ at small $x$ which is observed in
 the experimental data. On the other hand,
when the total resummation of the logarithms is taken into account, the singular factors
in the fits can be dropped. This  allows one to simplify the parametrization of the initial parton densities.. \\
The Reggeon structure in the two approaches has been discussed in Sect. XIII, However,
in the case of the SA those Reggeons are, in a sense, fictitious: they are generated by the fits
for the initial parton densities and because of that are present at any $x$ instead of
appearing in the asymptotic expressions at $x \to 0$.  \\
The total resummation of the small-$x$ logarithms is important in the region \textbf{B}, but also
non-logarithmic
contributions are quite essential in the DGLAP region \textbf{A}, where they are accounted for
to  NLO accuracy.  Then the interpolating
expressions for $g_1$ are presented in Eqs.~(\ref{gnsab}) and (\ref{gsab}).
On one hand, they almost coincide with $g_1^{DGLAP}$ in the region \textbf{A} and on the other
hand,
with Eqs.~(\ref{gnsb},\ref{gsb}) in the  region \textbf{B}, and at the same time do not require
the use of the singular parameterizations of the parton densities.\\
The small-$Q^2$ region \textbf{C} is absolutely beyond the reach of DGLAP. On the other hand, the analysis of the
Feynman diagrams contributing to $g_1$ shows that a shift of $Q^2$ allows us to extend
Eqs.~(\ref{gnsb},\ref{gsb}) into the region \textbf{C}. Similarly, Eqs.~(\ref{gnsab},\ref{gsab})
can be extended into the  region \textbf{D}.  Eventually we arrived at the Eqs.~(\ref{gnsd},\ref{gsd})
which are the interpolation expressions which can  describe $g_1$ in the whole region
$\textbf{A} \oplus \textbf{B} \oplus \textbf{C} \oplus \textbf{D}$.  This was the subject of Sects.~XIV-XVI. \\
This shift of $Q^2$,  given in Eq.~(\ref{shift}), inevitably causes the appearance of power $1/Q^2$ -corrections.
$Q^2$ power corrections were found earlier phenomenologically by confronting $g_1^{DGLAP}$ and the experimental data.
They were attributed to the  impact of  higher twists.
 In Sect.~XVII we argued that the role played by
 the higher twists can be estimated reliably only after accounting for
the pertubative power corrections. \\
Finally, in Sect.~XVIII we have used the small-$Q^2$ description of the singlet $g_1$ in Eq.~(\ref{gsd}) for the
interpretation
of the recent COMPASS data. First  we have shown that $g_1$ in the COMPASS kinematic
region does not depend on $x$, even at very small $x$. Then, we have suggested that
the COMPASS data are compatible with positive gluon densities . We also
 argued in favor of studying the dependence of $g_1$ on $2pq$  in the COMPASS experiment rather than on
$x$, in order to estimate the ratio $\delta g/\delta q$.

\section{Acknowledgments}
We are grateful to G.~Altarelli, A.V.~Efremov, S.~Jadach, W.~Schafer and O.V.~Teryaev for useful discussions.
The work is partly supported by Grant RAS 9C237,  Russian State Grant for Scientific School
RSGSS-3628.2008.2 and by an EU Marie-Curie Research Training Network under Contract
No. MRTN-CT-2006-035505 (HEPTOOLS).

\appendix

\section{Simplification of the color structure of the forward scattering amplitudes}

We consider below in more detail the color structure of the Born amplitude
$A^{Born}$ defined in Eq.~(\ref{aborn}). As the external partons
in $A^{Born}$ are quarks or gluons,

\begin{equation}\label{aborncolor}
 A^{Born} = - C^{(col)}4\pi\alpha_s  \frac{\bar{u}(-p_2) \gamma_{\mu}u(p_1)
 \bar{u'}(p_1) \gamma_{\mu}u'(-p_2)}{s + \imath \epsilon}
 \end{equation}
 where the $SU(3)$ -matrix $C^{(col)}$ describes the color structure of $A^{Born}$. When all external partons
 are quarks, $C^{(col)} = t^a t^a$, with $t^a~(a=1,..,8)$ being the $SU(3)$ -generators in the fundamental
(three-dimensional) representation; when the quarks are replaced by gluons, $t^a$ are replaced by the $SU(3)$
 -generators $T^a$ in the vector representation. Each of the initial and final
 color two-parton states in  Eq.~(\ref{aborncolor}) corresponds to a reducible representation
 of $SU(3)$ and can be expanded into a sum of irreducible states. It is convenient to do it in
 the $t$ -channel where the  amplitude $A^{Born}$ describes the
 quark-antiquark annihilation $q \bar{q} \to q' \bar{q}'$ and therefore the irreducible
 initial $q_i \bar{q}^j$ and final $q_p \bar{q}^q$ color states (with $i,j,p,q = 1,2,3$)
 are the singlet ($S$) and octet ($V$). The  initial $\textbf{3}\otimes\textbf{3}$
 color state  in the $t$ -channel
 state is $q_i \bar{q}^j$. It can be expanded
 into the sum of the singlet and octet, each one is the
 irreducible state:
 $\textbf{3}\otimes\textbf{3} = \textbf{1 }\oplus \textbf{8}$.
 We denote them $(q_i \bar{q}^j)_S$ and $(q_i \bar{q}^j)_V$ respectively. It
 can be done by applying the projection operators $P_S$ and $P_V$ to
 the quark-antiquark states:

 \begin{equation}\label{colsum}
 (q_i \bar{q}^j)_S = (P_S)_{ij'}^{i'j} q_{i'} \bar{q}^{j'},~~
 (q_i \bar{q}^j)_V = (P_V)_{ij'}^{i'j} q_{i'} \bar{q}^{j'}
 \end{equation}
 where
 \begin{equation}\label{psv}
 P_S)_{ii'}^{jj'} = \frac{1}{N} \delta_i^{i'} \delta^j_{j'},~~
 (P_V)_{ii'}^{jj'} = 2 (t^a)_i^{i'} (t^a)^j_{j'}~.
 \end{equation}
 Obviously, these operators are orthogonal to each other and the factors $1/N$
 (with $N=3$) and 2 in Eq.~(\ref{psv}) are introduced to guarantee
 the property $P^2 = P$ for each of $P_S$ and $P_V$. Let us notice
 that $||P_S||^2 = Tr[P^+_S P_S]~=~1,~~||P_V||^2 = Tr[P^+_V P_V] =2N C_F$ where
 $C_F = (N^2-1)/2N$. Applying the projection
 operators $P_S$ and $P_V$ to the color factor $C^{(col)}$ in Eq.~(\ref{aborncolor})
 allows us to write down $C^{(col)}$ as the sum of the scalar $C^{(col)}_S$
 and octet $C^{(col)}_V$ color factors:
 \begin{equation}\label{colexp}
 (C^{(col)})_{iq}^{jp} = C^{(col)}_S (P_S)_{iq}^{jp} + C^{(col)}_V (P_V)_{iq}^{jp}~,
 \end{equation}
 where
\begin{eqnarray}\label{csv}
&&C^{(col)}_S = \frac{(P_S)^{iq}_{jp}
(t^a)_i^j(t^a)_q^p}{||P_S||^2} = \frac{1}{N} Tr[t^a t^a] = C_F~,
\nonumber \\ &&C^{(col)}_V = \frac{(P_V)^{iq}_{jp}
(t^a)_i^j(t^a)_q^p}{||P_V||^2} = \frac{2 Tr[t^a t^b t^a t^b]}{2 N
C_F} = - \frac{1}{2N}~.
\end{eqnarray}
Substituting Eq.~(\ref{colexp}) into Eq.~(\ref{aborncolor}), we rewrite it as

\begin{equation}\label{abornexp}
A^{Born} = A^{Born}_S P_S + A^{Born}_V P_V~,
\end{equation}
with
\begin{eqnarray}\label{abornsv}
&&A^{Born}_S= - C^{(col)}_S 4\pi\alpha_s \frac{\bar{u}(-p_2)
\gamma_{\mu}u(p_1)
 \bar{u'}(p_1) \gamma_{\mu}u'(-p_2)}{s + \imath
 \epsilon}~, \nonumber \\
&&A^{Born}_V= - C^{(col)}_V 4\pi\alpha_s \frac{\bar{u}(-p_2)
\gamma_{\mu}u(p_1)
 \bar{u'}(p_1) \gamma_{\mu}u'(-p_2)}{s + \imath \epsilon}~.
\end{eqnarray}
When the external quarks in Eq.~(\ref{aborncolor}) are replaced by gluons,
generators $t^a$  in the factor $C^{(col)}$ are replaced by the SU(3) -generators $T^a$ in the vector
representation. It allows one to generalize Eq.~(\ref{abornsv}) to the gluons case, expanding
the initial $t$ -channel gluon state $\textbf{8} \otimes \textbf{8}$ into scalar and octet
(see Refs.~\cite{ber,egtsns} for detail). The projection operators $P^{(gg)}_S$
and $P^{(gg)}_V$ projecting the two-gluon
$t$ -channel state on the scalar and octet states are:
\begin{equation}\label{pggsv}
P^{(gg)}_S = \frac{1}{N^2-1}\,\delta_{a'b'}\delta_{ab}~,\qquad
P^{(gg)}_V = \frac{1}{N}(T_c)_{a'b'} (T_c)_{ab}~.
\end{equation}
Strictly speaking, the normalization for projector operators with the gluon-quark transitions
can be arbitrary but in order to match DGLAP it can be chosen as follows:

\begin{eqnarray}\label{pqgsv}
&&P^{(qg)}_S = \frac{1}{N^2 - 1}~\delta_i^j\delta_{ab}~,\qquad
P^{(qg)}_V = \frac{1}{N} (t_c)_i^j (T_c)_{ab}~, \\ \nonumber
&&P^{(gq)}_S = \frac{1}{N} \delta_{ab} \delta_{ij}~,\qquad\qquad\;
P^{(gq)}_V = 2 (T_c)_{ab}(t_c)_i^j~.
\end{eqnarray}
In Eqs.~(\ref{pggsv},\ref{pqgsv}) we have kept the notation $i,j$ for the quark color
states while $a,b,a',b'$ denote the gluon states.
In contrast to the quark-quark case, the expansion of $\textbf{8} \otimes \textbf{8}$
into the sum of the irreducible $SU(3)$ -representations
includes the singlet $\textbf{1}$, the antisymmetric $\textbf{8}_{\textbf{A}}$ and symmetric
$\textbf{8}_{\textbf{S}}$ octets and
 other contributions which cannot be organized out of the gluon fields and therefore
 can be left out, so for the gluons
$\textbf{8} \otimes\textbf{8} = \textbf{1}
\oplus \textbf{8}_{\textbf{A}} \oplus \textbf{8}_{\textbf{S}}$. The
 symmetric octet $\textbf{8}_{\textbf{S}}$ does not contribute to $g_1$ with the leading logarithmic accuracy
 (see Ref.~\cite{ber} for detail). An additional argument in favor of neglecting the amplitudes with
 high color dimensions is that they die out quickly with energy.

\section{Non-singlet contribution to the structure function $F_1$}

The technology for calculating the leading logarithmic contributions to $F^{NS}_1(x,Q^2)$
is quite similar to the one for $g^{NS}_1(x,Q^2)$ (see Refs.~\cite{egtsns} for detail).
Similarly to $g_1^{NS}$, $F^{NS}_1$ is expressed through
the forward Compton amplitude $T^{(+)}_{NS}$ in the following way:

\begin{equation}\label{fnst}
 F^{NS}_1 = \frac{1}{2\pi}\;\Im T^{(+)}_{NS}~.
 \end{equation}
The superscript $(+)$ in Eq.~(\ref{fnst}) stands for
the positive signature. Then it is convenient to define the Mellin amplitude
$F^{NS}_1(\omega,y)$ related to $T^{(+)}_{NS}$
through the Mellin transform (\ref{mellintf}).  As a consequence, the
IREE for $F^{NS}_1$ is almost identical to Eq.~(\ref{eqfnssl}):

\begin{equation}\label{eqfplus}
 \omega F^{NS}_1(\omega,y) + \frac{\partial F^{NS}_1 (\omega,y)}{\partial y}=
\frac{1}{8 \pi^2}\,(1 + \lambda_{qq} \omega)\,
L^{(+)}_{qq}(\omega)\, F^{NS}_1 (\omega,y)~.
\end{equation}
 The amplitude $L^{(+)}_{qq}(\omega)$ again corresponds to
the quark-quark scattering but its signature is now positive. It should be found independently.
Eq.~(\ref{eqfplus}) can be solved similarly to Eq.~(\ref{eqfnssl}). The
only difference between them is replacement of  $h_{NS}$ by $h^{(+)}$.  The amplitude $h^{(+)}$
obeys the following IREE:

\begin{equation}\label{eqhplus}
\omega h^{(+)} = b^{(+)} + (1 + \lambda_{qq} \omega)\, (h^{(+)})^2
.
\end{equation}
The difference between Eq.~(\ref{eqhplus}) and Eq.~(\ref{eqfnssl}) is the
inhomogeneous term $b^{(+)}$. It is expressed through $a_{qq}$ defined
in Eq.~(\ref{aik}) and $V^{(+)}_{qq}$:
\begin{equation}\label{bplus}
b^{(+)} = a_{qq} + V^{(+)}_{qq} .
\end{equation}
Similarly to $V_{qq}$ introduced in Eq.~(\ref{vik}), $V^{(+)}_{qq}$ is expressed
in terms of $m_{qq}$ defined in Eq.~(\ref{mik}) and a new quantity $D^{(+)}$ instead of
$D$:

\begin{equation}\label{vplus}
V^{(+)}_{qq} = m_{qq} D^{(+)}
\end{equation}
with
\begin{equation}
\label{dplus} D(\omega) = \frac{1}{2 b^2} \int_{0}^{\infty} d \rho
e^{- \omega \rho} \ln \big( (\rho + \eta)/\eta \big) \Big[
\frac{\rho + \eta}{(\rho + \eta)^2 + \pi^2} - \frac{1}{\rho +
\eta}\Big]~.
\end{equation}

So, the expression for $F^{NS}_1$ in region \textbf{B } is

\begin{equation}\label{fnsb}
F_1^{NS} (x, Q^2) = \frac{e^2_q}{2} \int_{- \imath \infty}^{\imath
\infty} \frac{d \omega}{2 \pi \imath} x^{- \omega} C^{(+)}_{NS}
(\omega) \delta q (\omega) e^{h^{(+)}_{NS}(\omega)
\ln(Q^2/\mu^2)}~.
\end{equation}
where
\begin{equation}\label{cnsplus}
C_{NS}^{(+)} = \frac{2 \omega}{\omega + \sqrt{\omega^2 -
B^{(+)}_{NS}(\omega)}}~,
\end{equation}
\begin{equation}\label{hnsplus}
h^{(+)}_{NS}(\omega) = (1/2) \Big[\omega - \sqrt{\omega^2 -
B^{(+)}_{NS}(\omega)}\Big]
\end{equation}
and
\begin{equation}\label{bnsplus}
B^{(+)}_{NS} = 4(1 + \lambda_{qq} \omega) b^{(+)}_{qq}~.
\end{equation}

The small- $x$ asymptotics of $F_1^{NS} (x, Q^2)$ is also of the
Regge type but the value of the intercept $\Delta^{(+)}_{NS}$ is
smaller than the one of $g_1^{NS}$~:

\begin{equation}\label{intplus}
\Delta^{(+)}_{NS} = 0.38~.
\end{equation}

\section{Convolution of two amplitudes}

Let us consider the $t$- channel
convolution of two amplitudes: $Q = A_1^{(p_1)} \bigotimes A_2^{(p_2)}$ of the
scattering amplitudes $A_1^{(p_1)}$ and $A_2^{(p_2)}$ where
$p_{1,2} = \pm$ stand for the signatures. It is convenient to
describe amplitudes $M_a^{(p_a)}$ in terms of the invariant
amplitudes $M_a$. For example, the invariant amplitude $M^{(\pm)}$
for the quark-antiquark forward annihilation $q(p_1) +
\bar{q}(p_2) \to q(p'_1) + \bar{q}(p'_2)$ are introduced as
follows:
\begin{equation}\label{am}
A^{(\pm)} = \frac{j_{\nu} j_{\nu}}{s} M^{(\pm)}
\end{equation}
where $j_{\nu}$ are the quark currents. We remind we use the Feynman gauge.

It is also convenient to use the asymptotics of the
Sommerfeld-Watson transform (often called the Mellin
representation) for each of those amplitudes in the following
form:
\begin{equation}\label{mellin}
M_a^{(p)}(s, \mu^2) = \int_{- \imath \infty}^{\imath \infty}
\frac{d \omega}{2 \pi \imath} \Big( \frac{
s}{\mu^2}\Big)^{\omega} \xi^{(p_a)}(\omega)
F_a^{(p_a)}(\omega)~,
\end{equation}
with $a = 1,2$~. The signature factor
\begin{equation}\label{sign}
\xi^{(\pm)}(\omega_a) = -\frac{e^{- \imath \pi \omega} \pm1}{2}
\approx \frac{(1 \pm1) + \imath \pi \omega}{2}
\end{equation}
and the transform inverse to Eq.~(\ref{mellin}) is
\begin{equation}\label{invmel}
F^{(\pm)}(\omega) = - \frac{1}{\pi\omega} \int_{\mu^2}^{\infty}
\frac{ds}{s} \Big( \frac{s}{\mu^2}\Big)^{-\omega} ~\frac{\Im_s M
\pm \Im_u M}{2}~.
\end{equation}

 Using Eqs.~(\ref{am},\ref{mellin}) and skipping the overall factor
$j_{\nu}j_{\nu}/s$  allows us to write the convolution $Q$ through
the invariant convolution $Q_{inv}$ as
\begin{equation}\label{invq}
Q = \frac{j_{\nu}j_{\nu}}{s} Q_{inv}^{p_1 p_2}
\end{equation}
with\footnote{The factor $\imath$ is the product of  the overall factor
$-\imath$ and $(\pm \imath)^2$ from the $t$-channel quark or gluon
propagators.}
\begin{eqnarray}\label{conv}
Q_{inv}^{p_1 p_2} =  \imath \int_{- \imath \infty}^{\imath \infty}
\frac{d \omega_1}{2 \pi \imath} \frac{d \omega_2}{2 \pi \imath}
\xi^{(p_1)}(\omega_1)\xi^{(p_2)}(\omega_2) f_1^{(p_1)}(\omega_1)
f_2^{(p_2)}(\omega_2) \int \frac{d^4 k}{16 \pi^4}\frac{2
k^2_{\perp}}{(k^2 - m^2 + \imath
\epsilon)^2}\Big(\frac{s_1}{|k^2|}
\Big)^{\omega_1}\Big(\frac{s_2}{|k^2|} \Big)^{\omega_2}
\frac{s}{s_1}\frac{s}{s_2}
\end{eqnarray}
where the factor $2 k^2_{\perp}$  appears as the result of
simplifying the spinor structure, $s_1 = 2p_1 k,~s_2 = 2p_2 k$.
Both of them are understood as $s_{1,2} + \imath \epsilon$.

For integration over $k$ in Eq.~(\ref{conv}) we use the Sudakov
variables (\ref{sud}):
\begin{eqnarray}\label{convpsi}
Q_{inv}^{p_1 p_2} =  \frac{\imath}{16 \pi^4} \int_{- \imath
\infty}^{\imath \infty} \frac{d \omega_1}{2 \pi \imath} \frac{d
\omega_2}{2 \pi \imath} \xi^{(p_1)}(\omega_1)\xi^{(p_2)}(\omega_2)
f_1^{(p_1)}(\omega_1) f_2^{(p_2)}(\omega_2)
\Psi^{(p_1,p_2)}(\omega_1, \omega_2, s/\mu^2)
\end{eqnarray}
with
\begin{equation}\label{psi}
\Psi^{(p_1,p_2)}(\omega_1, \omega_2, s/\mu^2) = \int d \alpha d
\beta d^2 k_{\perp} \frac{k^2_{\perp}}{(s\alpha\beta - k^2_{\perp}
+ \imath \epsilon)^2} \Big(\frac{s\alpha}{|k^2|}
\Big)^{\omega_1}\Big(\frac{s\beta}{|k^2|} \Big)^{\omega_2}
\frac{s}{s\alpha}\frac{s}{s\beta}~.
\end{equation}
Let us first integrate Eq.~(\ref{psi}) over $\alpha$. The
integration can be done in the complex plane by applying the
Cauchy formula. The singularities in the complex $\alpha$ -plane
are the double pole $s\alpha \beta - k^2_{\perp} + \imath \epsilon
= 0$ and the cut from the Mellin factor $(s\alpha)^{\omega_1}$.
The integration yields a non-zero result when the pole and the cut
have opposite imaginary parts. The imaginary part of the cut is
positive while the imaginary part of the pole is negative
provided $\beta>0$. When $\beta < 0$, both singularities have
positive imaginary parts and therefore the integration over
$\alpha$ yields zero. Closing up the integration contour in the
lower hemi-plane and taking the residue of the pole
\begin{equation}\label{pole}
\alpha = (k^2_{\perp} - \imath \epsilon)/s\beta
\end{equation}
we perform the integration over $\alpha$ and arrive at
\begin{eqnarray}\label{psiint}
&&\Psi^{(p_1,p_2)}(\omega_1, \omega_2, s/\mu^2) = -2\pi^2 \imath
\int_{\mu^2}^{s} \frac{d k^2_{\perp}}{k^2_{\perp}}
\Big(\frac{s}{k^2_{\perp}}
\Big)^{\omega_2}\int_{k^2_{\perp}/s}^{1}d \beta \beta^{\omega_2 -
\omega_1 - 1} = \\ \nonumber &&\frac{-2\pi^2 \imath}{\omega_2 -
\omega_1} \int_{\mu^2}^{s} \frac{d k^2_{\perp}}{k^2_{\perp}}\Big[
\Big(\frac{s}{k^2_{\perp}} \Big)^{\omega_2}\ -
\Big(\frac{s}{k^2_{\perp}} \Big)^{\omega_1} \Big] =\frac{-2\pi^2
\imath}{\omega_2 - \omega_1}\Big[
\frac{1}{\omega_2}\Big(\frac{s}{\mu^2} \Big)^{\omega_2}\ -
\frac{1}{\omega_1} \Big(\frac{s}{\mu^2} \Big)^{\omega_1} \Big]
\end{eqnarray}
Therefore we obtain the following expression for $Q_{inv}$:

\begin{equation}\label{convint}
Q_{inv}^{p_1 p_2} = \frac{1}{8\pi^2} \int_{- \imath
\infty}^{\imath \infty} \frac{d \omega_1}{2 \pi \imath} \frac{d
\omega_2}{2 \pi \imath} \xi^{(p_1)}(\omega_1)\xi^{(p_2)}(\omega_2)
\frac{f_1^{(p_1)}(\omega_1) f_2^{(p_2)}(\omega_2)}{\omega_2 -
\omega_1}
\Big[\frac{1}{\omega_2}\Big(\frac{s}{\mu^2}\Big)^{\omega_2}
-\frac{1}{\omega_1}\Big(\frac{s}{\mu^2}\Big)^{\omega_1}\Big]~.
\end{equation}
Eq.~(\ref{convint}) involves two integrations, however one of them
can be done easily. The integration lines over $\omega_{1,2}$ in
Eq.~(\ref{convint}) are parallel to the imaginary axes and lie to
the right of the rightmost singularities of $f_{1,2}$. Let us
assume that additionally to it

\begin{equation}\label{lines}
0 < \Re \omega_1  < \Re \omega_2~.
\end{equation}
The opposite case $\Re \omega_1  > \Re \omega_2$ can be discussed
similarly. The integrand of Eq.~(\ref{convint}) includes two
similar terms in the squared brackets. Let us focus on integrating
the first term, $\big(s/\mu^2\big)^{\omega_2}$ and let us
integrate this part of Eq.~(\ref{convint}) with respect to
$\omega_1$. In this case closing up the $\omega_1$ -integration
contour to the left involves accounting for singularities of
$f_1(\omega_1)$. On the contrary, when we close up the $\omega_1$
-contour  to the right, the only singularity inside the contour is
the pole $1/(\omega_2 - \omega_1)$, so we can do this integration
without considering $f_1$, just by taking the residue at $\omega_1
= \omega_2$. At the same time, integrating the remaining,
proportional to $\big(s/\mu^2\big)^{\omega_1}$ part of
Eq.~(\ref{convint}) with respect to $\omega_1$ yields zero.
Therefore Eq.~(\ref{convint}) is reduced to the simpler form:

\begin{equation}\label{convres}
Q_{inv}^{p_1 p_2} = \frac{1}{8\pi^2} \int_{- \imath
\infty}^{\imath \infty} \frac{d \omega}{2 \pi \imath}
\big(s/\mu^2\big)^{\omega} \xi^{(p_1)}(\omega)\xi^{(p_2)}(\omega)
\;\frac{f_1^{(p_1)}(\omega) f_2^{(p_2)}(\omega)}{\omega}
\end{equation}
Obviously,
\begin{equation}\label{xis}
\xi^{(p_1)}(\omega)\xi^{(p_2)}(\omega) \approx (1/4)[(1 + P_1 P_2)
- \imath \pi\omega (P_1 + P_2)]  =
\xi^{(p_1)}(\omega)\,\delta_{p_1p_2}
\end{equation}
where $P_{1,2} = \pm 1$~. It means that the leading contribution
to $Q_{inv}^{p_1 p_2}$ is diagonal in the signatures. Strictly
speaking, this should be checked in advance, with using the
Sommerfeld-Watson transform where analytical properties of the
involved amplitudes are explicitly accounted for.


\section{Off-shell invariant amplitude $M$ in Eq.~(\ref{em})}

The invariant amplitude $M$ in Eq.~(\ref{em}) is a generic notation for the following
invariant off-shell amplitudes: $M^{NS}(2pk,k^2)$
contributing to the Bethe-Salpeter for $g_1^{NS}$ and the
flavor singlet amplitudes $M_{qq}(2pk,k^2),~M_{qg}(2pk,k^2)$.
They are related to the amplitudes $A_{qq}(2pk,k^2),~A_{qg}(2pk,k^2)$ of the forward
quark-quark and quark-gluon scattering:
\begin{eqnarray}
\label{amoff} A_{qq} &=& - \frac{\bar{u}(-k) \gamma_{\lambda} u(p)
\bar{u}(p) \gamma_{\lambda} u(-k)}{(p+k)^2 + \imath \epsilon}
\;M_{qq}(2pk,k^2)~,   \\\nonumber A_{qg} &=& -  e_{\lambda}
e^*_{\mu} \frac{\bar{u}(-k) \gamma_{\lambda} (\hat{p} - \hat{k})
\gamma_{\mu} u(-k)}{(p + k)^2 + \imath \epsilon}\;
M_{qg}(2pk,k^2)~.
\end{eqnarray}
The outgoing momenta $-k$ in Eq.~(\ref{amoff}) are assigned to the final (upper) off-shell
quarks whereas the initial quarks and gluons
have momenta $p$~; $e_{\lambda}$ and $e_{\mu}$ are the polarization vectors of the gluons.
We remind that the amplitudes $M_{qq}(2pk,k^2),~M_{qg}(2pk,k^2)$ are off-shell and therefore they differ from the
amplitudes $M_{ik}$ introduced
in Eq.~(\ref{ireetint}): $M_{qq}$ and
$M_{qg}$ logarithmically depend on two arguments, $2pk$ and $k^2$, i.e.
\begin{equation}\label{marg}
M_{qq}= M_{qq} (\rho, z);~~M_{qg} = M_{q,g} (\rho, z)
\end{equation}
where $\rho = \ln(2pk/\mu^2),~z= \ln (k^2/\mu^2)$~.

It is convenient to introduce the Mellin amplitudes $\varphi_{ik} (\omega, z)$ conjugate
to $M_{ik} (\rho, z)$ through Eq.~(\ref{mellintf}).
The IREE for $\varphi_{ik} (\omega, z)$ is quite similar to Eqs.~(\ref{eqfns},\ref{eqfs}),
and  we put  here  all $\lambda_{ik} = 0$ for the sake of simplicity:

\begin{eqnarray}\label{eqfic}
&&\frac{\partial \varphi^{NS}}{\partial z} + \omega \varphi^{NS} =
\frac{1}{8 \pi^2} \varphi^{NS} L_{qq} (\omega)~, \\ \nonumber
&&\frac{\partial \varphi_{qq}}{\partial z} + \omega \varphi_{qq} =
\frac{1}{8 \pi^2} \varphi_{qq} L_{qq} (\omega) + \frac{1}{8 \pi^2}
\varphi_{qg} L_{gq} (\omega)~, \\ \nonumber
&&\frac{\partial\varphi_{qg}}{\partial z} + \omega \varphi_{qg} =
\frac{1}{8 \pi^2} \varphi_{qq} L_{qg} (\omega) + \frac{1}{8 \pi^2}
\varphi_{qg} L_{gg} (\omega)~.
\end{eqnarray}
The amplitudes $L^{NS},~L_{ik}$ in Eq.~(\ref{eqfic}) are on-shell, so in accordance with
Eq.~(\ref{fl}) they can be expressed in terms of
$h^{NS}(\omega) = (1/8 \pi^2) L^{NS}(\omega),~
h_{ik}(\omega) = (1/8 \pi^2) L_{ik}(\omega)$ which are obtained in  Eqs.~(\ref{hns},\ref{hik}).
General solutions to the linear equations (\ref{eqfic}) can easily be found (cf Eq.~(\ref{gengs})):
\begin{eqnarray} \label{ficgen}
&&\varphi^{NS} =  \Phi^{NS} (\omega)e^{z [-\omega + h^{NS}]}~,\\
\nonumber &&\varphi_{qq} = \Psi_1 (\omega) e^{- \omega z + z
\Omega_{(+)}} + \Psi_2 (\omega) e^{- \omega z + z \Omega_{(-)}}~,
\\ \nonumber &&\varphi_{qg} = \Psi_1 (\omega) \frac{X +
\sqrt{R}}{2 h_{gq}} e^{- \omega z + z \Omega_{(+)}} + \Psi_2
(\omega)\frac{X - \sqrt{R}}{2 h_{gq}} e^{- \omega z + z
\Omega_{(-)}}~,
\end{eqnarray}
with $\Omega_{(\pm)},~X$ and $R$ defined in Eqs.~(\ref{omegapm}, \ref{x},\ref{r}) respectively
whereas $\Psi_{1,2}$ should be specified. Therefore, $E(2pk.k^2)$ used in Eq.~(\ref{bsregc})
can be any of $E^{NS}(2pk.k^2),~E_{qq}(2pk.k^2),~E_{qg}(2pk.k^2)$ given by the following
expressions:

\begin{eqnarray} \label{e}
&&E^{NS}(2pk.k^2) = \int_{- \imath \infty}^{ \imath \infty}
\frac{d \omega}{2 \pi \imath} \Big(\frac{2pk}{k^2}\Big)^{\omega}
\omega \Phi^{NS} (\omega)e^{z h^{NS}}~, \\ \nonumber
&&E_{qq}(2pk.k^2) = \int_{- \imath \infty}^{ \imath \infty}
\frac{d \omega}{2 \pi \imath} \Big(\frac{2pk}{k^2}\Big)^{\omega}
\omega \Big[\Psi_1 (\omega) e^{z \Omega_{(+)}} + \Psi_2 (\omega)
e^{z \Omega_{(-)}} \Big]~,\\ \nonumber &&E_{qg}(2pk.k^2) = \int_{-
\imath \infty}^{ \imath \infty} \frac{d \omega}{2 \pi \imath}
\Big(\frac{2pk}{k^2}\Big)^{\omega} \omega \Big[\Psi_1 (\omega)
\frac{X + \sqrt{R}}{2 h_{gq}} e^{z \Omega_{(+)}} + \Psi_2
(\omega)\frac{X - \sqrt{R}}{2 h_{gq}} e^{z \Omega_{(-)}}\Big]~.
\end{eqnarray}

In order to specify $\Psi_{1,2}$, we use the obvious
matching condition:
\begin{equation}\label{matchpsi}
\varphi^{NS} (\omega, z =0) = 8 \pi^2
h^{NS}(\omega)~,\qquad\varphi_{qq} (\omega, z =0) = 8 \pi^2
h_{qq}(\omega)~,\qquad\varphi_{qg}(\omega, z =0) = 8 \pi^2
h_{qg}(\omega)~.
\end{equation}
It immediately fixes $\varphi^{NS}$:
\begin{equation}\label{phiins}
\varphi^{NS} = 8 \pi^2 h^{NS}(\omega) e^{z [-\omega + h^{NS}(\omega)]}
\end{equation}
and leads to the explicit expressions for $\Psi_{1,2}$:
\begin{equation}\label{psi12}
\Psi_1 = 8 \pi^2 \frac{\big[2 h_{qg}h_{gq} - h_{qq}(h_{gg} -
h_{qq} - \sqrt{R})\big]}{2 \sqrt{R}}~,\qquad\Psi_2 = 8 \pi^2
\frac{\big[- 2 h_{qg}h_{gq} + h_{qq}(h_{gg} - h_{qq} +
\sqrt{R})\big]}{2 \sqrt{R}}~.
\end{equation}

\section{Calculating the small-$x$ asymptotics of the non-singlet $g_1$.}

Eq.~(\ref{gnsb}) for $g_1^{NS}1(x,Q^2)$ in region \textbf{B} can be written as follows:
\begin{equation}\label{gnsbphi}
g_1^{NS}(x,Q^2) = \int_{- \imath \infty}^{\imath \infty} \frac{d \omega}{2 \pi \imath} e^{\Phi(\omega, x, Q^2)}
\end{equation}
where the phase $\Phi$ is

\begin{equation}\label{phi}
\Phi(\omega, x, Q^2) = \omega \ln (1/x) + \ln C_{NS}(\omega) + y h_{NS}(\omega) =
\omega \xi + \ln C_{NS}(\omega) - \frac{y}{2} \sqrt{\omega^2 - B_{NS}(\omega)}
\end{equation}
where we have used the expression (\ref{hns}) and denoted $\xi = \ln (w/\sqrt{Q^2 \mu^2})$. We remind that $y = \ln (Q^2/\mu^2)$.
 We are going to calculate the asymptotics of $g_1^{NS}$ at $x \to 0$ and fixed $Q^2$, i.e. at
 fixed $Q^2$ and $w \to \infty$. The standard way to calculate asymptotics is to apply the
saddle-point method to Eq.~(\ref{gnsbphi}). According to it,
\begin{equation}\label{saddlephase}
g_1^{NS} \sim \Pi_{NS}(\omega_0, w,Q^2) e^{\Phi_0(\omega_0, w, Q^2)} ~,
\end{equation}
with the stationary phase $\Phi_0 = \Phi(\omega_0, w, Q^2 )$ and the stationary point $\omega_0$
is defined from the requirement $d \Phi/d \omega = 0$, i.e. $\omega_0$ is a solution to the equation
\begin{equation}\label{sadpointeq}
\xi + \frac{C'_{NS}(\omega)}{C_{NS}(\omega)} - \frac{y}{4} \frac{(2 \omega - B'_{NS}(\omega))}{\sqrt{\omega^2 - B_{NS}(\omega)}} = 0~.
\end{equation}

Obviously, Eq.~(\ref{sadpointeq}) can have much more the one solution. In this case $\omega_0$ is the
solution with the largest $\Re \omega$. It is often called the rightmost stationary point. Substituting the
explicit expressions (\ref{cns}) for $C_{NS}$, we transform Eq.~(\ref{sadpointeq}) into

\begin{equation}\label{sadpointeq1}
\xi \omega B_{NS} \sqrt{\omega^2- B_{NS}} = (B_{NS} - \omega B'_{NS}/2) (\omega - \sqrt{\omega^2- B_{NS}}) +
\frac{y}{2} \omega B_{NS}(\omega - B'_{NS}/2)
~.
\end{equation}
Obviously this equation cannot be solved analytically. The analytical solution can be found for the particular
case when $B$ does not depend on $\omega$ (it corresponds to the case of fixed $\alpha_s$) and $y=0$.
In this case Eq.~(\ref{sadpointeq}) can be reduced to the
algebraic equation
\begin{equation}\label{bfixeq}
\omega^4 + (2/\rho) \omega^3 -  \omega^2 B_{NS} - (2 B_{NS}/\rho) \omega - (B_{NS}/\rho^2) = 0
\end{equation}
where $\rho = \ln (w/\mu^2)$. Eq.~(\ref{bfixeq}) has four roots which can be found with using the
known from the literature Ferrari formulae but only two of them, namely
$\omega = \pm \sqrt{B_{NS}}$ do not go to zero when $w \to \infty$. Obviously, in this case
the rightmost root is $\omega_0 = \sqrt{B_{NS}}$. It is easy to make this conclusion,
without solving Eq.~(\ref{bfixeq}). Indeed, Eq.~(\ref{bfixeq}) can be written as
\begin{equation}\label{bfixeq1}
\omega^2 (\omega^2 - B_{NS}) + (2/\rho)\omega( \omega^2 -  B_{NS}) - (B_{NS}/\rho^2) = 0~.
\end{equation}
When the terms $\sim 1/\rho^2$ and $\sim 1/\rho$ are dropped, Eq.~(\ref{bfixeq1}) can be solved
immediately and the rightmost root can easily be found.  Applying the same arguments allows one
to solve Eq.~(\ref{sadpointeq}) at $w \to 0$ drives us to conclude that the rightmost
and non-vanishing at $w \to \infty$ root of
Eq.~(\ref{sadpointeq})
does not depend on $y$ and it can be found as the rightmost root of the much simpler equation
$\omega^2 = B_{NS}(\omega)$ as is stated in Eq.~(\ref{eqbrns1}). This leads to the Regge
asymptotics of Eq.~(\ref{asscal}) quite different from the well-known DGLAP asymptotics
(\ref{asdglap}).  Let us notice that, in the perfect agreement with the concepts of the
phenomenological Regge theory, this root corresponds
to the branching point singularity of Eq.~(\ref{sadpointeq}).

\section{The DGLAP small-$x$ asymptotics}
Let us remind how the DGLAP asymptotics of $g_1^{NS}$ in Eq.~(\ref{asdglap}) was obtained.
As this topic is well-known, for the sake of simplicity we consider $g_{1~DGLAP}^{NS}$ with the
LO accuracy. When the singular term in $\delta q$ is absent, the small-$x$ asymptotics of
$g_{1~DGLAP}^{NS}$ can also be obtained with the saddle-point method.
Similarly to Eq.~(\ref{gnsbphi}), $g_{1~DGLAP}^{NS}$ can be written as

\begin{equation}\label{gnsdglapphi}
g_{1~DGLAP}^{NS}(x,Q^2) = \int_{- \imath \infty}^{\imath \infty} \frac{d \omega}{2 \pi \imath}
e^{\Phi_{DGLAP}(\omega, x, Q^2)}
\end{equation}
where the phase $\Phi_{DGLAP}$ is

\begin{equation}\label{phidglaptot}
\Phi_{DGLAP} = \omega \ln(1/x) + \int_{\mu^2}^{Q^2} \frac{d k^2_{\perp}}{k^2_{\perp}}
\frac{\alpha_s(k^2_{\perp})}{2 \pi} \gamma^{(0)},
\end{equation}
with $\gamma^{(0)}$ being given by Eq.~(\ref{gnslo}).
The bulk of the integral in Eq.~(\ref{gnsdglapphi}) comes from the region of $\omega$
obeying
\begin{equation}\label{smallomega}
\omega \ln(1/x) \lesssim 1
\end{equation}
because the factor $e^{\omega \ln(1/x)}$ in Eq.~(\ref{gnsdglapphi})
strongly oscillates beyond this region.
So, the values of $\omega$ mainly contributing to the integral become small when $x \to 0$.
As a consequence, the most important term in $\gamma^{(0)}$ is now the singular in $\omega$
term $A_{DGLAP}(Q^2)/\omega $,
with
\begin{equation}\label{agdlap1}
A_{DGLAP}(Q^2) = \int_{\mu^2}^{Q^2} \frac{d k^2_{\perp}}{k^2_{\perp}}
\frac{\alpha_s(k^2_{\perp}) C_F}{2 \pi}~.
\end{equation}
Therefore, approximately

\begin{equation}\label{phidglap}
\Phi_{DGLAP} \approx \omega \ln(1/x) + A_{DGLAP}(Q^2)/\omega~.
\end{equation}

The equation for the stationary point is
\begin{equation}\label{dglapstatpoint}
\Phi'_{DGLAP} = \ln(1/x) - A_{DGLAP}(Q^2)/\omega^2 = 0~,
\end{equation}
which leads to the stationary point

\begin{equation}\label{dglapstatpoint}
\omega_0^{DGLAP} = \sqrt{A_{DGLAP}(Q^2)/\ln(1/x)}
\end{equation}
and eventually to the DGLAP
-asymptotic given in Eq.~(\ref{asdglap}). Contrary to the case considered in Appendix E,
the DGLAP stationary point depends on $\ln(1/x)$. Eq.~(\ref{dglapstatpoint}) shows that
the $Q^2$ -dependence in the DGLAP asymptotics follows from the DGLAP parametrization
$\alpha_s = \alpha_s (k_{\perp}^2)$ and mostly from keeping $Q^2$ as the upper limit of
the integration in Eq.~(\ref{agdlap1}). The latter takes place because of the use the DGLAP
ordering (\ref{dglapord}). However in the small-$x$ region the ordering (\ref{dglapord})
becomes unreliable and should be replaced by the ordering of Eq.~(\ref{dlord}) where
the upper limit is $w$. Obviously,
the $Q^2$ -dependence in Eq.~(\ref{dglapstatpoint}) vanishes after replacing $Q^2$ by $w$.
We remind that the DGLAP asymptotics (\ref{asdglap}) can be obtained only under the assumption that
the initial parton densities are not singular at $x \to 0$ otherwise the
asymptotics~(\ref{dglapstatpoint}) is changed for the Regge asymptotics~(\ref{leadpol}).

\section{The small-$x$ asymptotics of $g_1^{NS}$ with the truncated series for the
coefficient functions and anomalous dimensions.}

We consider here the case where the coefficient functions and anomalous dimensions are
calculated in high orders in $\alpha_s$, however without the total resummation of
those contributions. According to Eq.~(\ref{smallomega}) the essential values of $\omega$ in
Eq.~(\ref{gnsdglapphi}) are small at $x \to 0$, so the most important terms in
expressions for the non-singlet coefficient functions and anomalous dimensions
are the most singular terms in $\omega$, i.e. the double-logarithmic contributions.
They can be obtained, expanding Eqs.~(\ref{hns}, \ref{cns}) into series absolutely
in the same way as was done in
Eqs.~(\ref{cnsser}, \ref{gnsser}). The expansion for the coefficient function in expressions
(\ref{gnsb}) and (\ref{gnsdglap}) have the same form
\begin{equation}\label{cnsser1}
C_{NS} = 1 + \frac{a}{4 \omega^2} + 2 \Big(\frac{a}{4 \omega^2}\Big)^2 +
5 \Big(\frac{a}{4 \omega^2}\Big)^3 +...
\end{equation}
however, with different $a$. For $C_{NS}$ of Eq.~(\ref{gnsb})
\begin{equation}\label{adl}
a = B_{NS}(\omega),
\end{equation}
 with
 $B_{NS}$ given by Eq.~(\ref{bns}). Alternatively, when the DL contributions to
$C_{NS}$ are calculated in the
DGLAP framework, $a = a_{DGLAP}$:
\begin{equation}\label{adldglap}
 a_{DGLAP}= \alpha_s(Q^2) C_F/(2\pi)~.
\end{equation}
In contrast, the
expansions for the exponent in Eq.~(\ref{g1nsdglap}) involves integrations of
$\alpha_s(k^2_{\perp})$ and therefore those two cases look quite different.
As this difference
it is not essential for the topic we consider here, we will use the approximation of
fixed QCD coupling for the DGLAP description of $g_1^{NS}$. In this case
the most singular, i.e. DL contributions to the DGLAP expression for
the anomalous dimension of $g_1^{NS}$ can easily be
obtained from the series for $H_{NS}$. By doing so, we arrive at the following series for
the exponent in Eq.~(\ref{gnsb}):
\begin{equation}\label{hnsexp}
\Gamma^{DL} \equiv y H_{NS} = y\Big[\frac{a}{4 \omega} + \frac{a^2}{16 \omega^3} +
\frac{a^3}{32 \omega^5}\Big] + ...,
\end{equation}
with $a$ defined in Eq.~(\ref{adl}) and $y= \ln(Q^2/\mu^2)$, whereas in the DGLAP case the
DL contribution to the exponent in Eq.~(\ref{g1nsdglap}) can again be obtained
with replacement $a$ by $a_{DGLAP}$:
\begin{equation}\label{gnsexp}
\Gamma_{DGLAP}^{DL} \equiv \int^{Q^2}_{\mu^2}
\frac{d k^2_{\perp}}{k^2_{\perp}}\gamma (\omega, \alpha_s)   \frac{a_{DGLAP}y}{4\omega} +
c_2 \frac{a^2_{DGLAP}y^2}{ \omega^2} +
c_3 \frac{a^3_{DGLAP}y^3}{\omega^3}  + ...
\end{equation}
with $c_2,~c_3$ being numerical factors.
Therefore, in the $n$-th order in $\alpha_s$
the most singular contribution to $C_{NS}$
can be written as follows:
\begin{equation}\label{leadc}
C_{NS}^{(n)} = c_{(n)} \frac{a^n}{\omega^{2n}} + O(1/\omega^{2n-1})~
\end{equation}
where $c_{(n)}$ is a numerical factor. It can be obtained with further expansion of
Eq.~(\ref{cns}) into series. Similarly, the most singular contributions, $\Gamma^{DL~n}$ and
$\Gamma^{DL~n}_{DGLAP}$ are

\begin{equation}\label{leadgamma}
\Gamma^{(n)~DL} = y \tilde{c}_{n} \frac{a^n}{\omega^{2n-1}}~,
\Gamma^{(n)~DL}_{DGLAP} = c_{n} \frac{a^n_{DGLAP}y^n}{\omega^{n}}~,
\end{equation}
with $ \tilde{c}_{n},~c_n$ be numerical factors.

The phase $\Phi$ in the Mellin integrals (\ref{gnsbphi}) and (\ref{gnsdglapphi}) can now
be written as follows:
\begin{eqnarray}\label{phihigh}
\Phi &\approx& \omega \zeta + \ln C_{NS} +
y \tilde{c}_{n} \frac{a^n}{\omega^{2n-1}} ~, \\ \nonumber
\Phi_{DGLAP} &\approx& \omega \ln(1/x) + \ln C_{NS}^{DGLAP} +
c_{n} \frac{a^n_{DGLAP}y^n}{\omega^{n}} ~,
\end{eqnarray}
where we have denoted $\zeta = (1/2)\ln (Q^2/(x^2 \mu^2))$.
Let us first consider the asymptotics of $g_1^{NS}$ at $Q^2 \sim \mu^2$. In this case the last term in
each of the equations in (\ref{phihigh}) is zero and the stationary point is determined from the following equation:
\begin{equation}\label{eqstpointqzero}
\zeta + \frac{C'_{NS}}{C_{NS}} = 0.
\end{equation}
Obviously, this equation does not have solutions leading to the Regge behavior of $g_1^{NS}$ because
all terms in Eq.~(\ref{cnsser1}) are positive. \\
Let us consider now the case of large $Q^2$: $Q^2 \gg \mu^2$.
The equations for the stationary points of the phases in Eq.~(\ref{phihigh}) are:
\begin{eqnarray}\label{stpointtrunk}
\Phi' &=& \zeta +  \frac{C'_{NS}}{C_{NS}} -
y \tilde{c}^{(n)}(2n-1) \frac{a^n}{\omega^{2n}} =0 ~, \\ \nonumber
\Phi'_{DGLAP} &=& \ln(1/x) + \frac{C_{NS}^{'~DGLAP}}{C_{NS}^{DGLAP}} -
c_n n \frac{a^n_{DGLAP}y^n}{\omega^{n+1}} = 0~.
\end{eqnarray}

The use of Eq.~(\ref{leadc}) for the coefficient function makes easy to see that the term $C'_{NS}/C_{NS}$
and $C_{NS}^{'~DGLAP}/C_{NS}^{DGLAP}$ are
proportional to $1/\omega$, so these terms are much less singular
than the last terms in Eq.~(\ref{stpointtrunk}) and therefore they can be dropped. After that
solving Eq.~(\ref{stpointtrunk}) is easy and we arrive at the following
approximate expression for the stationary point:
\begin{equation}\label{stpointq}
\omega_0 \approx \big((2n -1) {c}^{(n)} a^{(n)} y/\zeta\big)^{1/2n},~~~\omega_0^{DGLAP} \approx
\big(n c_n a^{n}_{DGLAP} y^n/\ln(1/x))\big)^{1/(n+1)}
\end{equation}
and leads to the following asymptotics:
\begin{equation}\label{asq}
g_1^{NS} \sim \exp\Big[\Big(a^{(n)}\Big)^{1/2n} y^{1/2n} \zeta^{(1 - 1/2n)}\Big]~,
~~~g_1^{NS~DGLAP} \sim \exp\Big[\Big(a^n_{DGLAP}\Big)^{1/(n+1)} y^{1/(n+1)} \big(\ln(1/x)
\Big)^{(1 - 1/(n+1))}\Big]
\end{equation}
Obviously, Eq.~(\ref{asq}) coincides with the LO DGLAP asymptotics (\ref{asdglap}) at $n=1$.
Eq.~(\ref{asq}) demonstrates explicitly that the asymptotics of
$g_1^{NS~DGLAP}$ always depends on $Q^2$ and the Regge behavior of $g_1^{NS}$
and $g_1^{NS~DGLAP}$ cannot
be achieved
at any fixed $n$. On the other hand, the Regge behavior of $g_1^{NS}$ and $g_1^{NS~DGLAP}$
is approached closer and closer when $n$ grows, and is eventually achieved
when the total resummation is performed. It is interesting to notice that the "intercept" of
$g_1^{NS~DGLAP}$ in this case could depend on $Q^2$ through the $Q^2$ -dependence of $a^{(n)}_{DGLAP}$.
Such a dependence originates from
the $Q^2$ -dependence of $\alpha_s$. However, we
have shown in Sect.~IV that the parametrization
$\alpha_s = \alpha_s (Q^2)$ should not be used at small $x$.

\end{document}